\documentclass[a4paper,nobibnotes,nofootinbib,preprintnumbers]{revtex4}
\usepackage{amssymb}
\usepackage{amsmath}
\usepackage{epsfig}
\newcommand{\be}{\begin{equation}}
\newcommand{\ee}{\end{equation}}
\newcommand{\bea}{\begin{eqnarray}}
\newcommand{\eea}{\end{eqnarray}}

\newcommand{\bi}{\begin{itemize}}
\newcommand{\ei}{\end{itemize}}

\newcommand{\xp}{x_{I\!\!P}} 
\newcommand{\g}{\gamma}
\newcommand{\f}{\frac}

\newcommand{\cN}{{\cal N}}

\begin{document}
\title{Systematic Analysis of Scaling Properties in Deep Inelastic Scattering}
\author{Guillaume Beuf}\email{guillaume.beuf@cea.fr}
\affiliation{Institut de physique th{\'e}orique, CEA/Saclay, 91191 
Gif-sur-Yvette cedex, France
\\URA 2306, unit{\'e} de recherche associ{\'e}e au CNRS}
\author{Robi Peschanski}\email{robi.peschanski@cea.fr}
\affiliation{Institut de physique th{\'e}orique, CEA/Saclay, 91191 
Gif-sur-Yvette cedex, France
\\URA 2306, unit{\'e} de recherche associ{\'e}e au CNRS}
\author{Christophe Royon}\email{christophe.royon@cea.fr}
\affiliation{IRFU/Service de Physique des Particules, CEA/Saclay, 91191 
Gif-sur-Yvette cedex, France}
\author{David \v S\'alek}\email{salekd@mail.desy.de}
\affiliation{Institute of Particle and Nuclear Physics, Charles University, Prague, 
Czech Republic}


\begin{abstract}
Using the ``Quality Factor'' (QF) method, we analyse the scaling properties
of deep-inelastic processes
at HERA and fixed target experiments for $x\!\le \!10^{-2}.$ We look for scaling formulae of the form $\sigma 
^{\gamma^*p}(\tau),$ where $\tau (L\!=\!\log Q^2,Y)$ is a scaling variable 
suggested by the asymptotic properties of QCD evolution 
equations with rapidity $Y$. We consider four cases: ``Fixed Coupling'', 
corresponding to  the original  geometric scaling proposal and motivated by the  
asymptotic properties of the 
Balitsky-Kovchegov (BK) equation with {\it fixed} QCD coupling constant, two 
versions ``Running Coupling I,II'' of the scaling suggested by the BK 
equation with {\it running} coupling, and ``Diffusive Scaling'' suggested by the 
QCD evolution equation with Pomeron loops. The Quality Factors, quantifying  the 
phenomenological validity of the candidate scaling variables, are fitted on the 
total and 
DVCS cross-section data from HERA and predictions are made for the  elastic vector-meson 
and for the diffractive cross-sections at fixed small $\xp$ or $\beta.$ 
The first three scaling formulae have comparably good  QF while the fourth one 
is disfavored. Adjusting  initial conditions gives a significant  
improvement of the ``Running Coupling  II'' scaling.
\end{abstract}
\maketitle
\section{Introduction}
Geometric scaling \cite{Stasto:2000er} is a remarkable empirical 
property 
verified by data on high energy deep inelastic scattering (DIS) $i.e.$
virtual photon-proton cross-sections. 
It was realized that one can  represent with 
reasonable 
accuracy the cross section $\sigma^{\gamma^*p}$ by the formula
$\sigma^{\gamma^*p}(Y,Q)=\sigma^{\gamma^*}(\tau)\  ,$
where $Q$ is the virtuality of the photon,
$Y$ the total rapidity in the ${\gamma^*}$-proton system and 
\begin{equation}
\tau = \log Q^2-\log Q_s^2(Y) =  \log Q^2-\lambda 
Y\  ,
\label{tau}
\end{equation}
is the scaling variable. The value $\lambda \sim 0.3$ confirms the value found within
the Golec-Biernat and W\"usthoff model \cite{Golec-Biernat:1998js} 
where 
geometric scaling was explicitely used for the parametrization.

The scaling using the variable $\tau$ defined in Formula~\eqref{tau} was
intimately related to the concept of {\it saturation}~\cite{saturation}, i.e. 
the behaviour of perturbative 
QCD amplitudes when the density of partons becomes high enough to exercise nonlinear effects ensuring the 
unitarity bound. Indeed, there were many theoretical arguments  to 
infer that in a domain in $Y$ and  $Q^2$ where saturation effects  set in, geometric 
scaling 
is  expected to occur. Within this framework, the function  $Q_s (Y)$ can be 
called the 
saturation scale, since it  delineates the approximate lower bound of the 
saturation domain.

This type of geometric scaling is motivated by asymptotic properties of QCD 
evolution equations with rapidity. Using the nonlinear 
Balitsky-Kovchegov 
(BK) equation~\cite{Balitsky:1995ub} which represents the ``mean-field'' 
approximation of high energy 
(or high density) QCD, geometric scaling could be derived
from its asymptotic solutions~\cite{Munier:2003vc}.
This equation is supposed to capture some essential 
features of 
saturation effects. Considering the BK equation with $fixed$ 
coupling constant leads asymptotically to the original geometric scaling of 
Formula~\eqref{tau}. Considering a $running$  coupling leads to the following
scaling: 

\begin{equation}
\tau = \log Q^2-\log Q_s^2(Y) =  \log Q^2-\lambda 
\sqrt Y\  ,
\label{taurunI}
\end{equation}

Recently~\cite{gb}, it was noticed that the scaling 
solution~\eqref{taurunI} of the BK equation with running coupling is only 
approximate and not unique. Another equivalent approximation leads to a 
different scaling variable, namely 
\begin{equation}
\tau =   \log Q^2-\lambda ~ \f{Y}{\log Q^2}
\  .
\label{taurunII}
\end{equation}

The key ingredient to theoretically prove  geometric scaling of the asymptotic
solutions of the nonlinear BK equation
 is the {\it traveling wave} 
method~\cite{Munier:2003vc}. Indeed, the BK equation admits 
solutions in the form of {\it traveling waves} $\cN (L\! -\!\upsilon_c 
t)$. $L=\log Q^2$ has the interpretation of a space variable while 
$t$, interpreted as time, is an increasing function of 
rapidity $Y,$ namely $t \!\propto  Y$ for the fixed coupling case and  $t 
\!\propto  \sqrt Y$ for the running one. $\upsilon_c$ is the {\it critical 
velocity} of the wave. These properties confirmed results previously 
obtained~\cite{Mueller:2002zm}, for instance by replacing the nonlinear damping term by absorbing boundary 
conditions, and thus considering only the linear part of the BK equation, which is equivalent to the Balitsky Fadin Kuraev 
Lipatov (BFKL) equation~\cite{Lipatov:1976zz}.

The subsequent elaboration on QCD evolution 
equations led to go beyond the mean-field approximation. The effect of 
fluctuations was examined in Ref.~\cite{Hatta:2006hs} in the fixed coupling scheme 
and gives rise to a new ``diffusive scaling'', the scaling variable being
\begin{equation}
\tau =   \f{\log Q^2-\lambda {Y}}{\sqrt{Y}}
\  .
\label{diffuse}
\end{equation}

The aim of this paper from a theoretical point-of-view is to test and compare the 
different scaling behaviors, arising from different versions of 
QCD evolution, using the data available from HERA.
We shall study the phenomenological relevance of the four kinds of scaling and 
refer to them in the following as ``Fixed Coupling'' for the variable~\eqref{tau},
``Running Coupling I'' and ``Running Coupling II'' for the 
variables \eqref{taurunI} and \eqref{taurunII} respectively, and to  ``Diffusive 
Scaling'' for \eqref{diffuse}. Our method can be easily applied to any new
proposal of scaling.

From an experimental point-of view, one wants to include not only the newest 
published data on $\sigma^{\gamma^*p}$ from HERA, but also from other processes where the 
small-$x$ physics allow one to discuss scaling properties. We thus include the 
data on  DVCS cross sections and, following the geometric scaling analysis 
performed in Ref.~\cite{Marquet:2006jb},  we extend the analysis to elastic 
vector-meson production and to diffractive cross-sections at fixed and small 
$\xp.$ Moreover, and as a new aspect of scaling analysis, we are led to 
also include   diffractive cross-sections at fixed and small $\beta,$ since scaling 
should also be considered for the  $\gamma^*Pomeron$ cross section, following the 
Ingelman-Schlein interpretation of diffractive structure functions~\cite{Ingelman:1984ns}.

In order to 
define how good the scalings are, the authors of Ref.~\cite{usgeom} 
introduce a $Quality \ 
Factor$ (QF) as an estimator on the validity of scaling. The main property of the 
QFs 
is that it does not depend $a\ priori$ on a given parametrization of the scaling 
curve as in an ordinary $\chi^2$ approach. The aim of this paper is thus to
extend on the same footing the QF studies to all scaling laws,
including the new scaling Running Coupling II, 
and to compare them. We perform a systematic analysis using the QF techique 
investigating the scaling properties in deep-inelastic scattering using all
data available. It will 
include  both the largest available set of published data where scaling is 
expected to 
occur and the new status on theoretical scaling properties using QCD evolution 
equations.

The plan of the paper is as follows. In section 
{\bf II} we describe the theoretical motivation and the 
give the precise formulation of 
the various scaling hypotheses. In section {\bf III} we present the 
Quality Factor method we shall use in the analysis and the set of data we 
consider for the fitting procedure. We provide and comment our results in  
section {\bf IV}, and present our conclusions in the final section {\bf V}.

\eject


\section{Scaling Variables}

Let us sketch the theoretical motivation for the different forms of scaling 
(\ref{tau}-\ref{diffuse}) in deep-inelastic 
scattering 
and their extension to DVCS cross-section data, elastic vector-meson 
and for diffractive cross-sections at fixed and small $\xp$ or $\beta.$ We 
shall focus on the general arguments leading to scaling which could be 
independent from too specific theoretical predictions. 
The different types of scaling we refer to originate from 
properties of nonlinear QCD evolution equations, and the 
existence of asymptotic traveling wave solutions. We shall first introduce the 
general hint for these traveling wave solutions in the four cases and 
subsequently 
study the effect of varying the initial values of the phase space
variables $Y$ and $L$. This amounts to introduce some
natural non asymptotic effects. 

For the three first cases 
(\ref{tau},~\ref{taurunI},~\ref{taurunII}), let 
us start with a ``mean-field'' equation for the dipole-target amplitude of the 
type
\be
\partial_Y \cN (L, Y) = 
\bar\alpha\, \chi (-\partial_L)\: \cN (L, Y) - \bar\alpha\, \cN^{\,2} (L, Y),
\label{1}\ee
where $L \!= \!\log (k^2/k^2_0)$, $k$ is the gluon transverse momentum 
(Fourier conjugate to the dipole size), $k_0$ is an arbitrary constant
and $Y$ is the rapidity. In 
Eq.~\eqref{1}, the coupling constant will be considered in the 
following as fixed, or running 
such that $\bar\alpha(L)\! = \!{1}/{bL}\ ,
b = {11 N_c \!- \!2 N_f}/{12 N_c}$ where $N_f$ and $N_C$ are the numbers 
of flavours and colours respectively.
As we will show, for the sake of a general derivation of scaling, the 
differential operator kernel 
$\chi(-\partial_L)$ 
can be considered to be general, provided the function $\chi(\g)/\g$ admits a 
minimum value at some 
point 
$\g_c.$ Indeed, the kernel at the  
leading logarithm (LL) accuracy of the perturbative QCD expansion,  the BFKL 
kernel~\cite{Lipatov:1976zz} $
\chi(\gamma) \!=\! 2\psi(1) \!-\! \psi(\gamma)\! -\! \psi(1\!-\!\gamma)\ $ 
verifies 
this property but large corrections 
at next-to-leading logarithm (NLL) accuracy exist (see, for the initial references, 
\emph{e.g.} 
\cite{Fadin:1998py,Salam:1998tj,Ciafaloni:1999yw,us}). We will thus consider a 
general 
argument, 
instead of using a specific form of the QCD kernel.

Let us first consider the ``Fixed Coupling'' scheme $\bar\alpha = cst.$ The 
solution 
to the linear part of equation~(\ref{1}) 
corresponds to a linear superposition of waves:
\begin{equation} 
\cN (L, Y)=\int_{\cal C}
\frac{d\gamma}{2i\pi}\, \cN_0(\gamma)\,
\exp \left\{-\gamma \left(L-\f {\omega(\gamma)}{\g}Y\right)\right\}\ ,
\label{inicond}
\end{equation}
where $\omega(\gamma)=\bar\alpha\chi(\gamma)$ is the Mellin transform of the 
kernel.
In particular, one can interpret~\eqref{inicond} by stipulating that each 
partial wave of wave-number
$\gamma$ has a velocity
$\upsilon(\gamma)=\frac{\omega(\gamma)}{\gamma}.$
By contrast, the {\it critical velocity} is defined~\cite{GLR,Munier:2003vc} by the minimum 
$\gamma_c$
of the exponential
phase factor $
\upsilon=\left.\frac{d\omega}{d\gamma}\right|_{{\gamma_c}}= 
\frac{\omega(\gamma_c)}{\gamma_c}\equiv \upsilon_c\ .$
$\upsilon_c$ is thus the $minimal$ velocity of the 
waves. So, 
the balance between the initial velocity  (for fast enough initial condition 
with $\g_0>\g_c,$ see~\cite{Munier:2003vc}) 
and the 
damping due to the nonlinear term in Eq~\eqref{1} leads to an asymptotic  
$blocking$ 
of the velocity (and of the wave front) on the minimal and critical value 
$\upsilon_c.$ As a consequence, reporting this dominant value in formula 
\eqref{inicond}, one finds
\be
\cN (L, Y) \sim \exp \left\{-\gamma_c \left(L-\f 
{\omega(\gamma_c)}{\g_c}Y\right)\right\}\ ,
\label{N}
\ee
which is the expression of scaling with the variable~\eqref{tau}. 
 
For the running coupling case of Eq.~{5}, the situation is more subtle, 
since different approximate traveling wave solutions may coexist. In 
the original derivation of Ref.~\cite{Munier:2003vc}, one starts with the linear 
solution analogue to Eq.~(\ref{inicond}), namely
\begin{equation} \label{MellinRC}
{\cal N}(L,Y)=\int\frac{d\gamma}{2i\pi}{\cal N}_0(\gamma)
\exp\left\{-\gamma \left(L-\sqrt{Y}\,\sqrt{\frac{4X(\gamma)}{b\g}}
\right)\right\}\ ,
\end{equation}
where $X(\gamma)(\equiv \int^\gamma 
d\gamma^\prime\,\chi(\gamma^\prime))$ plays the role of the new effective 
kernel. 
The key point is that the equivalent of the $time$ variable is now $\sqrt{Y}$ 
instead of $Y.$ This $\sqrt{Y}$ comes from a saddle-point integration over 
$\omega,$ the conjugate variable to the rapidity \cite{Ciafaloni:1999yw}. 
Consequently, after the same ``blocking'' mechanism due to the 
nonlinear damping terms,  the scaling variable is $\tau(L,Y) = \log 
Q^2-\lambda 
\sqrt Y\  ,$ which is nothing else than the ``Running Coupling I'' case, see 
Eq.~\eqref{taurunI}. However, the scaling inferred by Eq.~\eqref{MellinRC}
is neither an exact nor unique approximate scaling solution. 

As an alternative to that Mellin transform method, it was proposed recently in 
Ref.~\cite{gb} to search directly for the running coupling analog of partial 
wave solutions, \emph{i.e.} scaling solutions of Eq.~\eqref{1} with running 
coupling 
$\bar\alpha(L)\! = \!{1}/{bL},$ with some generic velocity $\upsilon$.
Imposing the scaling of the left hand and right hand side of 
the equation as a function of the scaling variable $\tau(L,Y)$ leads to  two different 
constraints, namely the functions $ bL \partial_Y \cN [\tau(L,Y)]$ and
$\partial_L^n \cN [\tau(L,Y)]$ should be both functions of $\tau$ only
\bea
bL \partial_Y \cN [\tau(L,Y)]&=& f(\tau) \label{eq1BKRC}\\
\partial_L^n \cN [\tau(L,Y)]&=& g(\tau) \label{eq2BKRC}
\eea
which are  conditions to have a scaling 
solution
$\cN (L,Y) \equiv \cN [\tau(L,Y)].$  In 
fact conditions~\eqref{eq1BKRC} and~\eqref{eq2BKRC} cannot be 
fulfiled simultaneously (contrary to the 
`Fixed Coupling'' case, see~\cite{gb}) and can only be considered in an 
approximate way. 
If one choses to fulfil condition~\eqref{eq2BKRC} exactly, and~\eqref{eq1BKRC} 
only approximately, one finds again the scaling variable 
$\tau(L,Y)= \log 
Q^2-\lambda 
\sqrt Y\  ,$ which is the ``Running Coupling I'' case, see 
Eq.~\eqref{taurunI}. By this way one recovers the scaling variable previously 
found in 
Refs.~\cite{Mueller:2002zm,Munier:2003vc} in a different way.
If, conversely, one chooses to satisfy exactly~\eqref{eq1BKRC} and only 
approximately~\eqref{eq2BKRC}, 
one finds a new form of scaling, with the variable 
$\tau(L,Y)= 
L-\lambda \f{Y}{L}\  ,$ which corresponds to  the ``Running Coupling II'' case, 
see 
Eq.\eqref{taurunII}. As the ``blocking'' mechanism is more general than the 
approximations leading to the variables ``Running Coupling I'' or 
``Running Coupling II'', one expects that 
Eq.~\eqref{1} with running coupling has no exact scaling solution, but admits 
at least two different approximate solutions, namely ``Running coupling I" and ``Running
coupling II". One aim of our study is to make a phenomenological comparison
between both solutions.

 For the case of diffusive scaling, one considers the modification of 
Eq.~\eqref{1} due to Pomeron-loop contributions, which appear when  the system of 
QCD 
partons or dipoles is dilute. It may seem irrelevant to an equation describing 
the 
high density partonic  effects on the amplitude, but the structure of the 
equation, 
with an exponential increase of the linear regime may lead to important 
modifications~\footnote{Recent results on a toy model with running coupling and 
fluctuations~\cite{Dumitru:2007ew} 
seem not to give diffusive scaling. A more  
complete QCD study is still missing.}
w.r.t. the mean-field solutions of~\eqref{1}. 
Indeed, 
it was shown that, while traveling-wave solutions of the type~\eqref{N} are 
formed, the effect of  event-by-event fluctuations leads to a stochastic 
superposition of these waves around  an average solution which is governed by 
some 
diffusion coefficient and thus to a different scaling variable  $\tau(L,Y) =   
\f{\log Q^2-\lambda {Y}}{\sqrt{Y}}$ which corresponds to  the 
``Diffusive 
Scaling'' case, see Eq.\eqref{diffuse}.

The following  comments are in order. One would like to confront
the QCD predictions 
on the scaling form of the amplitude ($i.e.$ the wave front) with data. However, the 
precise determination of the wave front seems at present to be less reliable 
than the 
scaling variable ($i.e.$ the wave structure). The form of the 
linear 
kernel and of the non linear damping at next-to-leading logarithmic order have 
just been derived~\cite{BKNLL}. However, in the kinematical range where 
scaling is observed, higher order contributions may be relevant,
leading to a modified kernel in Eq.~(5). Hence we will 
stick to a rather general and model independent (for 
each case) prediction for the scaling variable. The phenomenological analysis 
may on 
contrary help the development of the theoretical investigation.

Another important comment concerns the non-asymptotic corrections. While some of 
them 
can be deduced from the $blocking$ mechanism itself (see $e.g.$~\cite{Munier:2003vc})
we shall consider the simplest and expected nonasymptotic corrections  due to 
the definition of the typical momentum and rapidity scales $\Lambda$ and $Y_0.$ 
Hence we 
will consider, if necessary the modification of all scaling variables 
(\ref{tau}-\ref{diffuse}), by a shift  $L\to 
L-\log\Lambda^2$ and $Y-Y_0.$ Note that the reference value (without shift) is 
always $Y_0=0$.
 
 Let us now consider diffractive, elastic vector-meson and DVCS cross-sections. 
As discussed in Ref.~\cite{Marquet:2006jb}, geometric scaling was shown to 
be valid and the ``Fixed Coupling'' was verified within  uncertainties for 
these reactions, using $\xp$ instead of the Bjorken $x$ in the scaling variable \eqref{tau}. 
 
In our phenomenological discussion, we shall thus extend the scaling properties 
to all four cases including the one studied in~\cite{Marquet:2006jb}, by 
considering  the various scaling variables (\ref{tau}-\ref{diffuse}).
 
We shall thus investigate  the following scaling properties.
The scaling prediction is  
\be
\sigma^{\g^*p\rightarrow\g p}_{DVCS}(x,Q^2)\!=\!
\sigma^{\g^*p\rightarrow\g p}_{DVCS}(\tau[x,Q^2])\label{test3}
\ee
for the DVCS cross-section, 
\be
\sigma^{\g^*p\rightarrow Vp}_{VM}(\xp,Q^2,M_V^2)=
\sigma^{\g^*p\rightarrow Vp}_{VM}(\tau_V[\xp,Q^2+M_V^2]))
\label{test2}
\ee
for the elastic vector meson cross sections,
where the hard scale is assumed to 
be $Q^2+M_V^2,$ from the known properties of the vector meson wave function, and
\be
\f{d\sigma^{\g^*p\rightarrow Xp}_{diff}}{d\beta}(\beta,\xp,Q^2)=
\f{d\sigma^{\g^*p\rightarrow Xp}_{diff}}{d\beta}(\beta,\tau_d[\xp,Q^2])
\label{test1}
\ee
for the diffractive cross-section at fixed $\beta$ and 
small $\xp.$ 

For completion, besides the cases studied in~\cite{Marquet:2006jb},  we may also 
consider the diffractive cross section at fixed $\xp$ and 
small $\beta,$ namely 
\be
\f{d\sigma^{\g^*p\rightarrow Xp}_{diff}}{d\beta}(\beta,\xp,Q^2)=
\f{d\sigma^{\g^*p\rightarrow Xp}_{diff}}{d\beta}(\xp,\tau_b[\beta,Q^2])\ 
,
\label{test1bis}
\ee
since in the hypothesis of factorisation of the Pomeron flux, it relates  to  
$\sigma^{\g^*{\cal P}\rightarrow X{\cal P}}(\beta,Q^2)$ ($i.e.$ the total 
``Pomeron-photon'' cross-section)  at small $\beta$ for which the scaling 
arguments should also hold.

The prediction for DVCS (see Eq.~\eqref{test3}) does not rely on any nonperturbative assumption.
This interesting feature of DVCS cross-sections will allow us to include the 
data in the fitting procedure. Note that we will not use the diffractive data 
nor the vector meson ones for the fit. Instead, we will consider the fit to total 
and DVCS cross-sections as scaling predictions for  diffractive and vector meson 
data.

Hence, in our phenomenological discussion, we shall extend the scaling 
properties to our four cases, by considering  the various scaling variables 
(\ref{tau}-\ref{diffuse}) in the equations (\ref{test3}-\ref{test1bis}). 

\section{The Quality factor method}

In order to fit the parameters in the scaling variables (\ref{tau}-\ref{diffuse}),
it is useful to use a quantity called \emph{quality factor}~\cite{usgeom} (QF). 
The quality factor is used  to find
the best parameters for a given scaling and to compare quantitatively
the different scalings.

Given a set of data points $(Q^2, x, \sigma=\sigma(Q^2,x))$ and a parametric
scaling variable $\tau = \tau(Q^2, Y=log(1/x); \lambda)$ we want to know
whether the cross-section can be parametrised as a function of the variable $\tau$ only. 
Since we do not know the function of the $\tau$ variable describing the data,
we need to define the QF independently of the form of that function.
Therefore the QF
is defined in such a way that it quantitatively describes how close the 
data points
$\sigma=\sigma(Q^2,x)$  are to the scaling law $\sigma=\sigma(\tau)$, for
a given $\tau=\tau(Q^2,Y; \lambda)$.

Let us consider a set of points $(u_i, v_i)$, where $u_i$'s are ordered,
and introduce the QF as follows~\cite{usgeom}

\be
QF(\lambda) = \biggl[ \sum_{i} \frac{(v_i-v_{i-1})^2}{(u_i-u_{i-1})^2+\epsilon^2} \biggr]^{-1}
\label{QF},
\ee
where $\epsilon$ is a small constant that prevents the sum from being infinite in case
of two points having the same value of $u$.
Using this definition, 
the contribution to the sum in (\ref{QF}) is large
when two successive points are close in $u$ and far in $v$. Thus, we expect a set of points
lying close to a unique curve to have large QF (small sum in (\ref{QF})) compared to a QF
of a set of points that are more scattered.

Since the cross-section in data differs by orders of magnitude and $\tau$ is
more or less linear in $log(Q^2)$ (see \ref{tau}-\ref{taurunII}),
we decided to take $v_i = log(\sigma_i)$ and $u_i = \tau_i(\lambda)$.
This ensures that the low $Q^2$ data points contribute to the QF with a similar weight as
the higher $Q^2$ data points. To complete the definition of the QF,
the set $(u_i, v_i)$ is also rescaled so that $0 \le u_i,v_i \le 1$,
and ordered in $u$ before entering the QF formula. All the QFs in this paper
are calculated with $\epsilon=0.01$.

In order to test a scaling law $\tau$, we search for the parameter $\lambda$ 
that minimises the 1/QF variable.
To do that, we fit 1/QF with the \verb+MINUIT+ package. Given the maximum value of the QF,
we are able to directly compare different scaling laws. In this paper, we test the scaling laws
(\ref{tau}-\ref{diffuse}). The different scaling variables are given in 
Table~\ref{scalings}.

\begin{table}
\begin{center}
\begin{tabular}{|c||c||c|c|} \hline
   & scaling & $\tau$ formula & parameters \\ 
\hline\hline
FC & ``Fixed Coupling" & $\log Q^2 - \lambda Y$ & $\lambda$ \\
\hline \hline
RCI & ``Running Coupling I" & $\log Q^2 - \lambda \sqrt{Y}$ & $\lambda$ \\ \hline
RCIb & ``Running Coupling Ibis" & $\log Q^2 - \lambda \sqrt{Y\!-\!Y_0}$ 
& $\lambda,~Y_0$ \\ \hline \hline
RCII & ``Running Coupling II" &  $\log (Q^2/\Lambda^2) - \lambda 
\frac{Y}{\log (Q^2/\Lambda^2)}$& $\lambda$ \\
 & & $\Lambda=$0.2 GeV & \\ \hline
RCIIb & ``Running Coupling IIbis" &  $\log (Q^2/\Lambda^2) - \lambda 
\frac{Y-Y_0}{\log (Q^2/\Lambda^2)}$& $\lambda,~Y_0,~\Lambda$ \\ \hline \hline
DS & ``Diffusive Scaling" & $\frac{\log (Q^2/\Lambda^2) - \lambda Y}{\sqrt{Y}}$ & $\lambda$ \\
& & $\Lambda=$1 GeV & \\ \hline
DSb & ``Diffusive Scaling bis" & $\frac{\log Q^2/\Lambda^2 - \lambda (Y-Y_0)}{\sqrt{
Y-Y_0}}$ & $\lambda,~Y_0,~\Lambda$
 \\
\hline \hline
\end{tabular}
\end{center}
\caption{Scaling variables used in the fits to deep inelastic scattering data.}
\label{scalings}
\end{table}

Some comments are in order. We always define two versions of the scaling
variables, the former depending only on a single parameter $\lambda$, and the
later on two or three parameters $\lambda,~Y_0,~\Lambda$. 
As an example, we define in this way the ``Running Coupling II"
and ``Running Coupling IIbis"  in this way respectively. The ``Fixed Coupling"
scaling does not depend on the scale $\Lambda$ and the rapidity $Y_0$ since
these additional variables disappear in the definition of the quality factor QF.
The conclusion remains the same for the $\Lambda$ dependence of the ``Running
Coupling I" scaling. Concerning the ``Running Coupling II" scaling, we fixed the
parameter $\Lambda$ to a typical non perturbative scale inside the proton,
$\Lambda = \Lambda_{QCD} = 0.2$~GeV.

The QF defined above raise two additional technical issues which
require further checks. 
First, it does not take into account
the errors on data points. The errors can be introduced to the QF in an error weight function
by modifying the definition (\ref{QF}) in the following way

\be
QF(\lambda) = \biggl[ \sum_{i} \frac{(v_i-v_{i-1})^2 W_{i,i-1}}{(u_i-u_{i-1})^2+\epsilon^2} \biggr]^{-1}
\label{QFer},
\ee
where $W_{i,j}$ is the error weight function of the data points $i,j$. In the simplest form,
we can take

\be
W_{i,j}=\rho_i \rho_j
,
\ee
where $\rho_i$ is the relative statistical or uncorrelated error of point $i$.
This definition allows to take into account the statistical scatter of the
different data points. This is particularly important if the statistical uncertainty
differs from one data point to the other, and is significantly large. The data points
we use in the following (mainly the structure function $F_2$ from the H1 and ZEUS
experiments) show very little statistical uncertainties.
We compared the results
using either formula (\ref{QF}) or formula (\ref{QFer}), and as expected,
we get similar values of the parameters for both definitions.
For sake of simplicity, we choose the QF (\ref{QF}) without any error weight function.

The second possible shortcoming is an ambiguity of the QF related to the $(u_i, v_i)$ point ordering
if two or more $u_i$'s are equal. The order of these points in the data set is not unique;
any permutation of these points satisfy the definition above. Obviously, we get diffrent QF for
different permutations. The comparison of the fit results with different data point
ordering was performed and
the results were found similar (both in the value of the fitted parameters and the QF
itself).


\section{Fits to $F_2$ and DVCS data}

The first natural data set to test the different scaling laws are the $F_2$ data measured
at low $x$ by the H1 \cite{H1}, ZEUS \cite{ZEUS}, NMC \cite{NMC} and E665 \cite{E665}
experiments. These data are very
precise and cover a wide range in $x$ and $Q^2$.

\begin{figure}[t]
\begin{center}
\epsfig{file=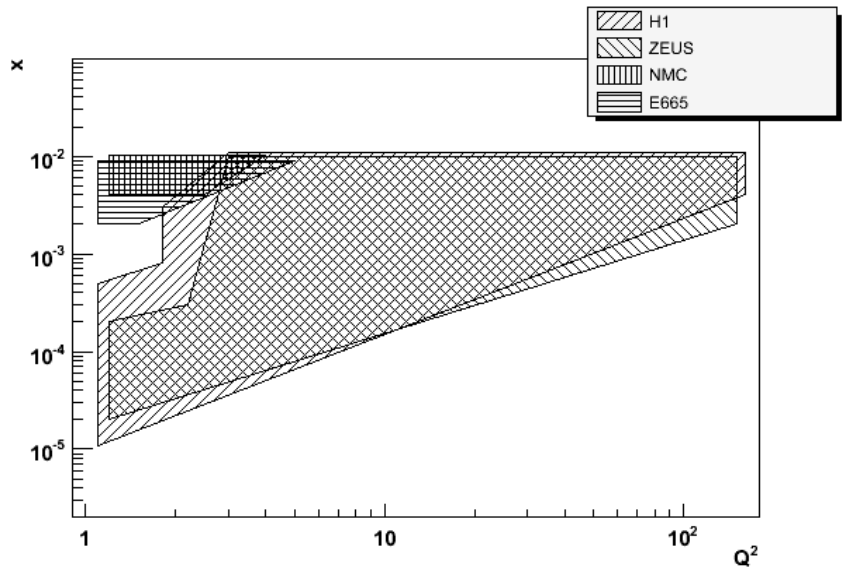,width=12.cm}
\caption{Kinematical domain in $x$ and $Q^2$ covered by the different data sets
from H1, ZEUS, NMC and E665 used in this anlysis. Only the kinematical
domain with 
$x<10^{-2}$ and $1<Q^2<150$~Gev$^2$ is shown.}
\label{xQ2}
\end{center}
\end{figure}

We choose to consider all available data from H1, ZEUS, NMC and E665 experiments
with $Q^2$ in the range $[1;150]$~GeV$^2$ and $x<0.01$. Fig.~\ref{xQ2} displays the
different kinematical domains reached by the H1, ZEUS, NMC and E665 experiments. While the
H1 and ZEUS kinematic coverage is similar (with a tendency for H1 to reach lower $x$ and
$Q^2$ values), the data from NMC and E665 cover mostly the low $Q^2$, higher $x$ kinematic
domain of our sample.
We choose not to consider the data with $x>10^{-2}$ since they are dominated by the valence
and sea quark densities, and the formalism of saturation does not apply in this kinematical
region. In the same  way, the upper $Q^2$ cut is introduced while the lower $Q^2$ cut
ensures that we stay away from the soft QCD domain. We will show
in the following that the data points with $Q^2<1$~GeV$^2$ spoil the fit stability.
Two kinds of fits to the scaling laws are performed, either 
in the full mentioned $Q^2$ range, or in a tighter $Q^2$ range $[3;150]$~GeV$^2$
to ensure that we are in the domain where perturbative QCD applies.
We will study the $Q^2$-dependence of the fit parameters, as a possible
indication of smooth scaling violations.

\subsection{Fits to $F_2$ data with $Q^2>3$ GeV$^2$}
\begin{figure}[t]
\begin{center}
\begin{tabular}{cc}
\epsfig{file=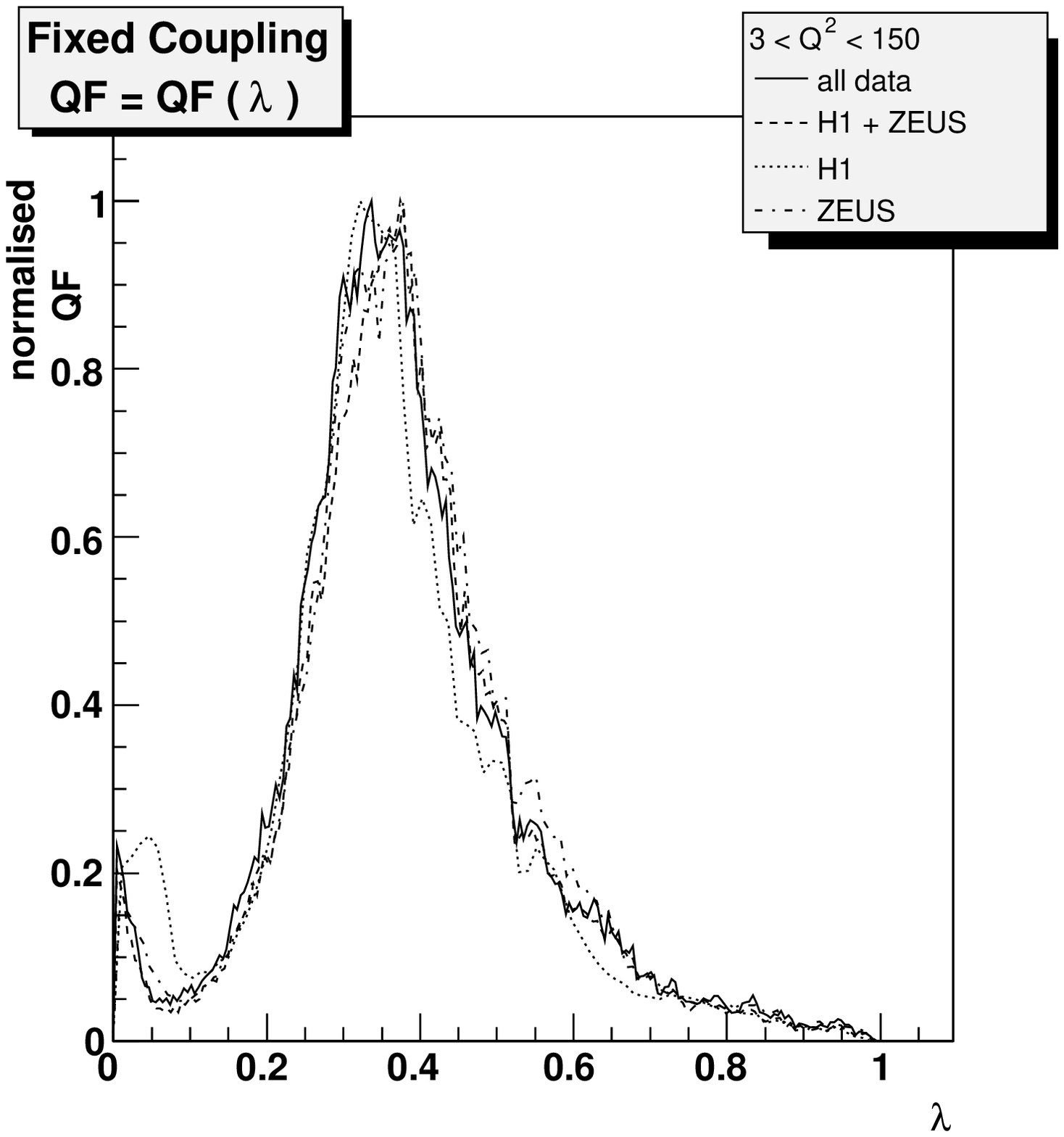,width=9.cm} &
\epsfig{file=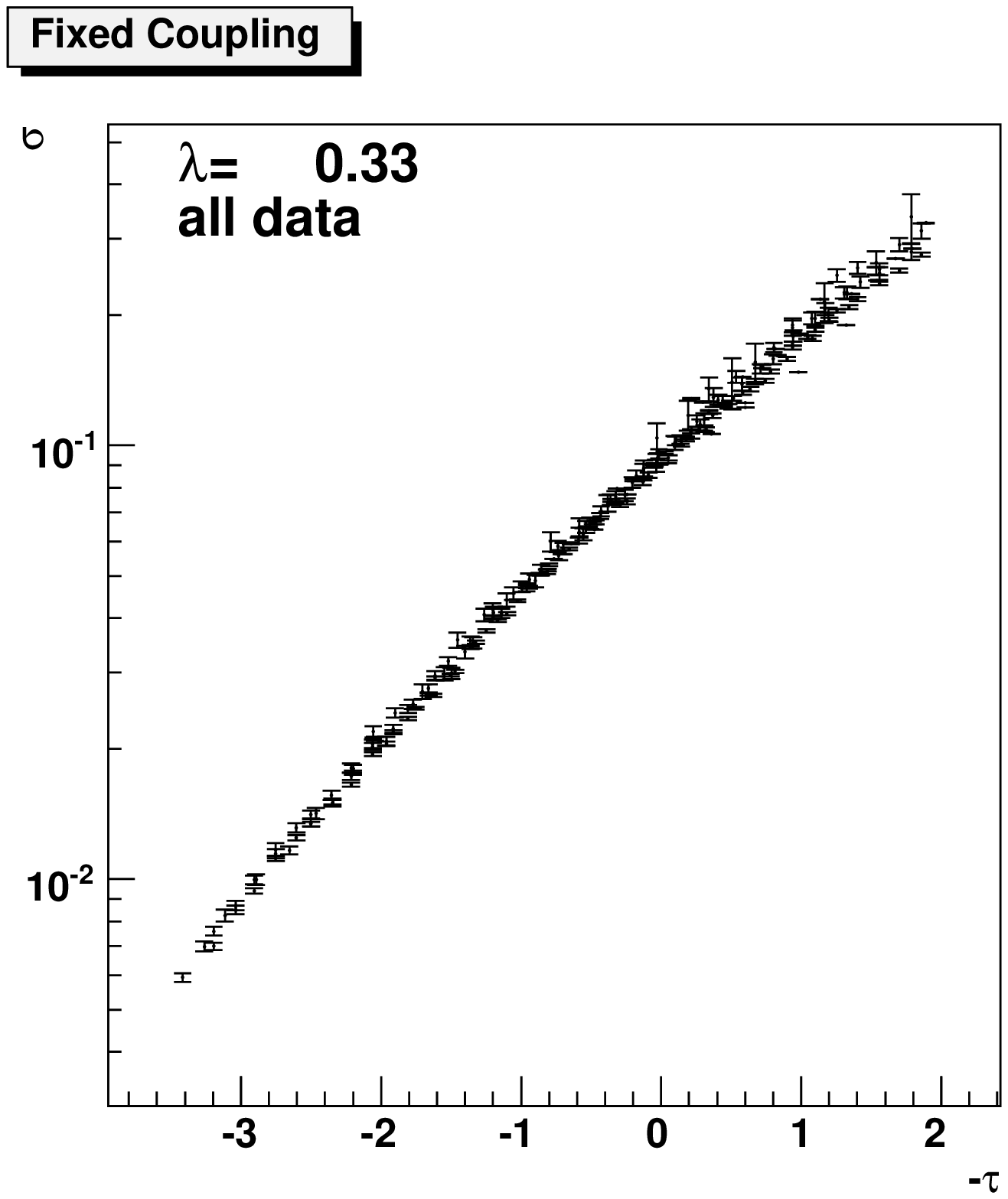,width=9.cm}
\end{tabular}
\caption{{\bf $F_2$ data:} Normalised QF to 1. as a function of $\lambda$ and scaling curve 
with $\lambda$ fixed to the value corresponding to the best QF 
for ``Fixed Coupling". A $Q^2>3$ cut was applied to the data.}
\label{F2_fixed_3}
\end{center}
\end{figure}

\begin{figure}[t]
\begin{center}
\begin{tabular}{cc}
\epsfig{file=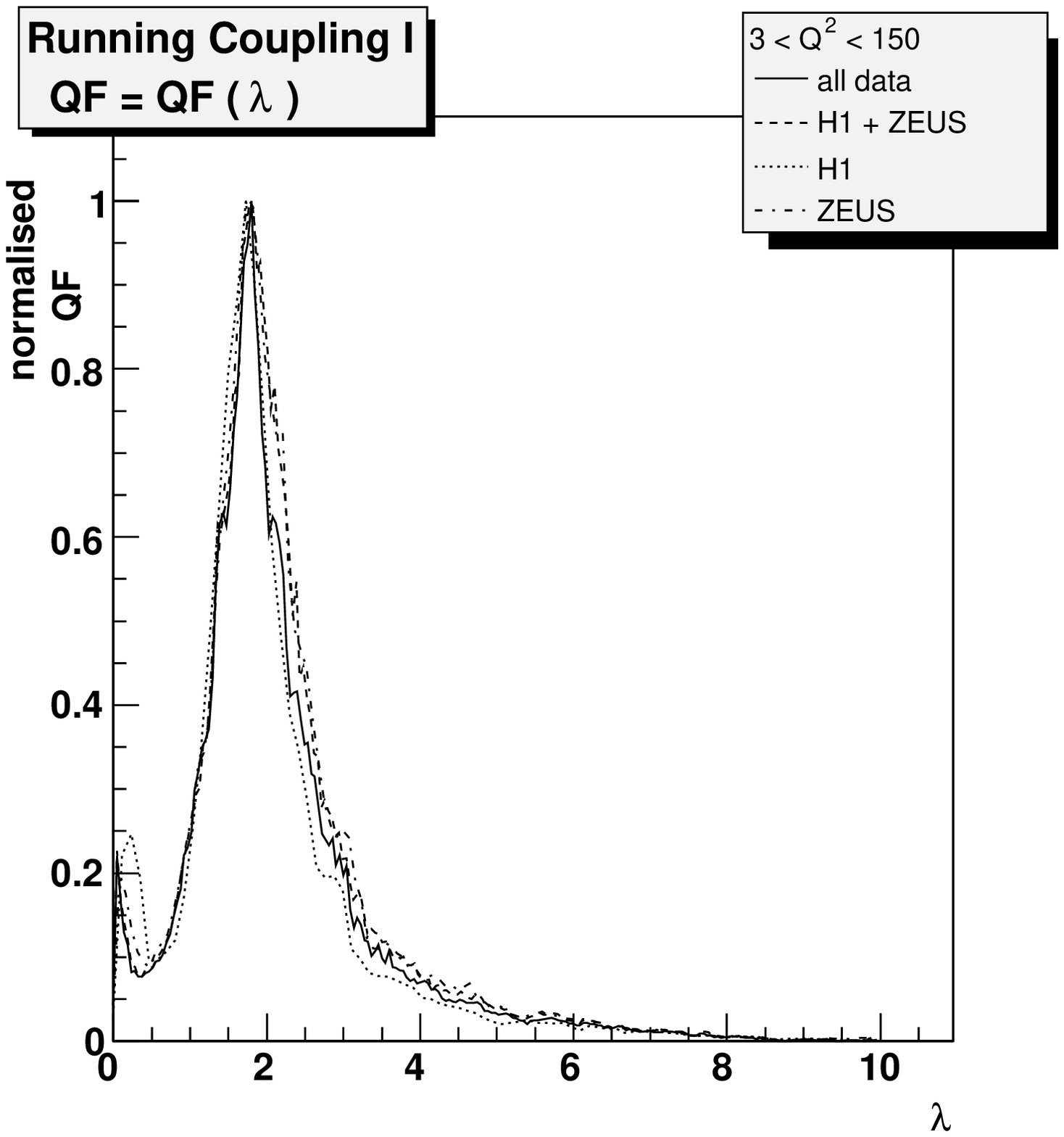,width=9.cm} &
\epsfig{file=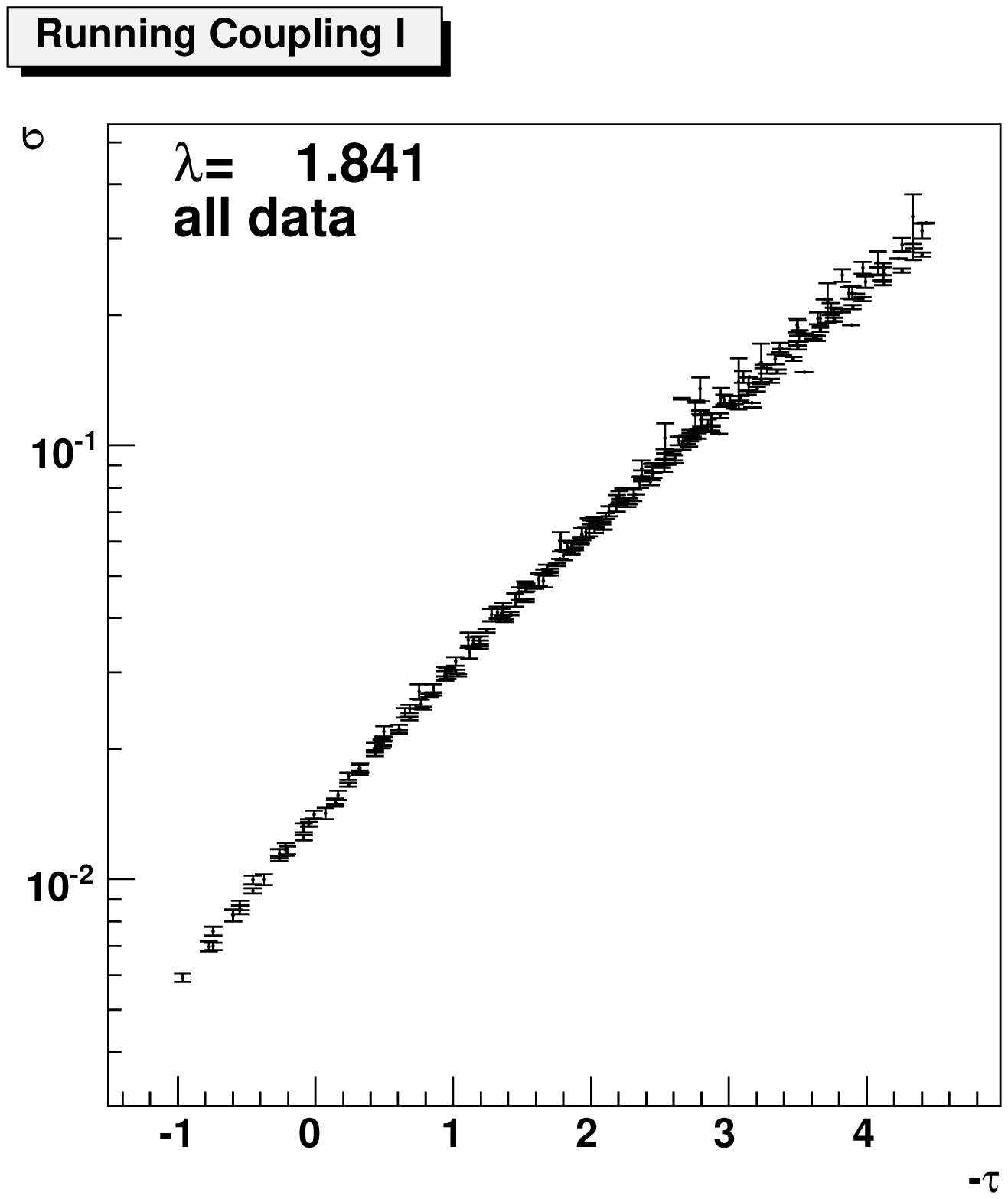,width=9.cm}
\end{tabular}
\caption{{\bf $F_2$ data:} Normalised QF to 1. as a function of $\lambda$ and scaling curve 
with $\lambda$ fixed to the value corresponding to the best QF
for ``Running Coupling I". A $Q^2>3$ cut was applied to the data.}
\label{F2_runI_3}
\end{center}
\end{figure}

\begin{figure}[t]
\begin{center}
\begin{tabular}{cc}
\epsfig{file=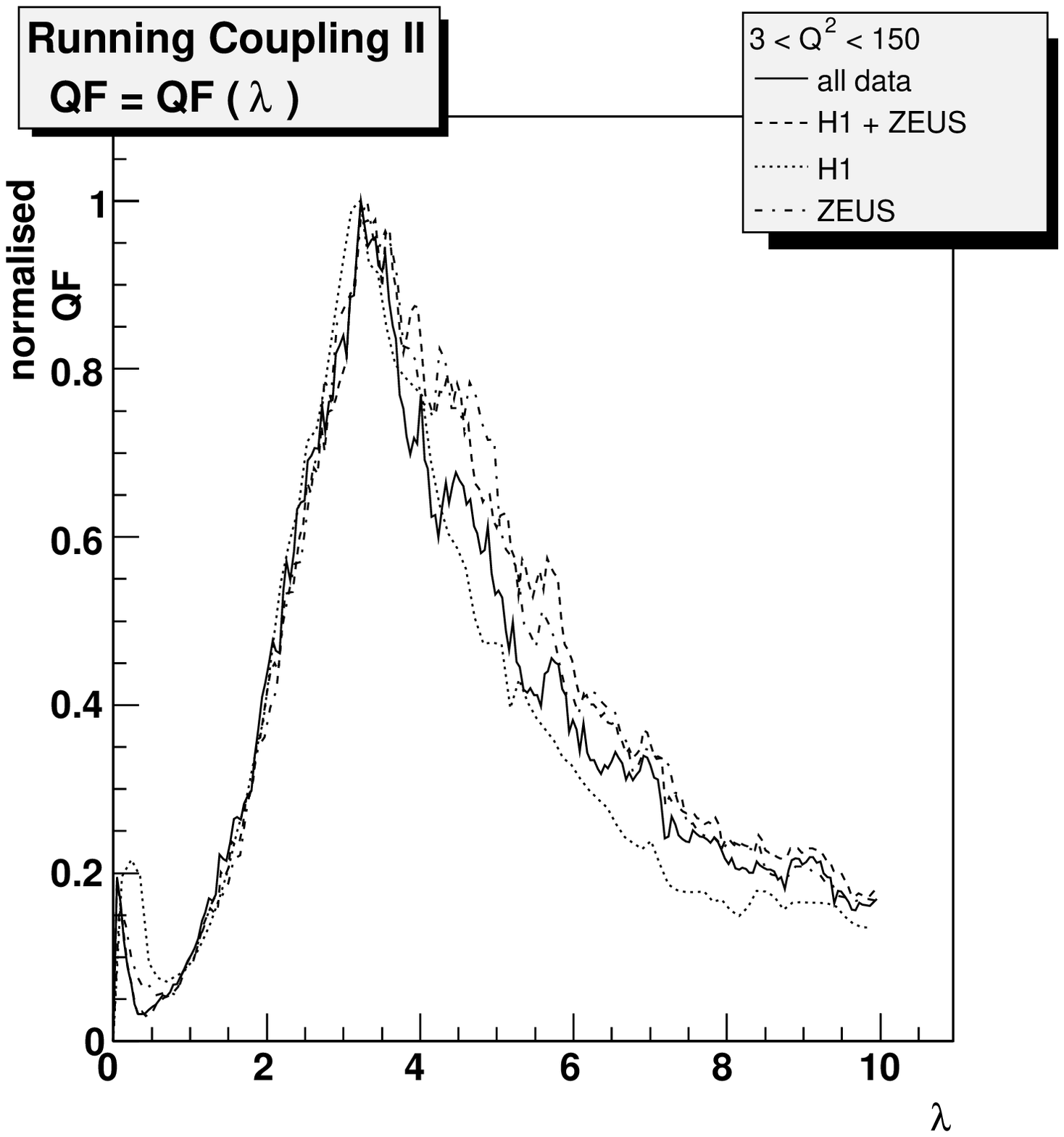,width=9.cm} &
\epsfig{file=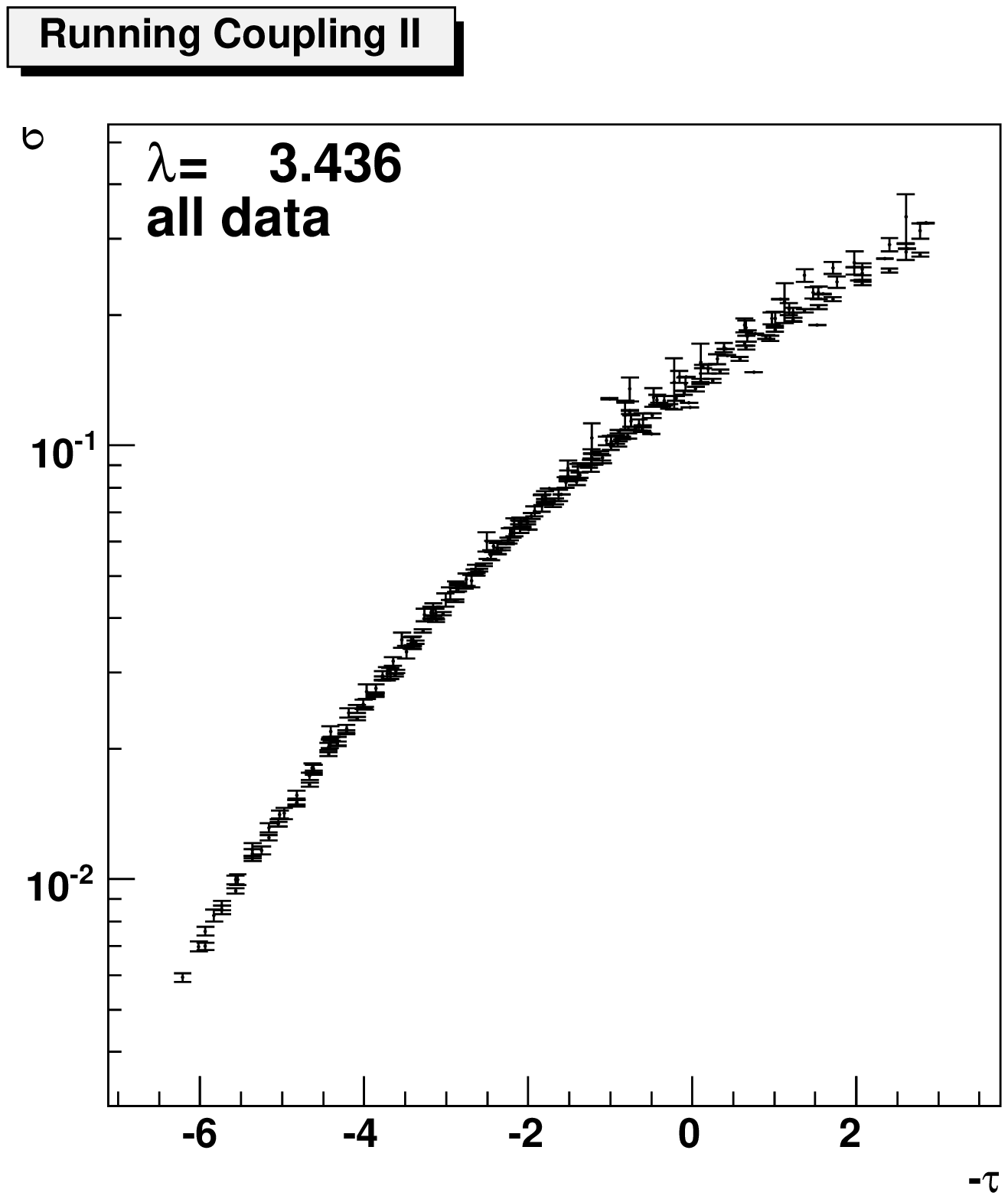,width=9.cm}
\end{tabular}
\caption{{\bf $F_2$ data:} Normalised QF to 1. as a function of $\lambda$ and scaling curve 
with $\lambda$ fixed to the value corresponding to the best QF
for ``Running Coupling II". A $Q^2>3$ cut was applied to the data.}
\label{F2_runII_3}
\end{center}
\end{figure}

\begin{figure}[t]
\begin{center}
\begin{tabular}{cc}
\epsfig{file=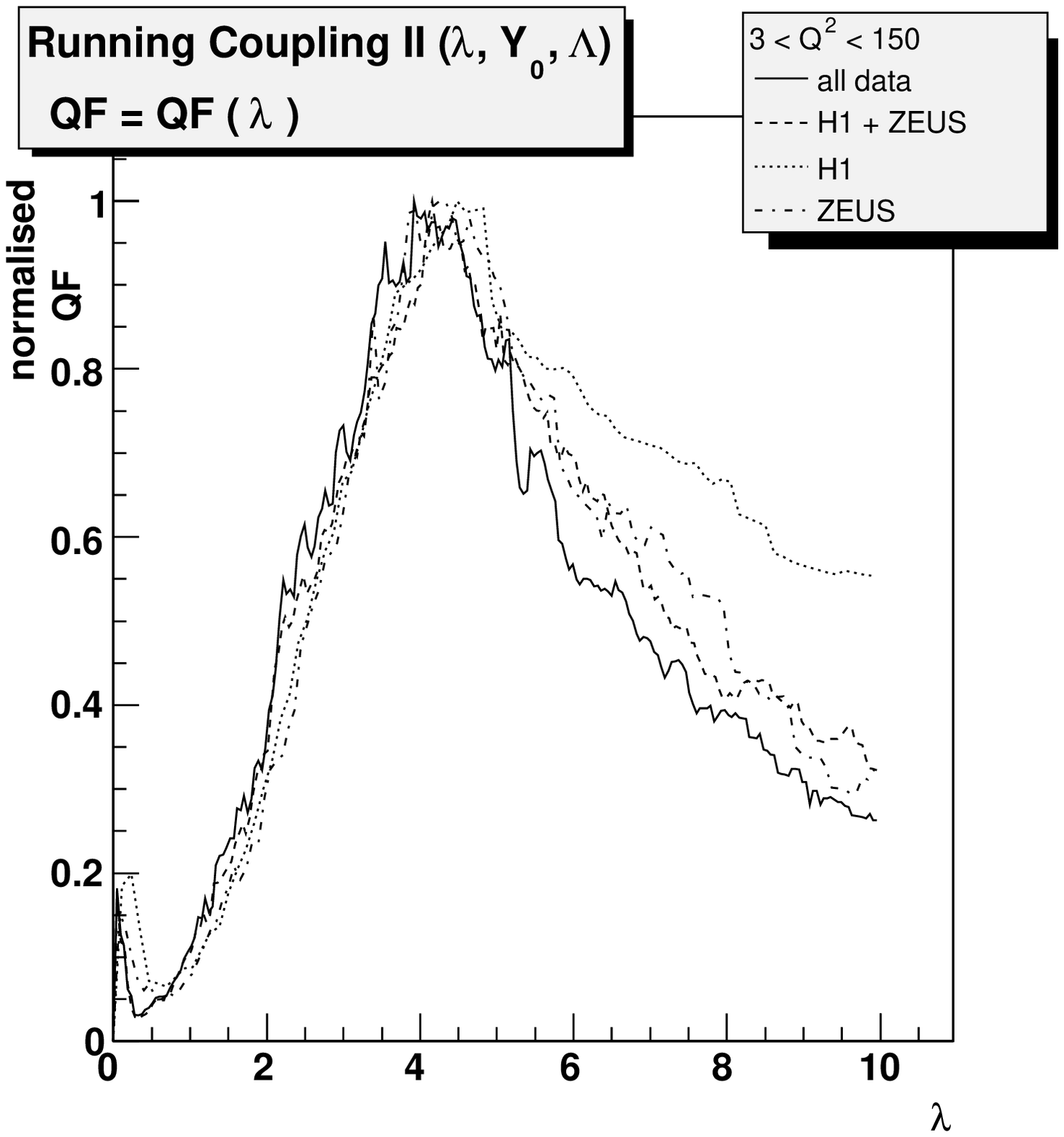,width=9.cm} &
\epsfig{file=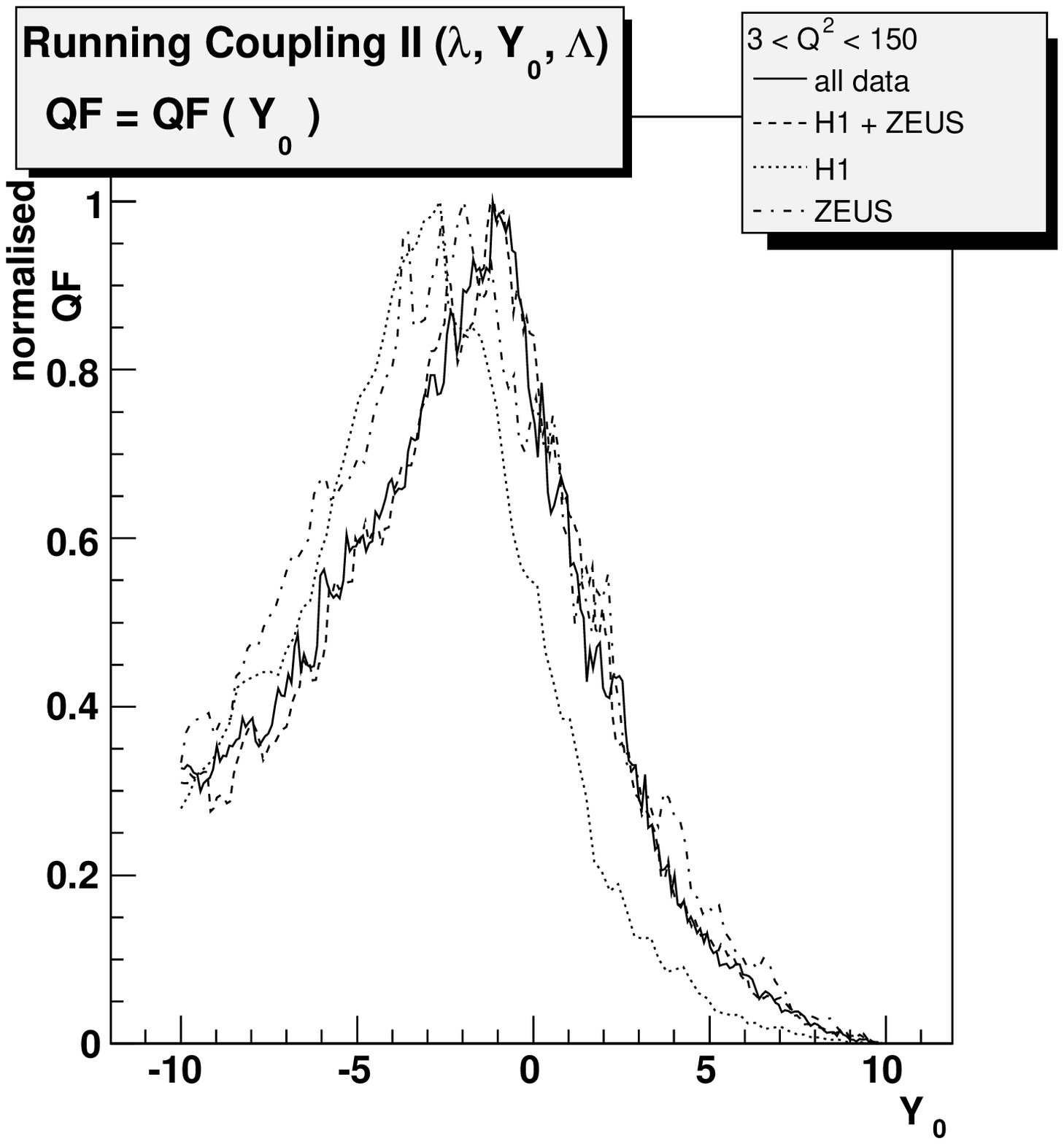,width=9.cm} \\
\epsfig{file=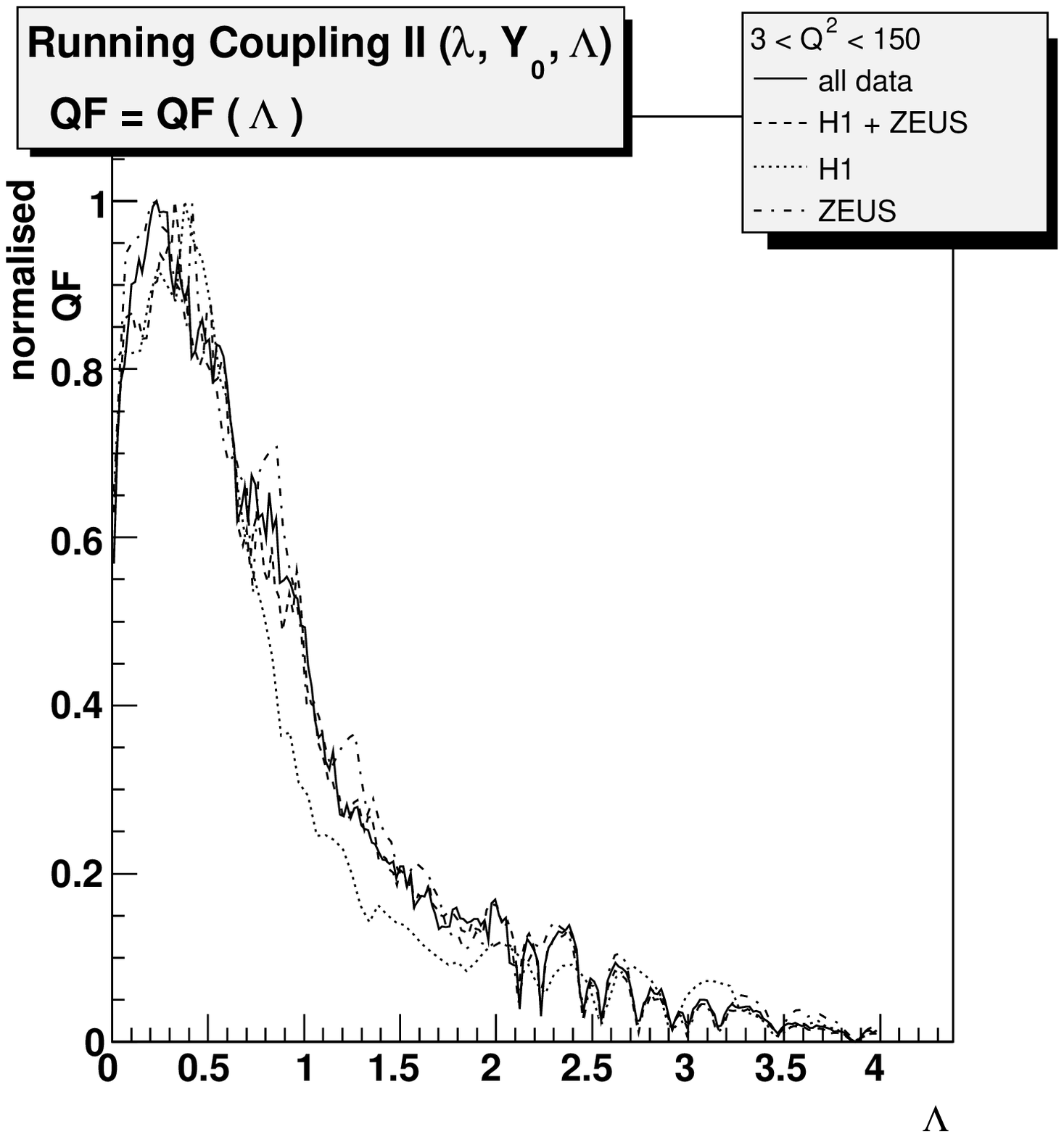,width=9.cm} &
\epsfig{file=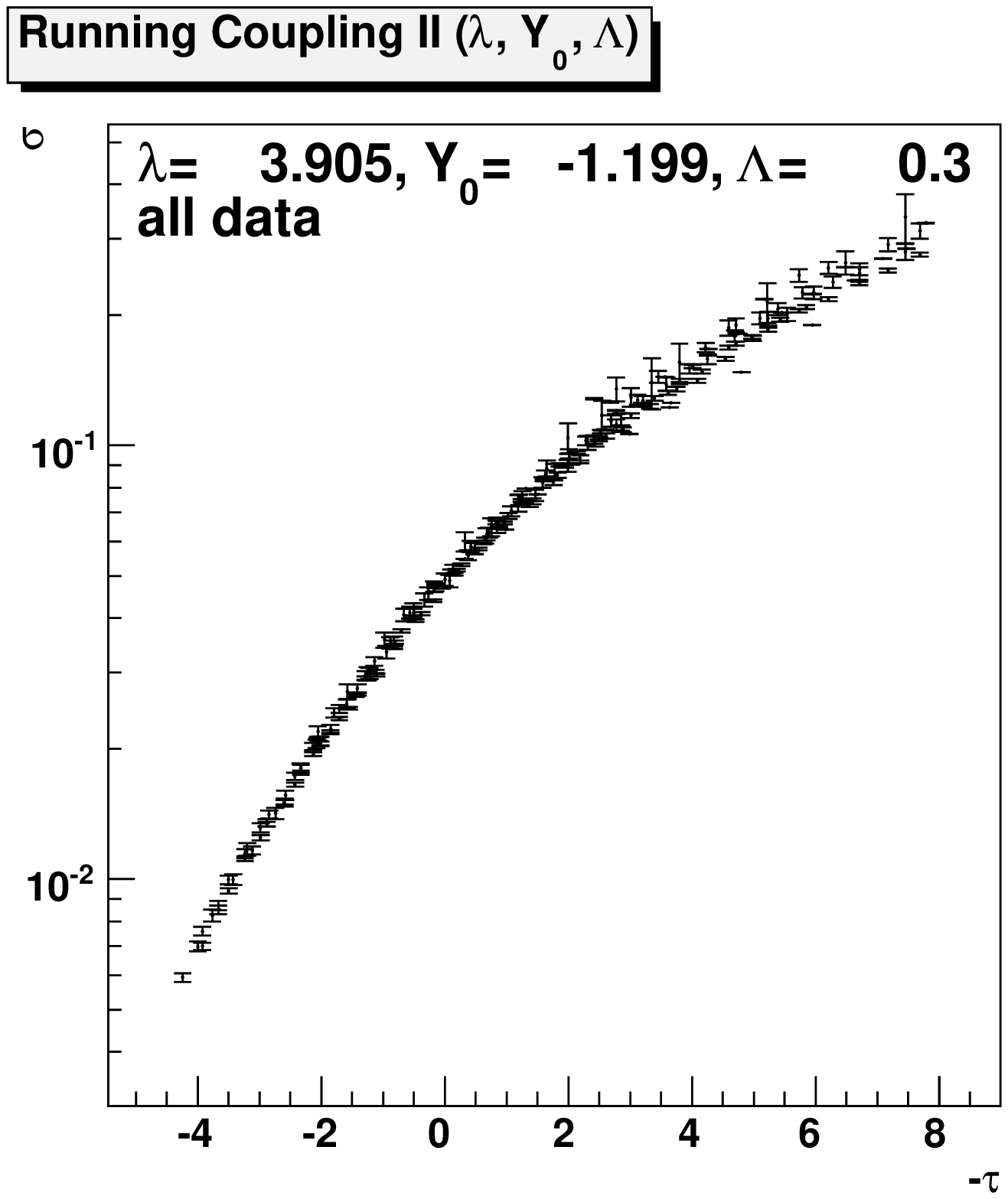,width=9.cm} 
\end{tabular}
\caption{{\bf $F_2$ data:} Normalised QF to 1. as a function of $\lambda$ and scaling curve 
with $\lambda$ fixed to the value corresponding to the best QF for ``Running
Coupling IIbis". A $Q^2>3$ cut was applied to the data.}
\label{F2_runIIext_3}
\end{center}
\end{figure}

\begin{figure}[t]
\begin{center}
\begin{tabular}{cc}
\epsfig{file=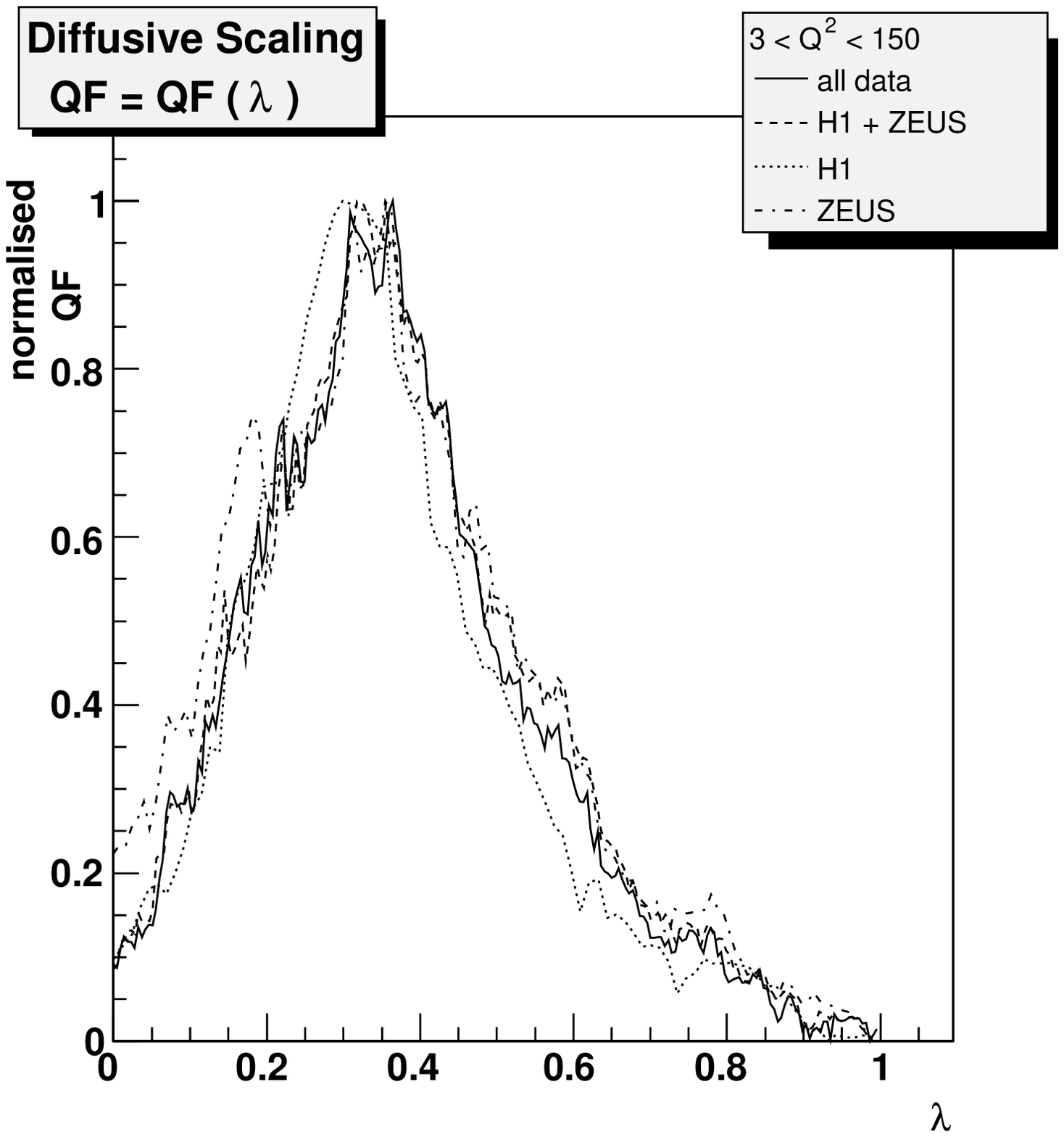,width=9.cm} &
\epsfig{file=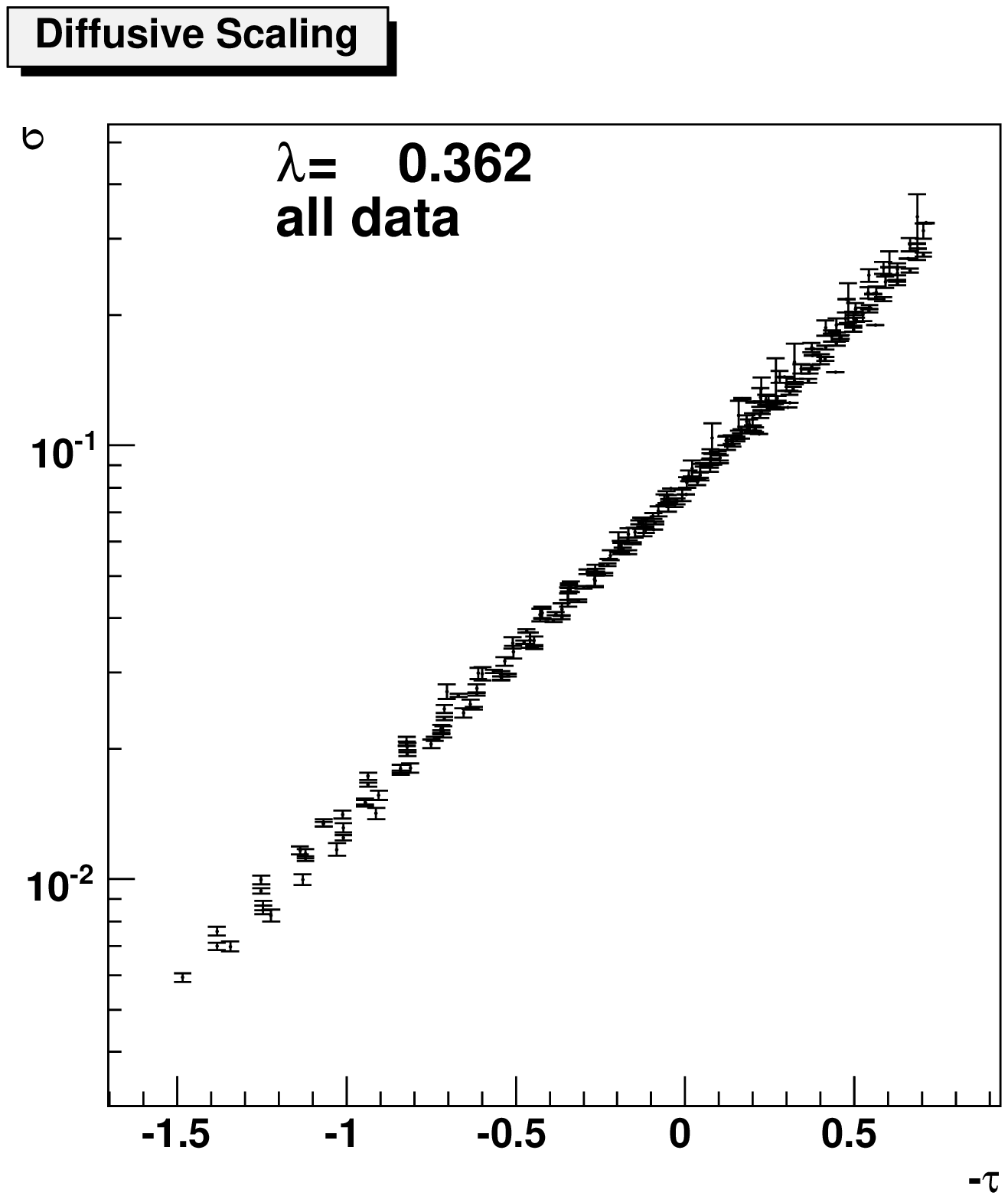,width=9.cm}
\end{tabular}
\caption{{\bf $F_2$ data:} Normalised QF to 1. as a function of $\lambda$ and scaling curve 
with $\lambda$ fixed to the value corresponding to the best QF 
for ``Diffusive Scaling". A $Q^2>3$ cut was applied to the data.}
\label{F2_diff_3}
\end{center}
\end{figure}

We first give the results of the different scalings for the structure function $F_2$ with
$Q^2>3$~GeV$^2$. Fig.~\ref{F2_fixed_3}-\ref{F2_diff_3}
show the normalised QF dependence on the value of the fitted parameters
and the scaling curves in the $Q^2>3$~GeV$^2$ range 
for the ``Fixed Coupling", ``Running Coupling I", ``Running Coupling II",
``Running Coupling II bis", and ``Diffusive Scaling" respectively.
To see the impact of the different data sets, we choose to divide the data points
in H1 alone, ZEUS alone, H1$+$ZEUS, and all data. In the figures,
we normalise the QF to 1. so that we are able to compare the values of
the $\lambda$, $Y_0$ and $\Lambda$ parameters for the different data sets.
Table \ref{F2_table} give the fit parameters and absolute QF values --- to
distinguish the quality of the different scalings ---
for the various scaling laws (see Table I).
On the different figures, we see that the $\lambda$ parameter is well determined in the fit
and the different data sets lead to values of $\lambda$ which are quite close to one
another. As an example, for fixed coupling, $\lambda$ ranges between 0.33 to 0.38,
for ``Running Coupling I" between 1.74 and 1.84, for ``Running Coupling II" between 3.23 and
3.44, and for ``Diffusive Scaling" between 0.31 and 0.37. The scaling plots (always on the
right of Fig.\ref{F2_fixed_3}, \ref{F2_runI_3}, \ref{F2_runII_3}, \ref{F2_runIIext_3}, and
\ref{F2_diff_3}) show similar behaviour for the different scalings while it is clear that
the scaling plot for diffusive scaling is slightly worse.

To improve the scaling quality, we checked the dependence of our results to additional
parameters which can be introduced in the scaling. Introducing 
a shift in rapidity $Y_0$ or a scale $\Lambda$ in the ``Fixed Coupling" scaling does not change the
results since these new parameters disappear in the definition of QF. This effect is
the same if one introduces a $\Lambda$ scale in the ``Running Coupling I" scaling.
Introducing a new parameter $Y_0$ to
shift the rapidity in the ``Running Coupling I" scaling does not lead to an improvement of the
scaling. The fit leads to a value of $Y_0$ compatible with 0 for all data sets considered.
The results are similar when one introduces the $Y_0$ parameter
and the scale $\Lambda$ in ``Diffusive
Scaling" and the parameters are found to be close to 0 and 1 after the fit.
We notice that ``Diffusive Scaling" including the additional parameters $Y_0$ and
$\Lambda$ admits also other solutions leading to better QF, but these solutions
are either instable from one data set to another or lead to unphysical values of
the parameters (large values of $\lambda$ and large negative values of $Y_0$).

The QF plots in figure~\ref{F2_runIIext_3} show the dependence on one of the parameters
while the remaining parameters are fixed to the fitted values.
It is important to notice that in the case of the ``Running
Coupling IIbis" version,
it is difficult to find
the absolute minimum since the $QF = QF(\lambda,Y_0,\Lambda)$ is not
smooth and shows several minima. 
There is even a plateau of minimal values in $\lambda$, $Y_0$
parameters and we picked the parameter values leading to the smallest values
of QF. This is why the ``Running Coupling IIbis" scaling law shows larger spread concerning the parameter values 
($\lambda$ varies between 3.91 and 4.89).
The typical values for
the $Y_0$ (respectively $\Lambda$) parameters vary between -1.2 and -2.5 (respectively 0.2
and 0.35). The values of $\Lambda$ are in a typical range of
non-pertubative scale in the proton as expected. 

We also notice in Table~\ref{F2_table} that the best quality factor is obtained for 
``Running Coupling IIbis"
for all data sets while the worst one is for the diffusive scaling. The comparison of the
different QF is also shown in Fig.~\ref{F2_QF}, left.

\subsection{Fits to $F_2$ data with $Q^2>1$ GeV$^2$}
\begin{figure}[t]
\begin{center}
\begin{tabular}{cc}
\epsfig{file=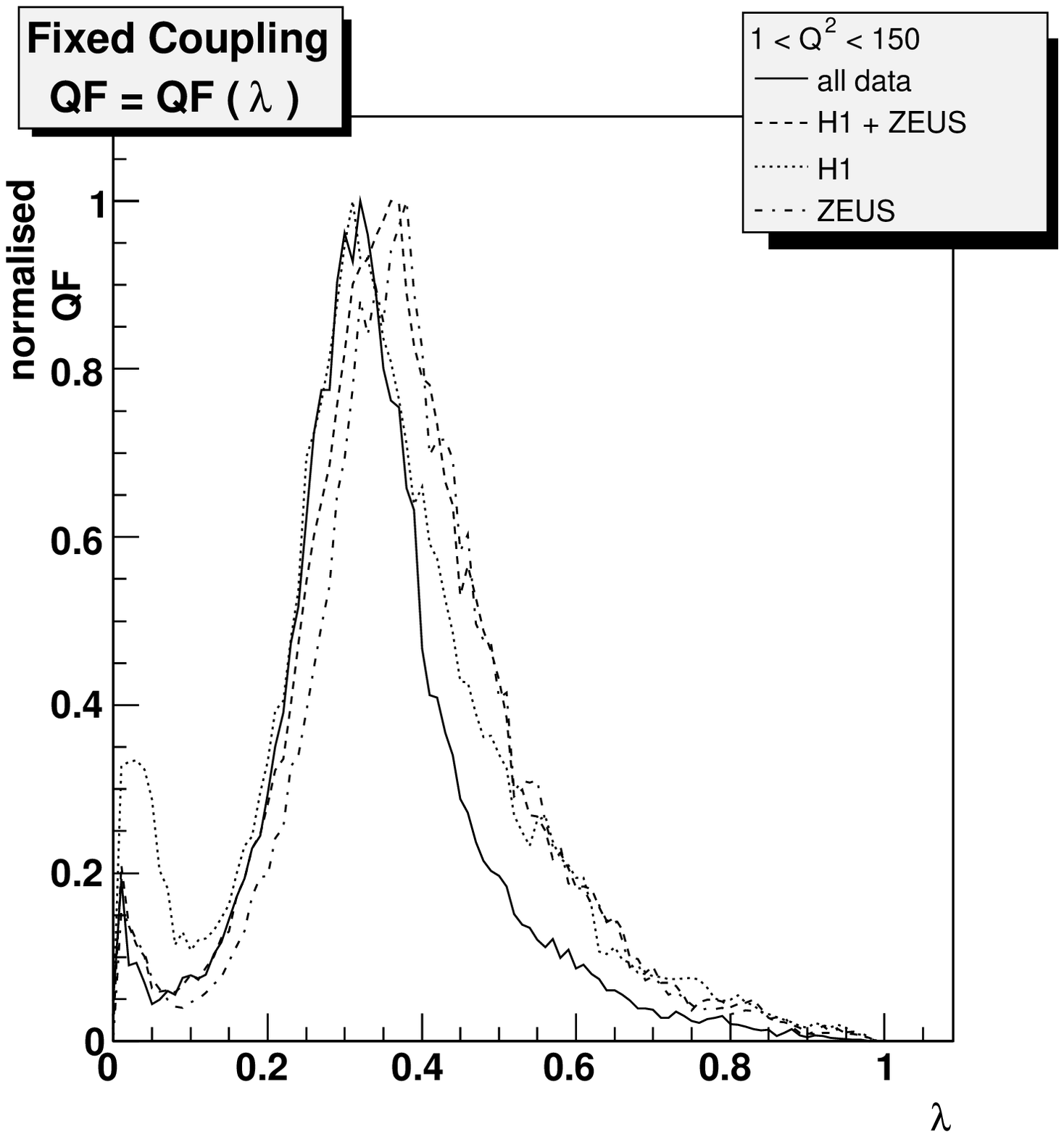,width=7.cm} &
\epsfig{file=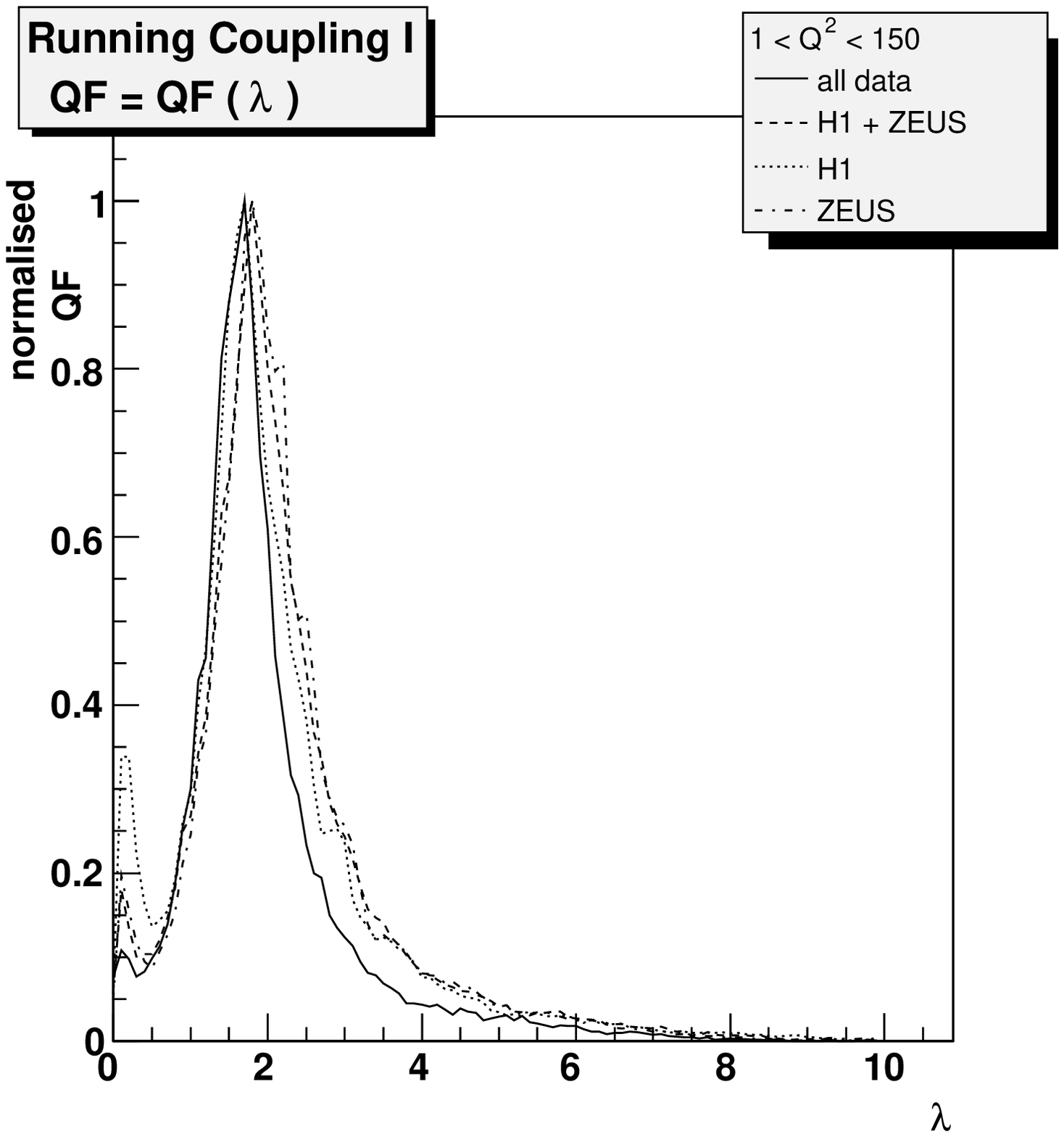,width=7.cm} \\
\epsfig{file=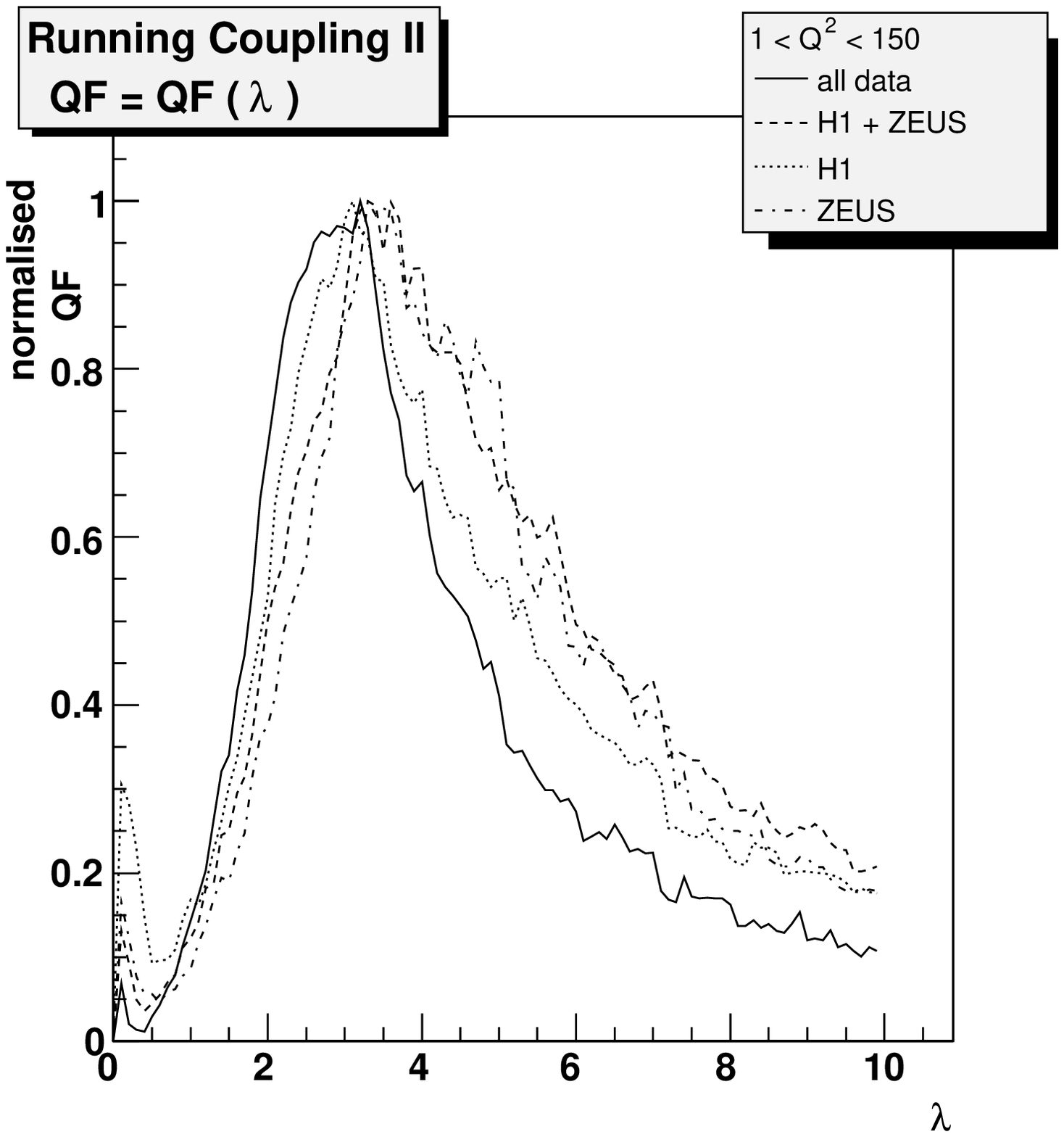,width=7.cm} &
\epsfig{file=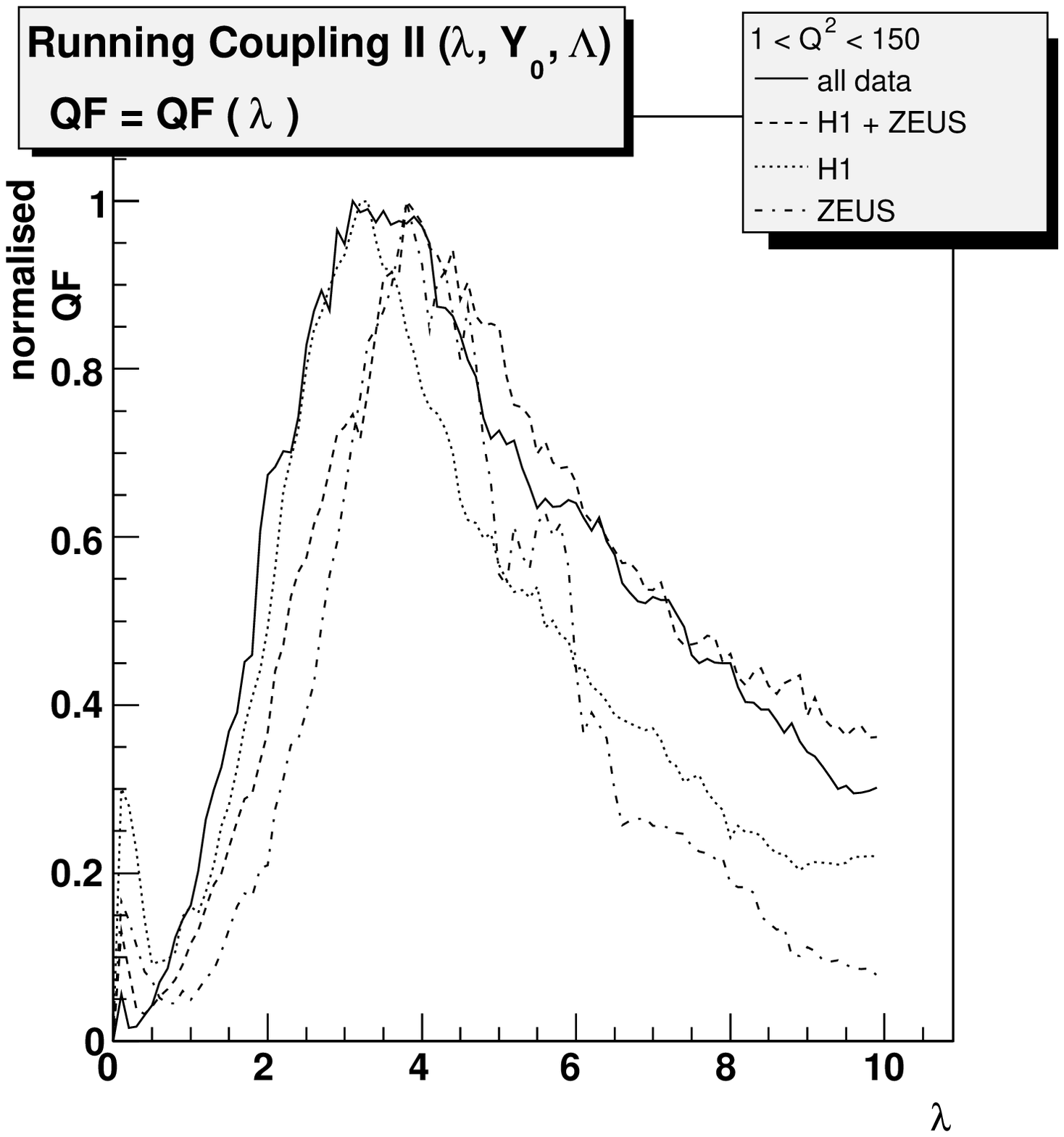,width=7.cm} \\
\epsfig{file=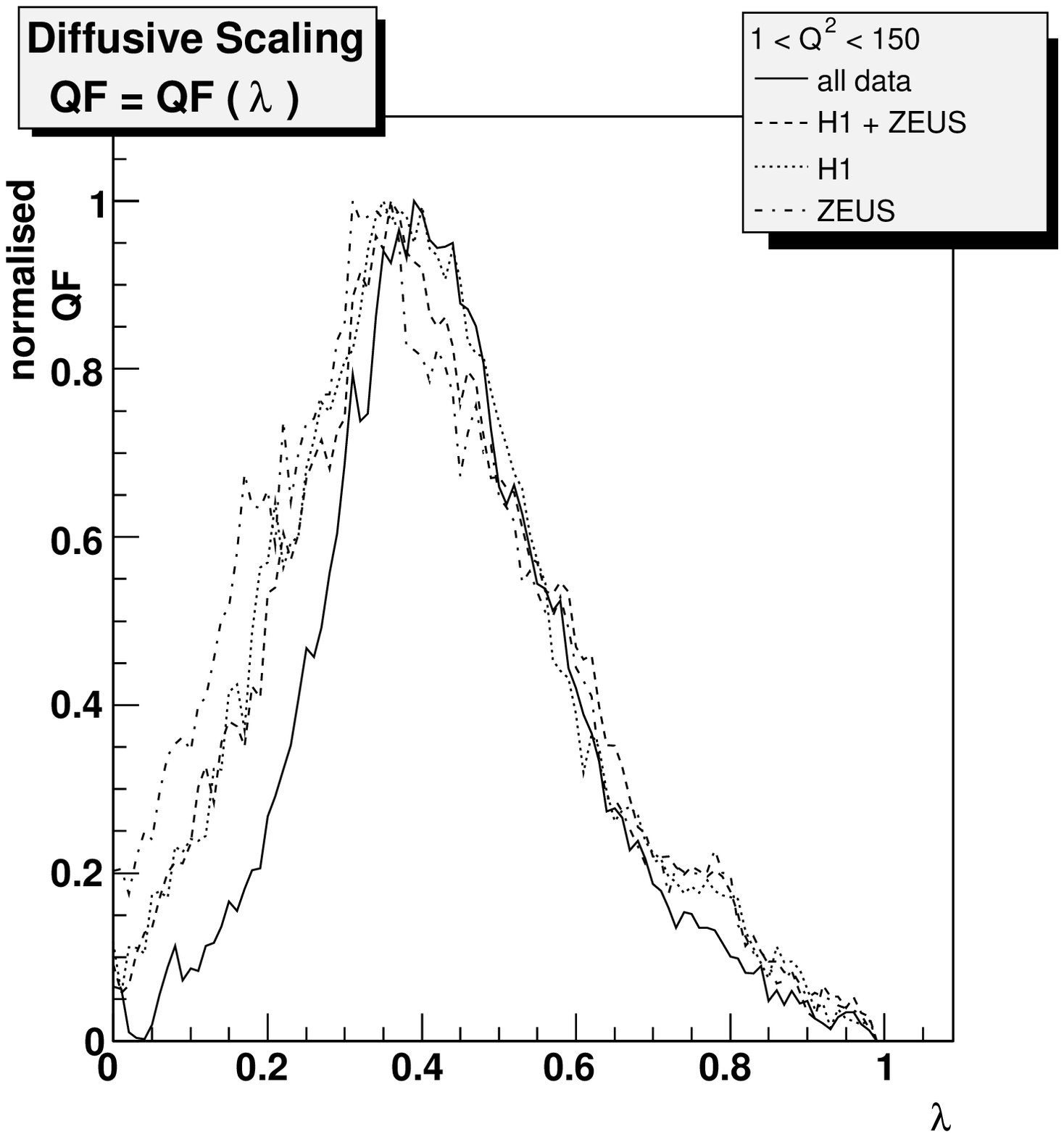,width=7.cm} &
\end{tabular}
\caption{{\bf $F_2$ data:} $\lambda$ dependence of the normalised QF to 1. for ``Fixed Coupling",
``Running Coupling I", ``Running Coupling II", ``Running Coupling IIbis" and
``Diffusive Scaling".
A $Q^2>1$ cut was applied to the data.}
\label{F2_QF_1}
\end{center}
\end{figure}

\begin{figure}[t]
\begin{center}
\begin{tabular}{cc}
\epsfig{file=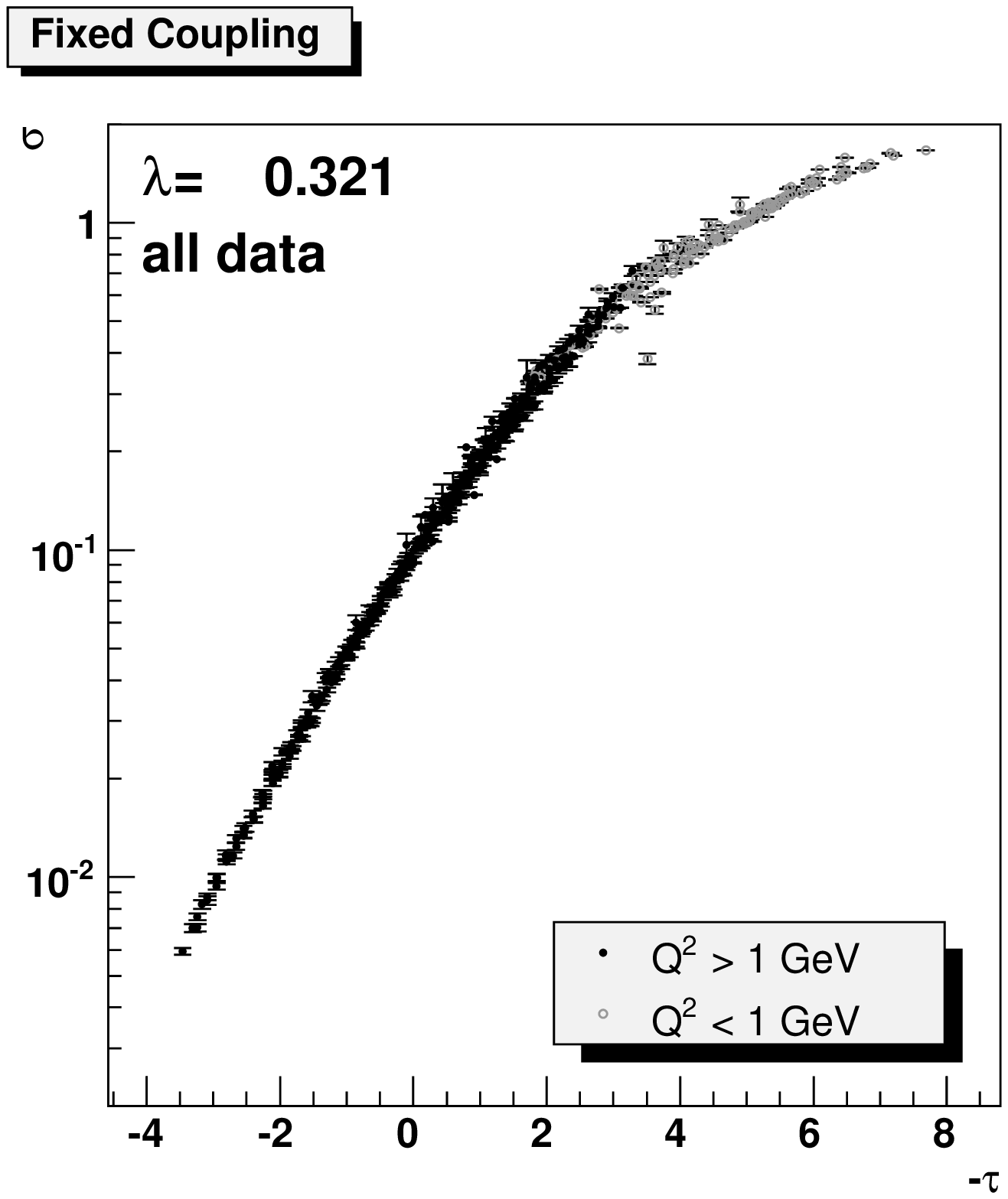,width=7.cm} &
\epsfig{file=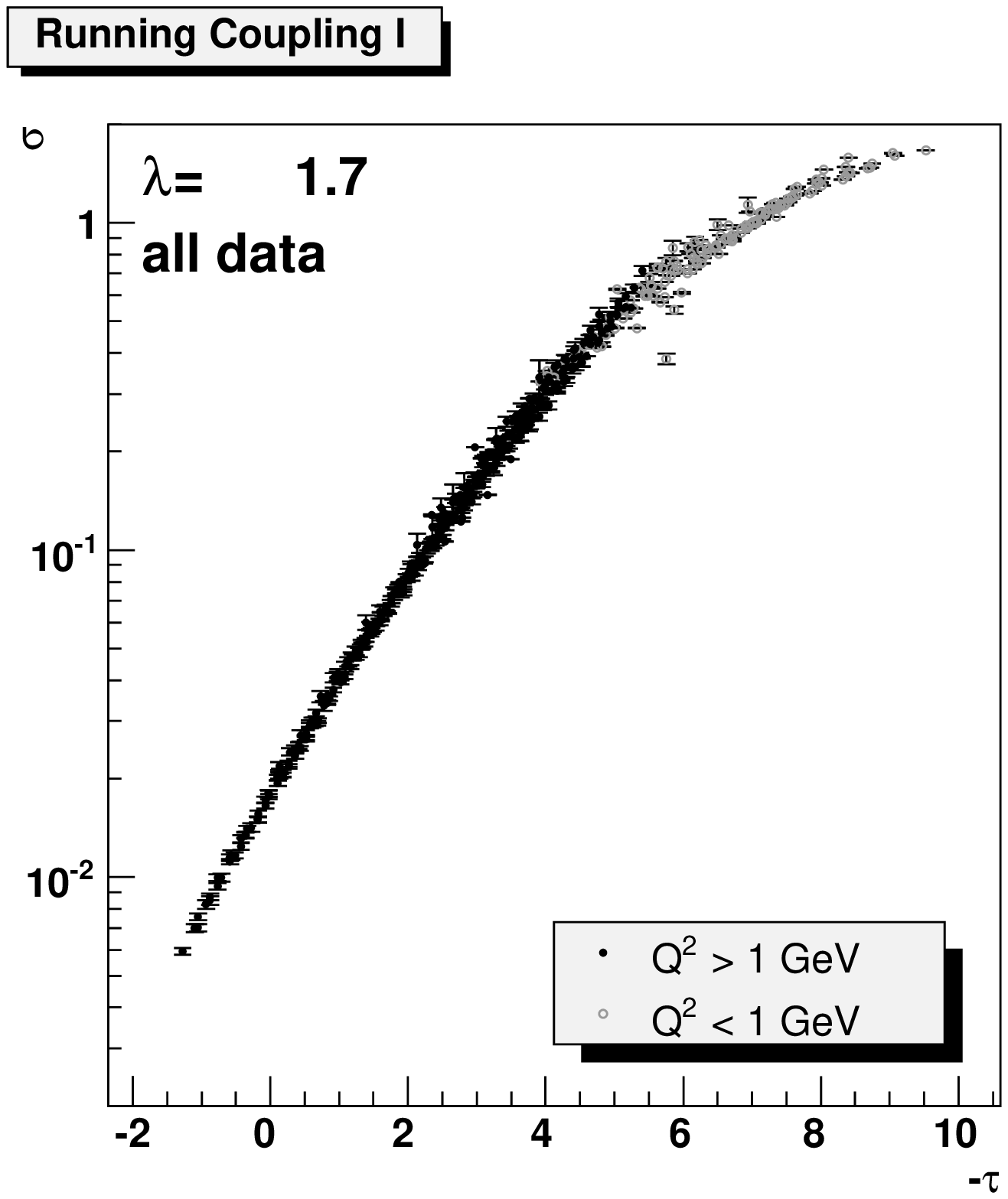,width=7.cm} \\
\epsfig{file=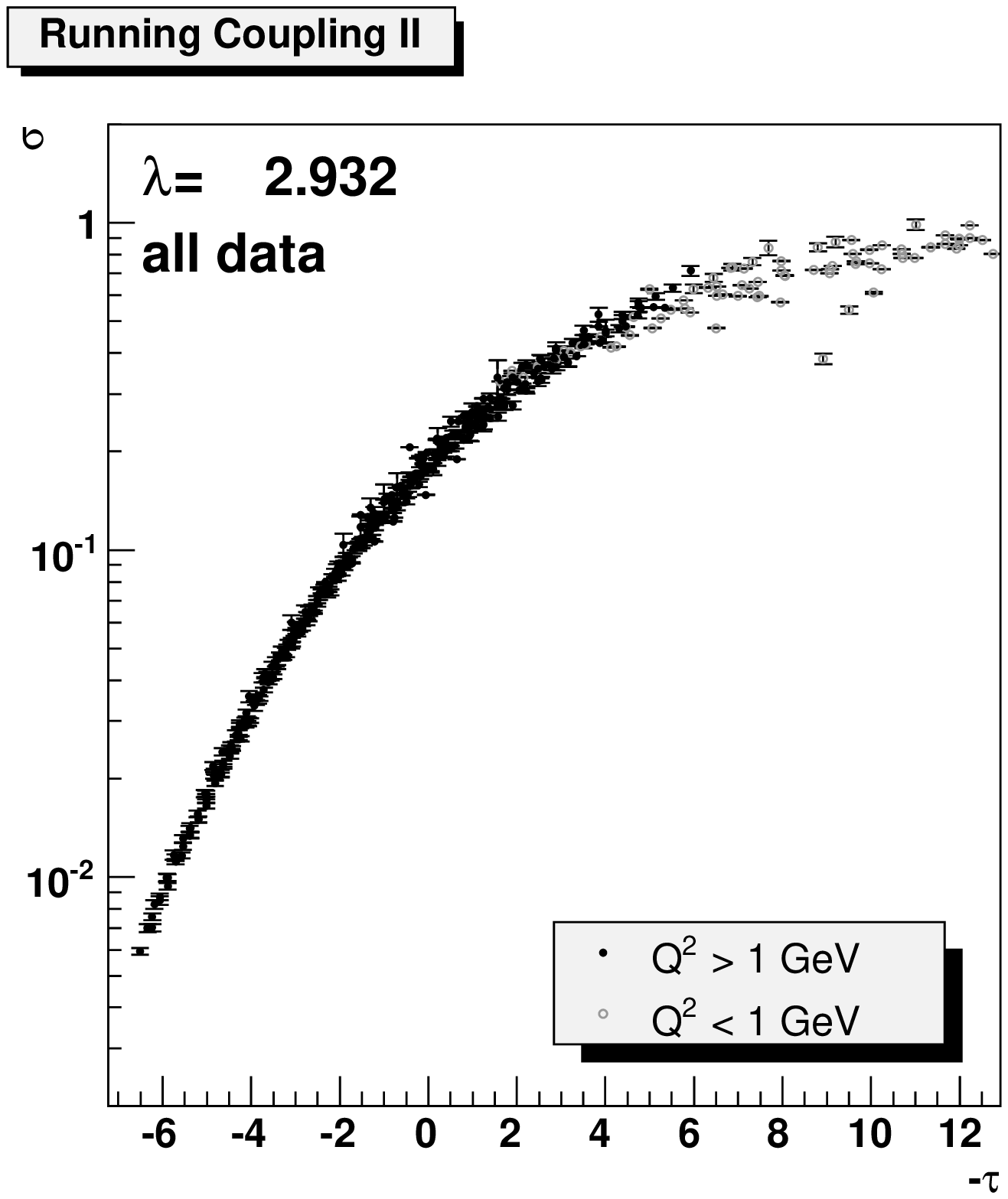,width=7.cm} &
\epsfig{file=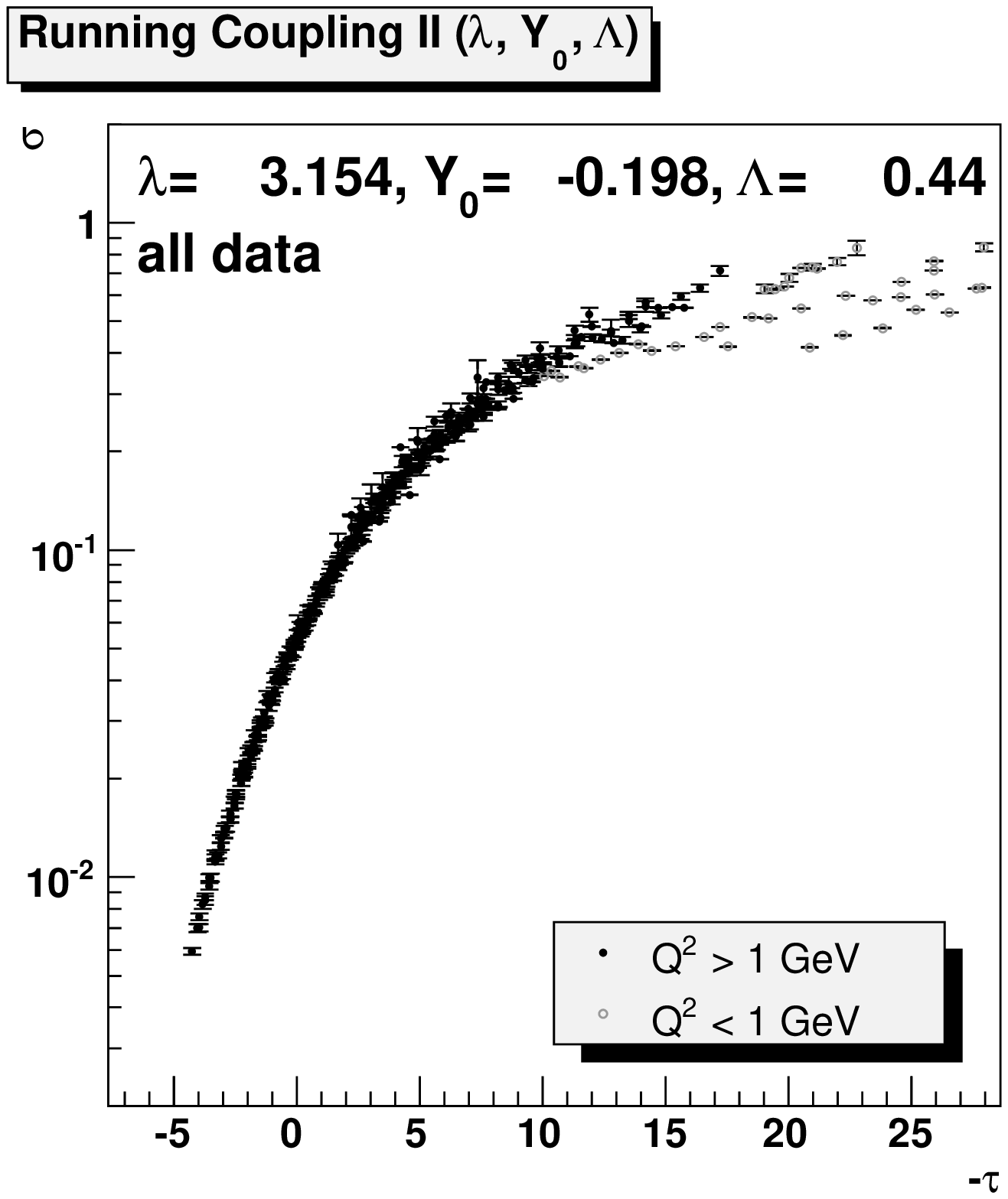,width=7.cm} \\
\epsfig{file=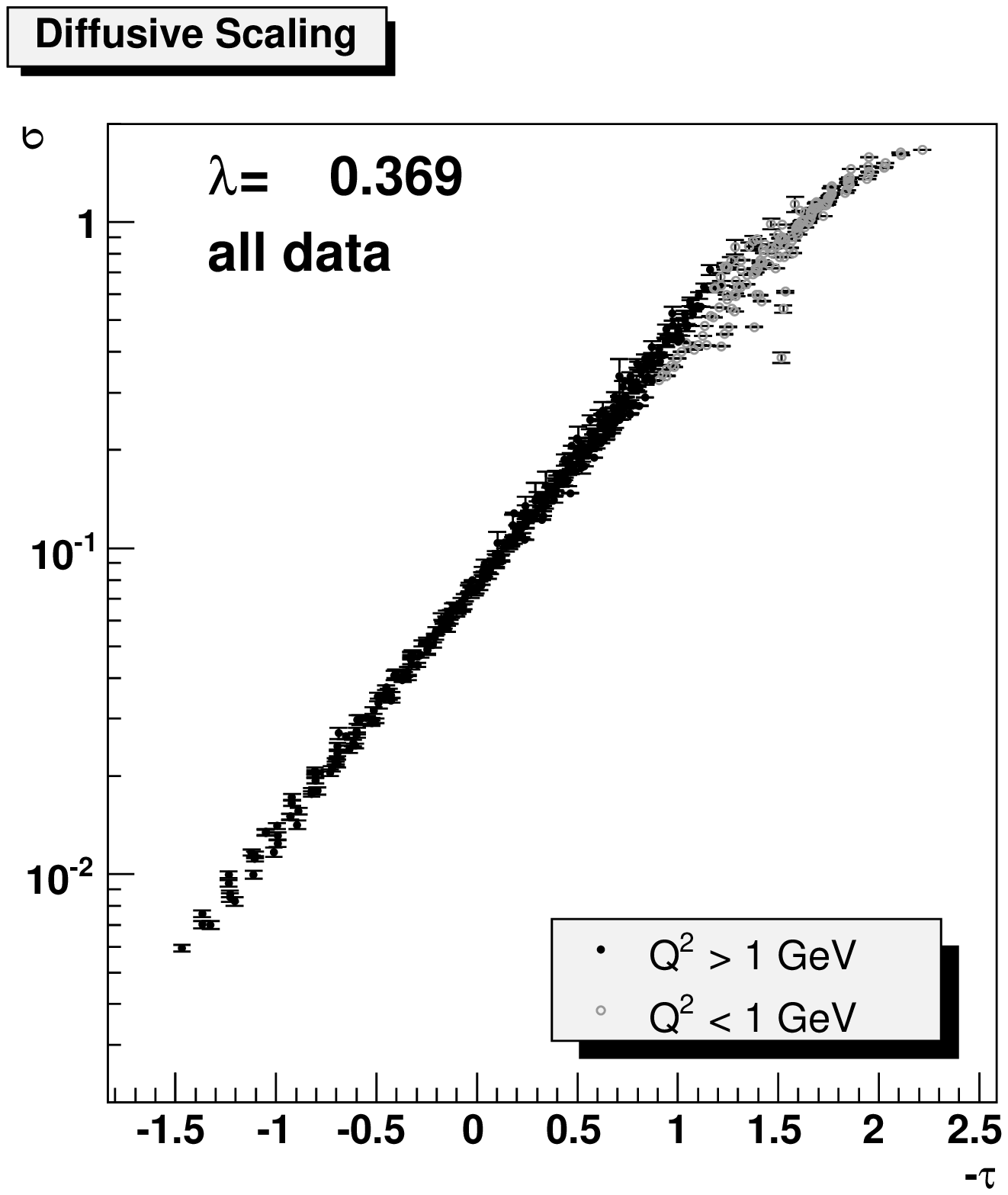,width=7.cm}
\end{tabular}
\caption{{\bf $F_2$ data:} scaling plot for ``Fixed Coupling",
``Running Coupling I", ``Running Coupling II", ``Running Coupling IIbis" and ``Diffusive
Scaling". 
A $Q^2>1$ cut was applied to the data to obtain the values of the
$\lambda$ parameter. The data with $Q^2<1$, not included
in the fit, are also shown as grey points.}
\label{F2_scaling_1}
\end{center}
\end{figure}

Figure~\ref{F2_QF_1} shows the QF plots for different scalings in the
$Q^2$ range $[1;150]$~GeV$^2$. We notice that the spread on the parameter values is larger than after the cut
on $Q^2>3$~GeV$^2$. The values of the parameters are given in 
Table~\ref{F2_table}. We also notice a tendency to find lower values of $\lambda$ for the
``Fixed Coupling", ``Running Coupling I" and ``Running Coupling II" scalings when one introduced
more low $Q^2$ data. The QF values are similar for the ``Fixed Coupling", ``Running Coupling I",
and ``Running Coupling IIbis" --- with a tendency to be slightly better for ``Running Coupling
IIbis" --- and is worse for diffusive scaling as before. 
The corresponding scaling curves are displayed in
Fig. \ref{F2_scaling_1}. They show similar behaviour with a tendency for 
``Diffusive Scaling" to show a larger spread of data.
These figures also depict the  $Q^2<1$~GeV$^2$ data points in grey,
which were not used in the fit. These data points tend to show the same kind of
scalings with a larger spread, and this is why they were not included in the fit.

\begin{figure}[t]
\begin{center}
\begin{tabular}{cc}
\epsfig{file=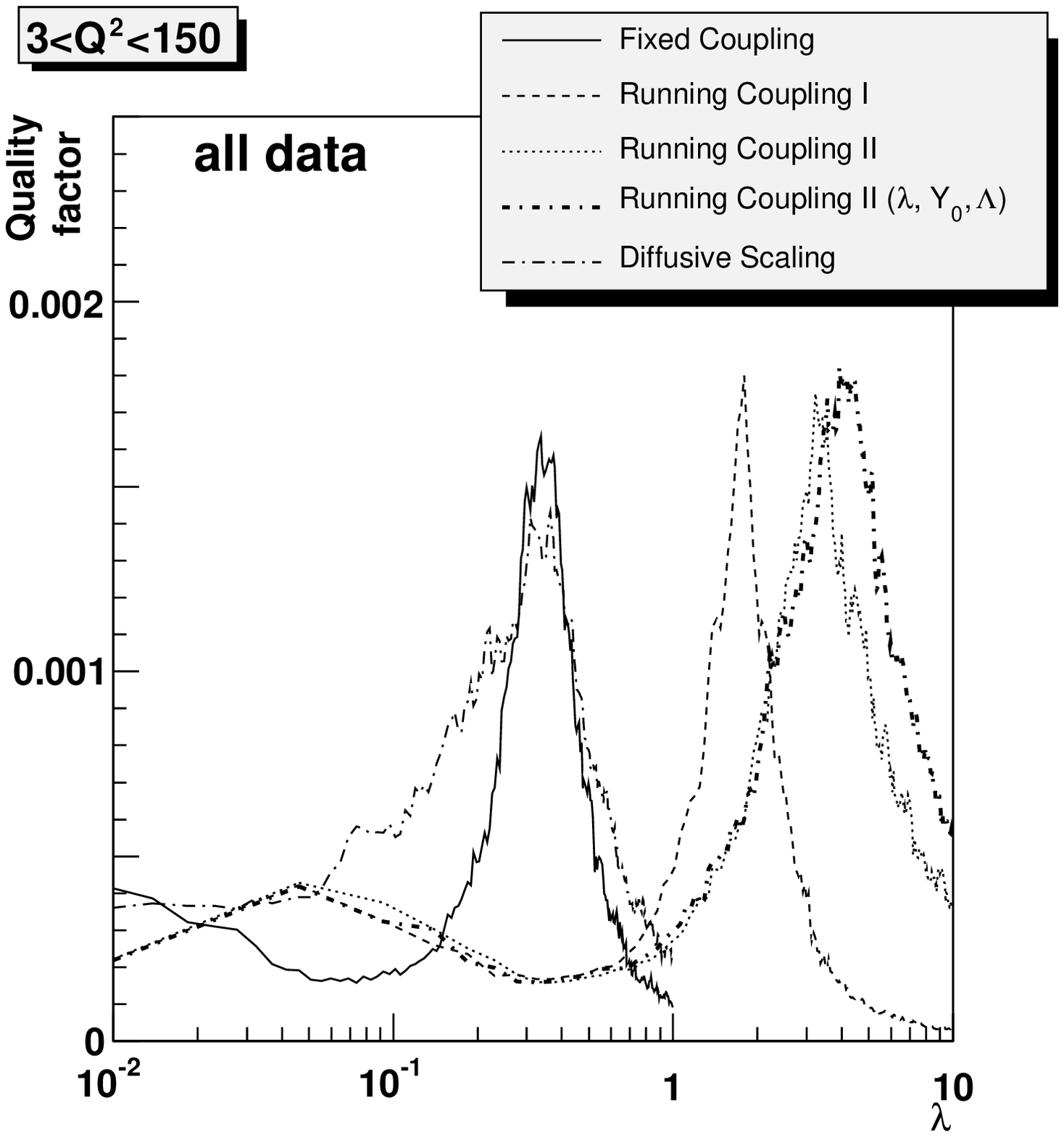,width=9.cm} &
\epsfig{file=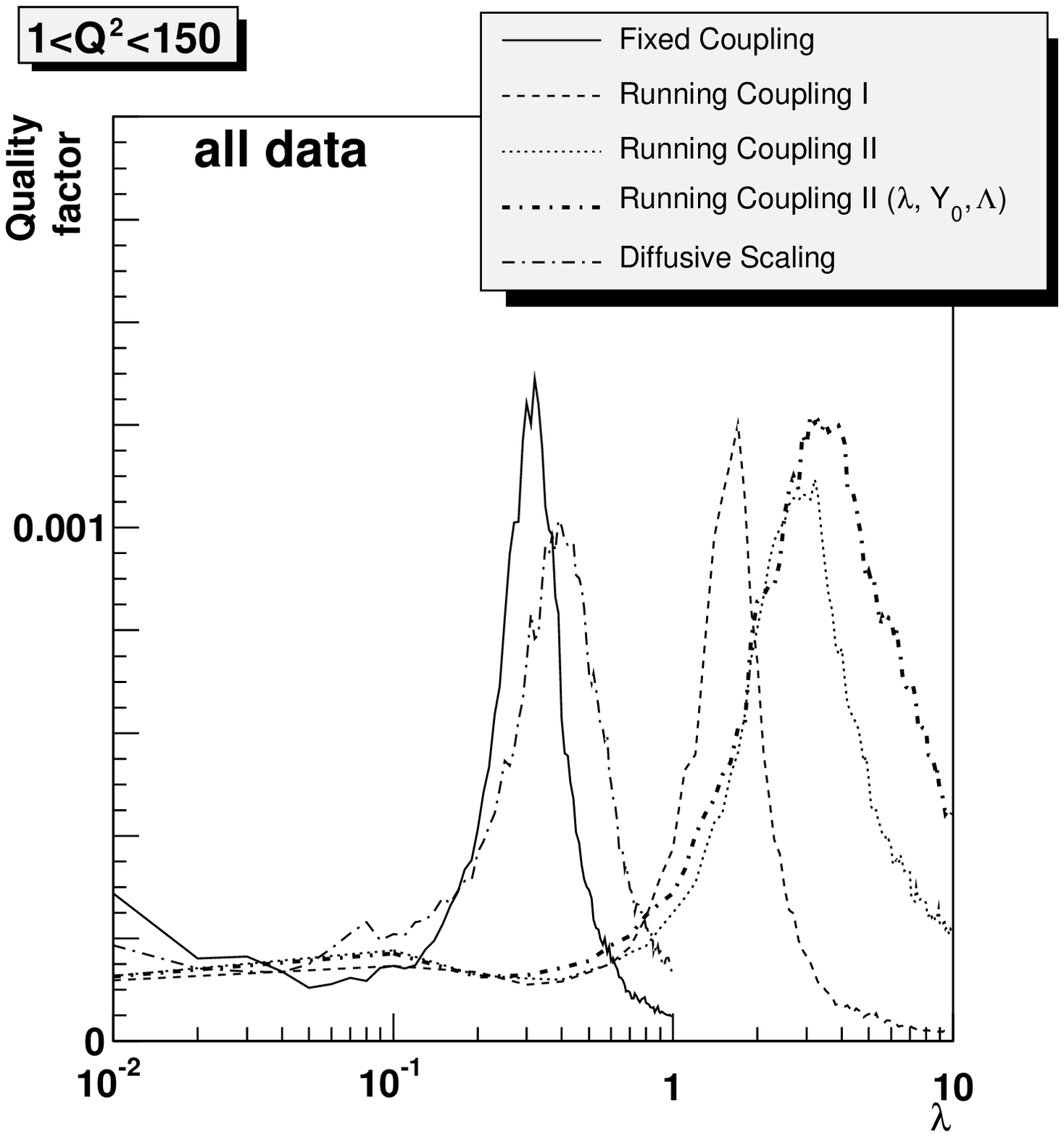,width=9.cm}
\end{tabular}
\caption{{\bf $F_2$ data:} 
Comparison of the QF for the different scaling as a function of $\lambda$. 
A $Q^2>3$ (left plot) or $Q^2>1$ (right plot) cut was applied to the data.}
\label{F2_QF}
\end{center}
\end{figure}

Fig.~\ref{F2_QF}, right, gives the $QF = QF(\lambda)$ curves for all data and
different scaling laws in the $Q^2$ regions $[1;150]$~GeV$^2$ (on Fig.~\ref{F2_QF}
left in the $[3;150]$~GeV$^2$ range) so that the different scalings can 
be compared easily.
As we mentionned already, in both $Q^2$ ranges, the ``Fixed Coupling",
the ``Running Coupling I", and the ``Running Coupling II" are quite good,
and the ``Diffusive Scaling" is disfavoured. For $Q^2>3$~GeV$^2$, the ``Running
Coupling IIbis"
scaling shows the best QF. 

\subsection{Dependence of the fitted parameters on $Q^2$}
\begin{figure}[t]
\begin{center}
\epsfig{file=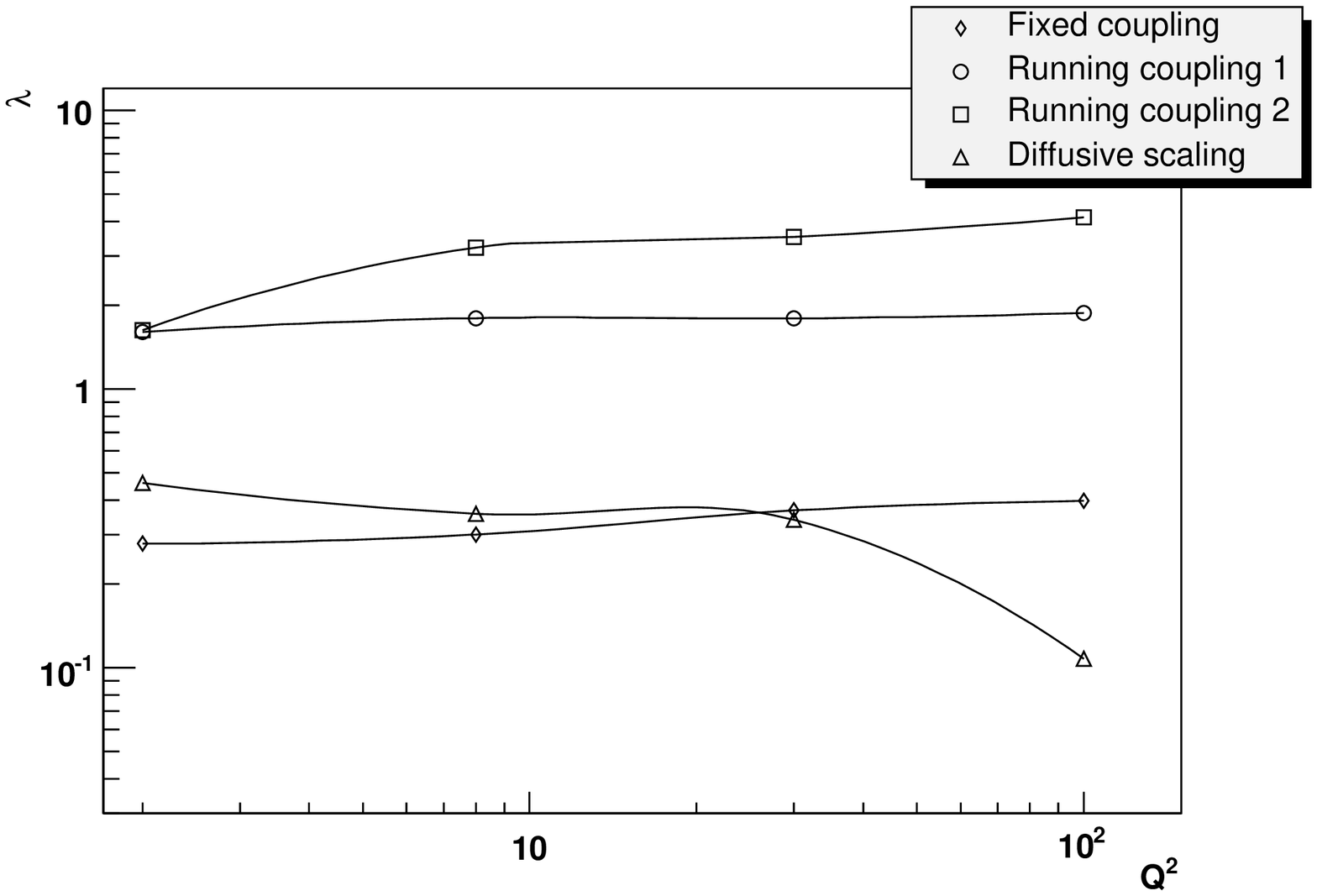,width=12.cm}
\caption{{\bf $F_2$ data:} 
Variation of lambda as a function of $Q^2$ for the different scalings.}
\label{lambdaQ2}
\end{center}
\end{figure}

To study the $Q^2$ dependence of the fitted parameters and to quantify
the amplitude of scaling violations,
the data points are divided into four separate $Q^2$ samples:
$[1;3]$, $[3;10]$, $[10;35]$, and $[35;150]$~GeV$^2$.
The number of points in each region is similar (see table~\ref{lambdaQ2_table}).
The lower and upper bounds of $1$~GeV$^2$ and $150$~GeV$^2$ are chosen 
for the same reason as before. The $Q^2$ dependence of the
parameter $\lambda$ is depicted in figure~\ref{lambdaQ2},
and the fitted parameters together with the QF values
are shown in table~\ref{lambdaQ2_table}. There is a slight increase
of the $\lambda$ parameter in the case of ``Fixed Coupling" while
``Running Coupling I" is quite flat. This can be easily understood
since ``Running Coupling I" shows a natural $Q^2$ evolution. 
We notice a stronger increase in the case of
``Running Coupling II", showing the breaking of this scaling as a function
of $Q^2$. Since this scaling gives already the best QF, it would be worth to study the
breaking of scaling and introduce it in the model to improve further the description of the
data. 
The $\lambda$ parameter decreases strongly
(especially in the last $Q^2$ bin) for the diffusive scaling. The fact that $\lambda$
depends strongly on $Q^2$ confirms the fact that this scaling leads to the worst
description of the data.

\subsection{Fits to DVCS data from H1 and ZEUS}
\begin{figure}[t]
\begin{center}
\begin{tabular}{cc}
\epsfig{file=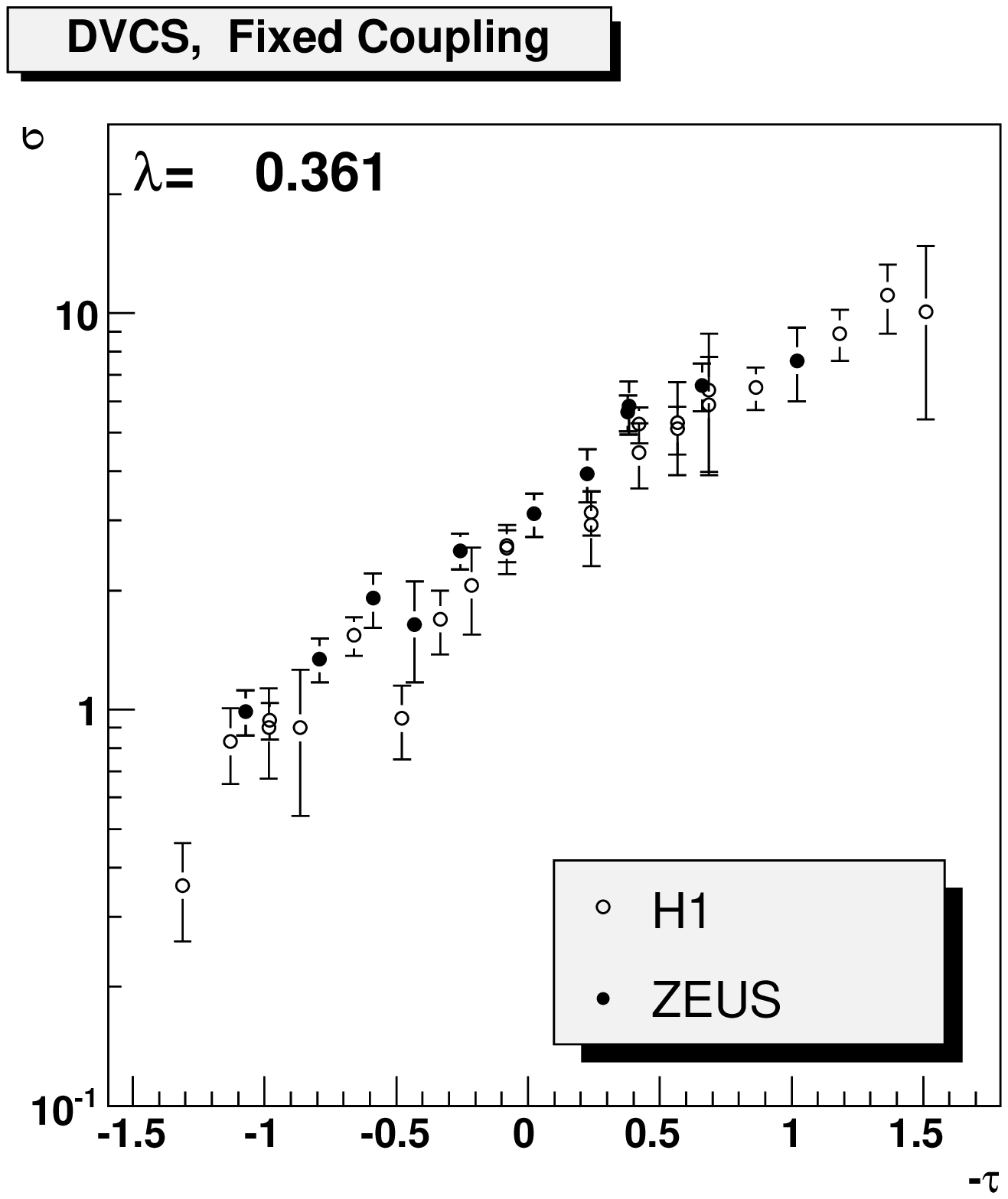,width=7.cm} &
\epsfig{file=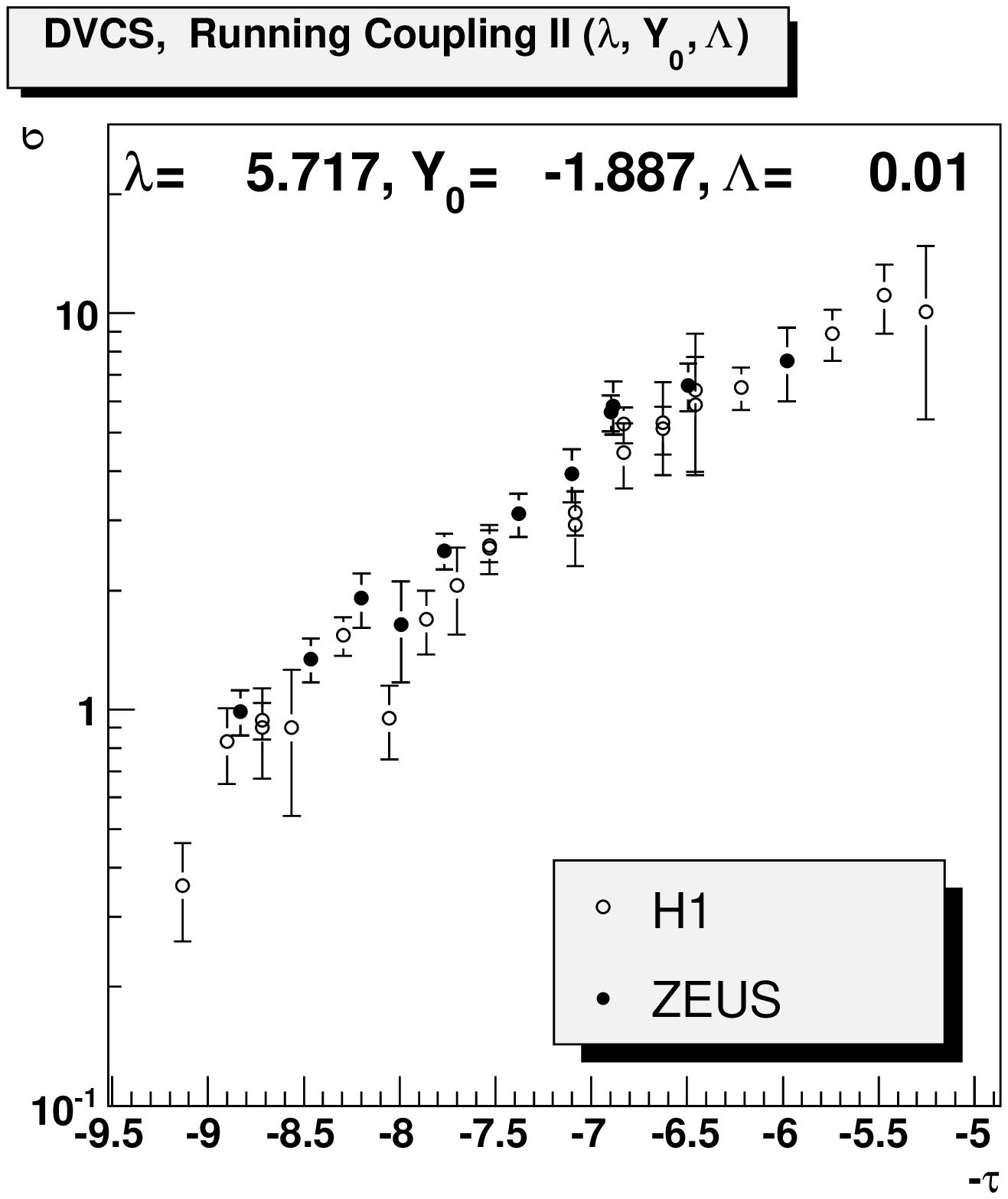,width=7.cm}\\
\end{tabular}
\caption{\label{DVCS_1} {\bf DVCS data:} Scaling curves for DVCS data and for ``Fixed Coupling"
and ``Running
Coupling IIbis".}
\end{center}
\end{figure}

\begin{figure}[t]
\begin{center}
\begin{tabular}{cc}
\epsfig{file=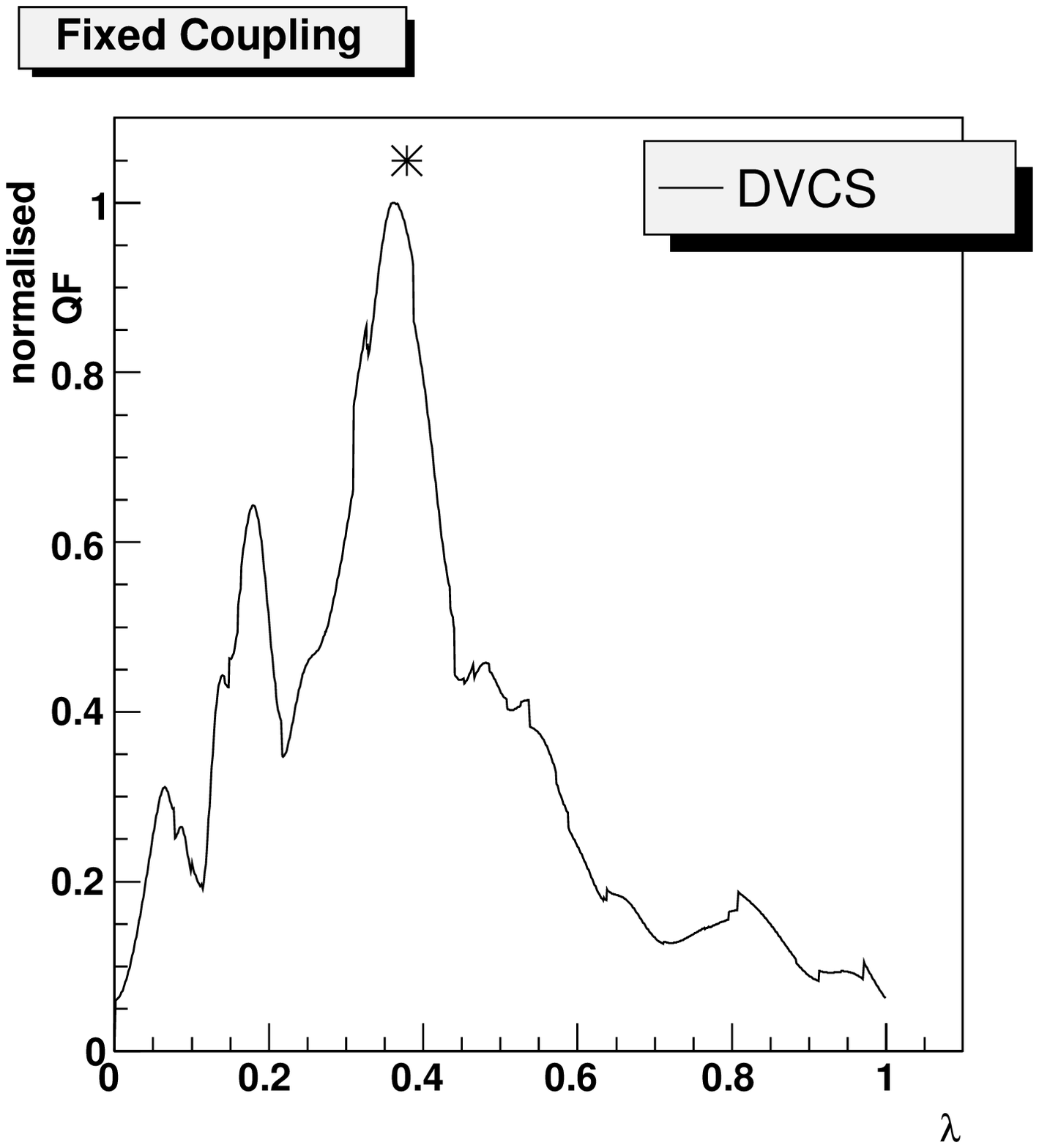,width=7.cm} &
\epsfig{file=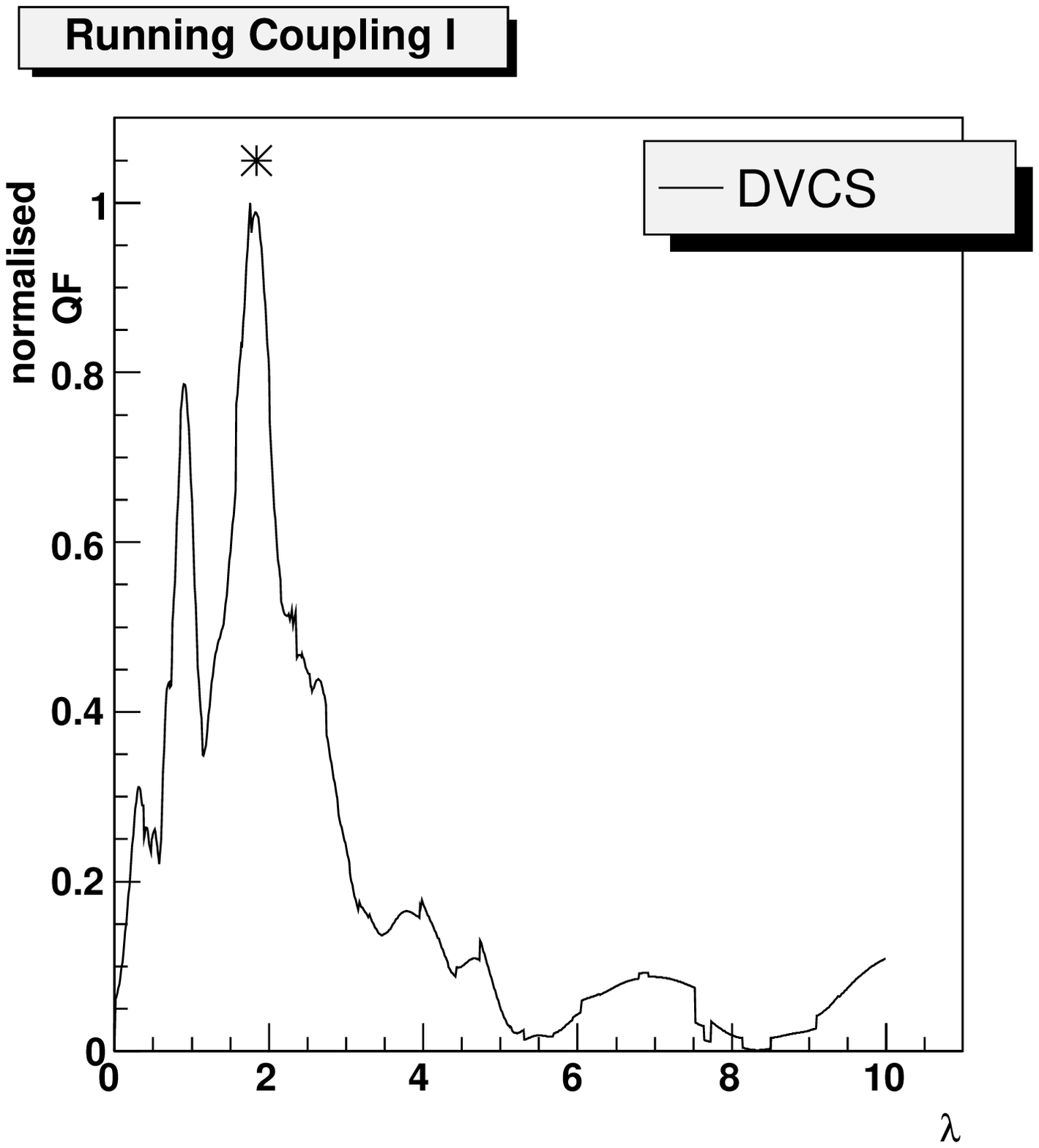,width=7.cm}\\
\epsfig{file=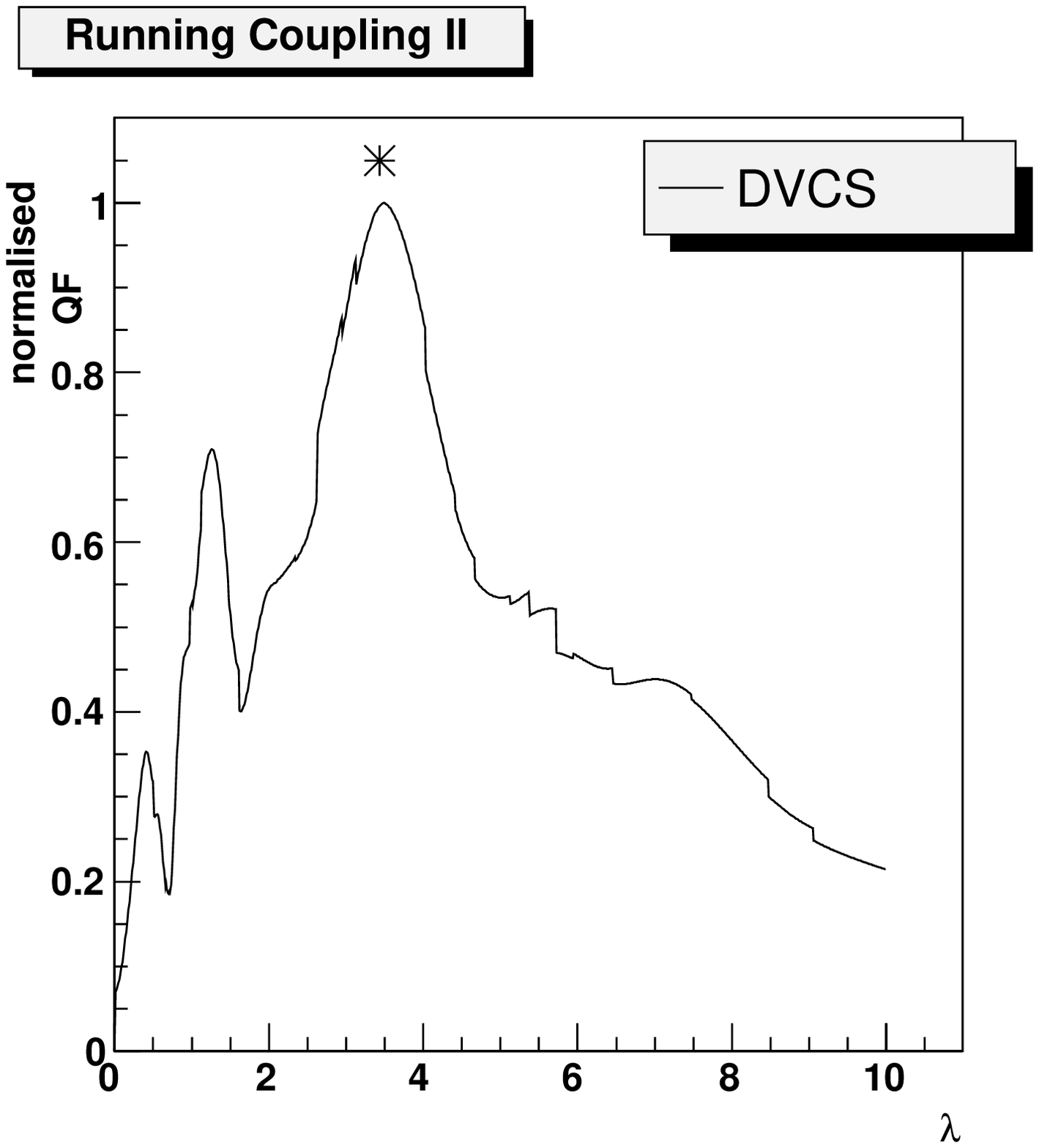,width=7.cm} &
\epsfig{file=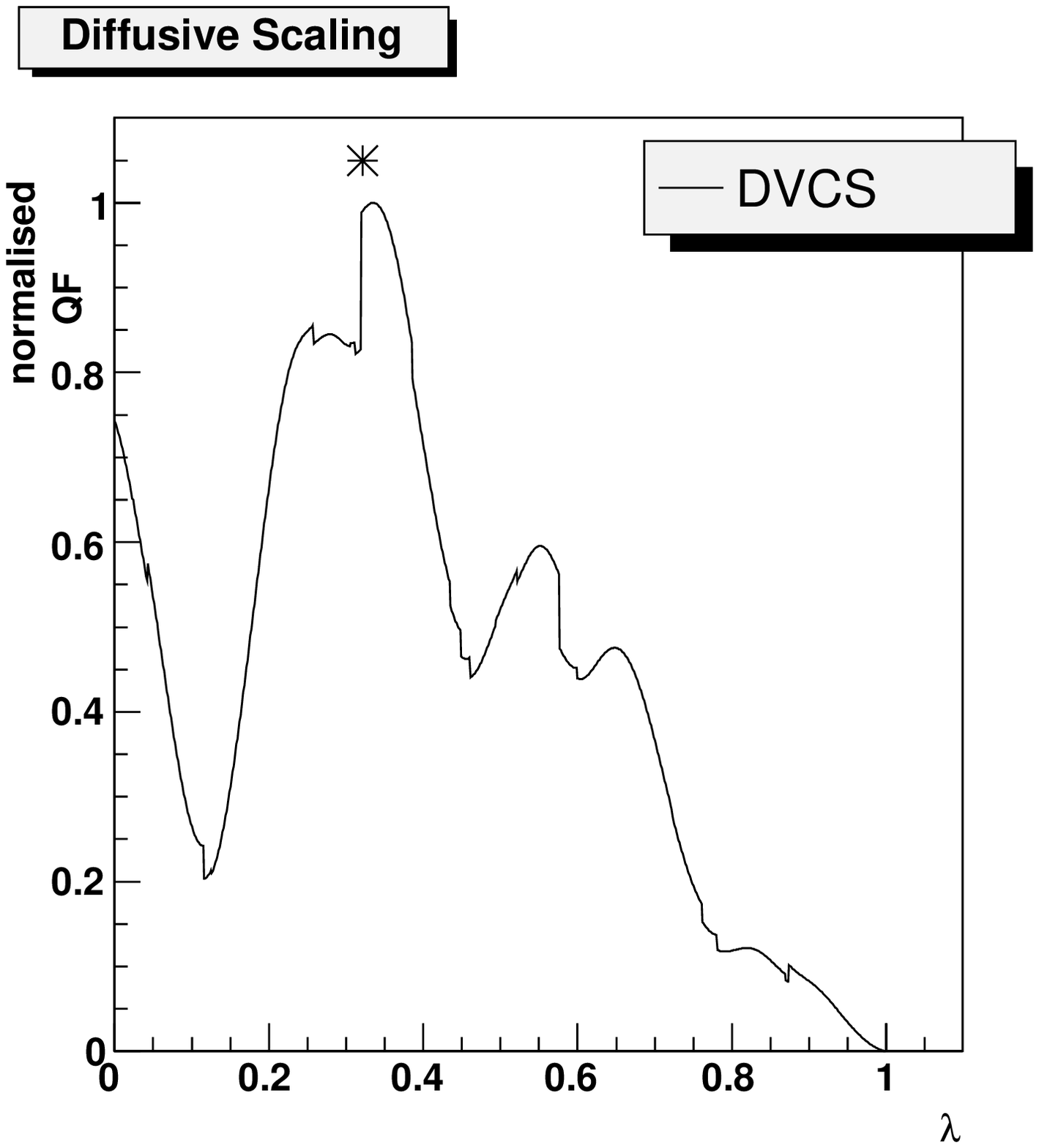,width=7.cm}\\
\end{tabular}
\caption{{\bf DVCS data:} $\lambda$ dependence of the normalised QF
to 1. for DVCS data and for ``Fixed Coupling",
``Running Coupling I", ``Running Coupling II" and ``Diffusive Scaling". The star indicated the
values of $\lambda$ obtained with a fit to $F_2$, $Q^2>3$ GeV$^2$.
The values of $\lambda$ obtained in a fit to DVCS data are similar to te ones obtained in a fit
to $F_2$ data.}
\label{DVCS_2}
\end{center}
\end{figure}

After fitting all H1 and ZEUS $F_2$ data, it is worth studying whether the DVCS data measured by
the same experiments~\cite{DVCS} lead to the same results. The amount of data is smaller
(34 points for H1 and ZEUS requiring $x \le 0.01$ as for $F_2$ data) and the precision
on the $\lambda$ parameter will be weaker.
The kinematic coverage of the DVCS data covers a smaller region in $x$ and $Q^2$ as $F_2$:
$4<Q^2<25$ GeV$^2$ and $5\cdot10^{-4}<x<5\cdot10^{-3}$.
The results of the fits can be found in table~\ref{DVCS_table} and
Fig.\ref{DVCS_1} and \ref{DVCS_2}. The scaling results are displayed only for ``Fixed
Coupling" and ``Running Coupling IIbis" since all plots look similar. 
To facilitate the comparison between the results of the fits to $F_2$ and DVCS
data, a star is put in Fig.~\ref{DVCS_2} 
at the position of the $\lambda$ value fitted to the H1+ZEUS $F_2$ data
with $Q^2$ in the range $[3;150]$~GeV$^2$. We note that 
the DVCS data lead to similar
$\lambda$ values to the $F_2$ data, showing the consistency of the 
scalings. The values of the QF show a tendency to favour fixed coupling, but all different
scalings (even ``Diffusive Scaling") lead to reasonable values of QF.

\section{Implications for Diffraction and Vector Mesons}

In this section, we check if the scalings found in the previous section can also describe
diffractive and vector meson data. Since these data are much less precise than the $F_2$ or
DVCS data
and depend more on non-perturbative inputs (meson wave function, diffractive
parton distribution inputs...), we choose to impose the same values of parameters found in the previous section
and check if the scaling is also observed using this value. A fit to the diffractive data
leads to large uncertainties on the parameter values since the number of data points is quite
small and their uncertainties large.

Let us first describe the diffractive data. We use the latest $t$-integrated
diffractive cross section measurements from H1 and ZEUS experiment~\cite{F2d}, derived
from the diffractive processes $e+p \rightarrow e+X+Y$, where the proton stays
either intact or turns into a low mass state $Y$. The diffractive cross section
in terms of the $t$-integrated reduced cross section
$\sigma_r^{D(3)}(x_{I\!\!P},x,Q^2)$ reads

\bea
\frac{d^3\sigma^{ep\rightarrow eXY}}{dx_{I\!\!P}dxdQ^2} &=&
\frac{4\pi^2\alpha_{em}}{xQ^4} \left( 1-y+\frac{y^2}{2} \right) \sigma_r^{D(3)}(x_{I\!\!P},x,Q^2), \\
\sigma_r^{D(3)} &=& F_2^{D(3)} - \frac{y^2}{1+(1-y)^2}F_L^{D(3)}.
\eea
The latter equation can be simplified to $\sigma_r^{D(3)} = F_2^{D(3)}$, since
it is a very good approximation anywhere but at large $y$. H1 large rapidity gap measurements
are realised with $M_Y<1.6$~GeV cut, whereas ZEUS measures with the $M_Y<2.3$~GeV cut using 
the $M_X$-method.
Both experiments do not measure exactly the same cross section. However,
the difference is a known constant factor; the ZEUS data points can be converted to the
same $M_Y$ range as H1 by multiplying ZEUS values by factor of 0.85. In case of the
$e+p \rightarrow e+X+p$ processes where the proton is tagged in Roman pot detectors (both H1 and ZEUS),
the data points have to be multiplied by 1.23.

In order to test the geometrical scaling properties we plot
the $\tau$ dependence of 
$\beta d\sigma_{diff}^{\gamma^*p\rightarrow Xp}/d\beta$, which
can be expressed in terms of diffractive structure function as follows
\be
\beta \frac{d\sigma_{diff}^{\gamma^*p\rightarrow Xp}}{d\beta} = \frac{4\pi^2\alpha_{em}}{Q^2}x_{I\!\!P}F_2^{D(3)}.
\ee
The definition of the $\tau$ variable is the same as in Table I replacing the
variable $x$ by $\xp$, $Q^2$ remaining the same.

\begin{figure}[t]
\begin{center}
\begin{tabular}{cc}
\epsfig{file=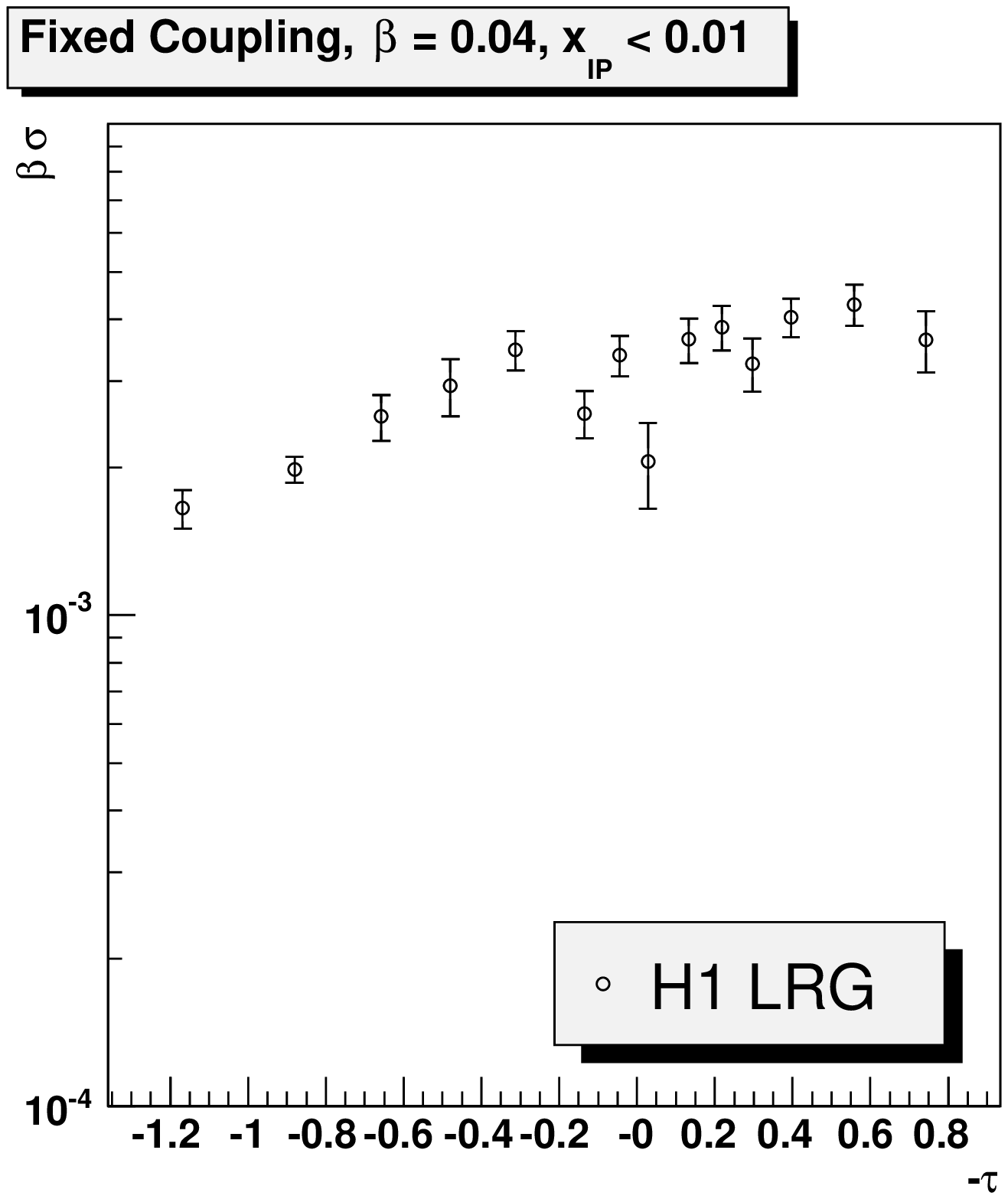,width=7.cm} &
\epsfig{file=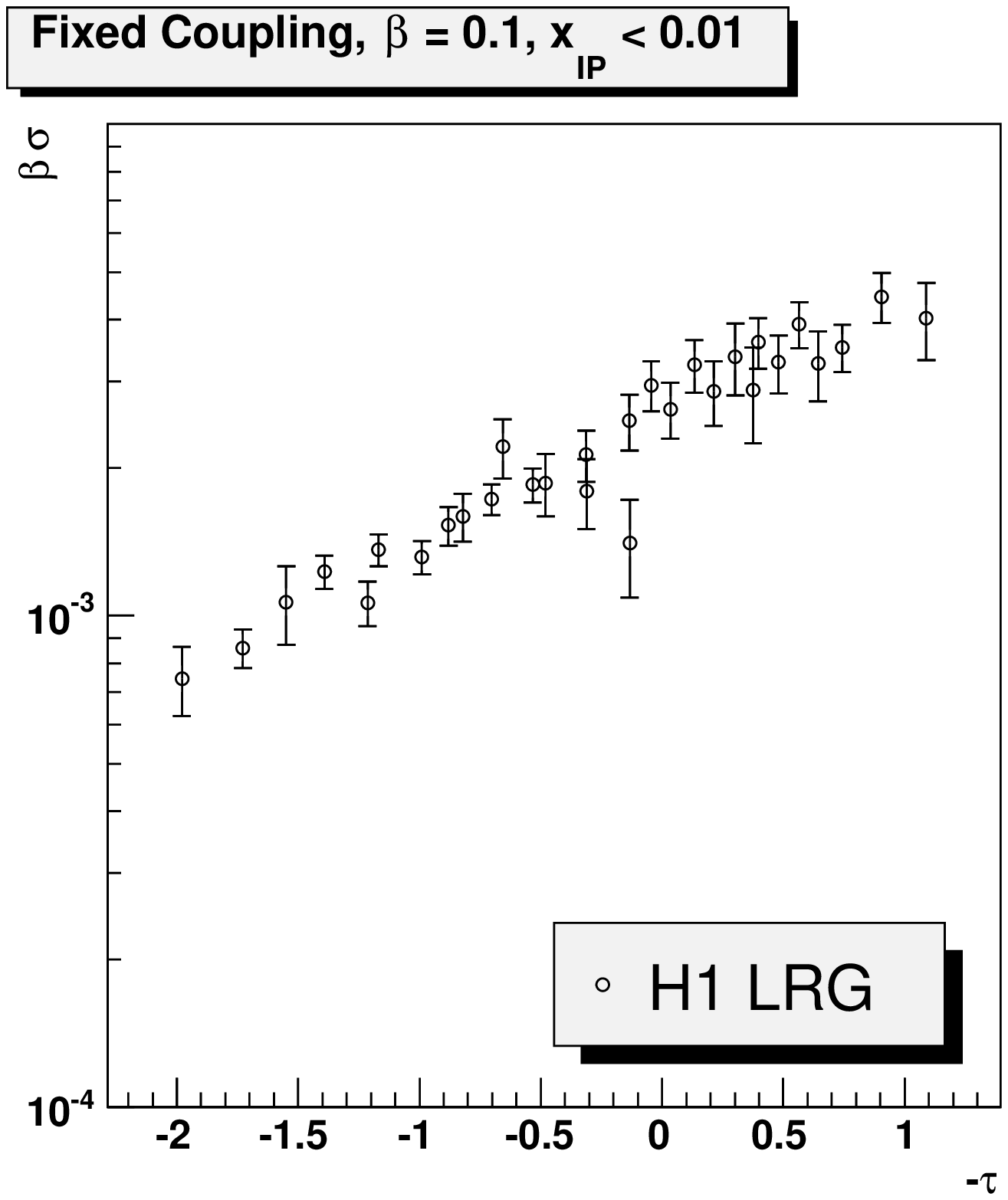,width=7.cm}\\
\epsfig{file=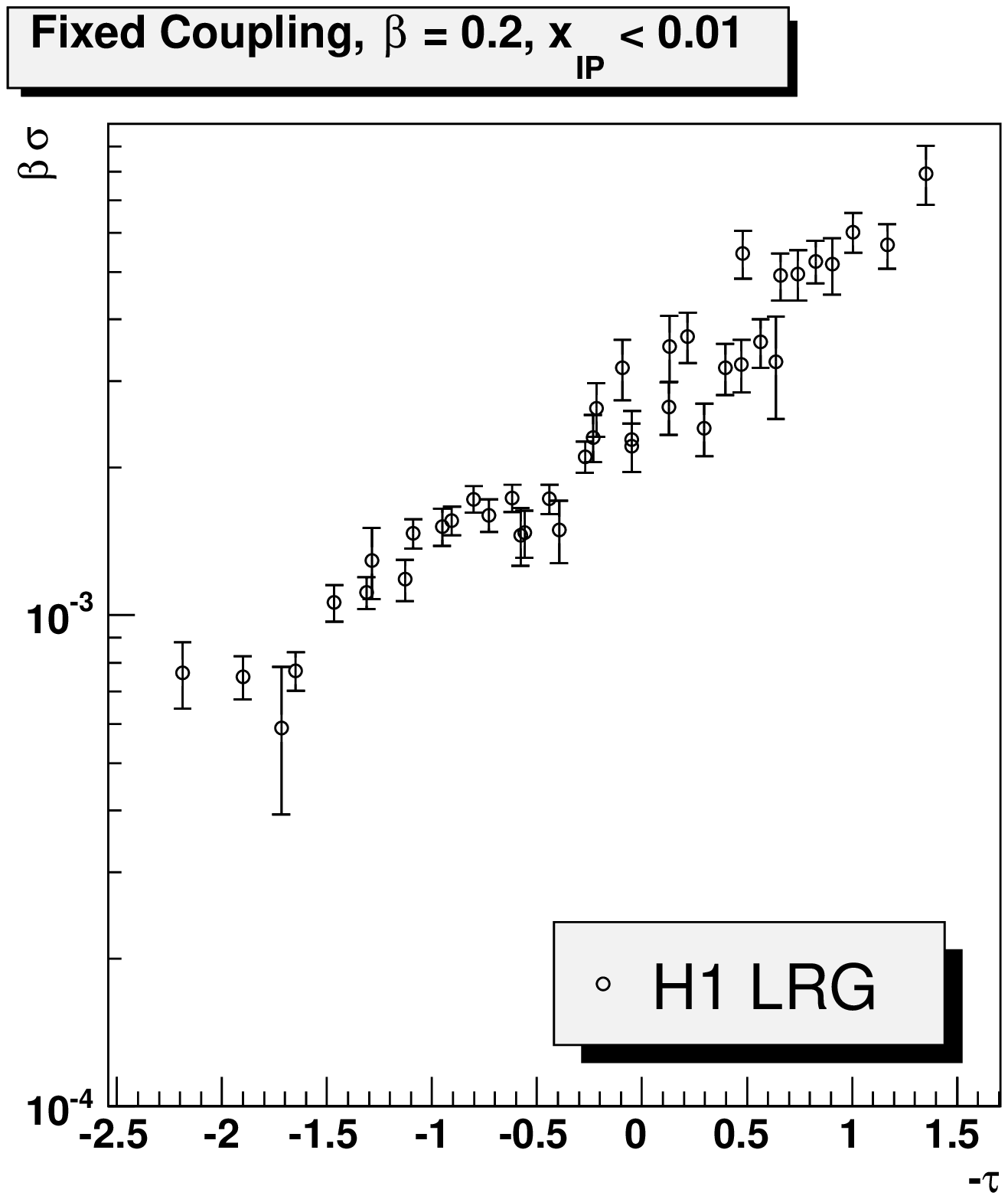,width=7.cm} &
\epsfig{file=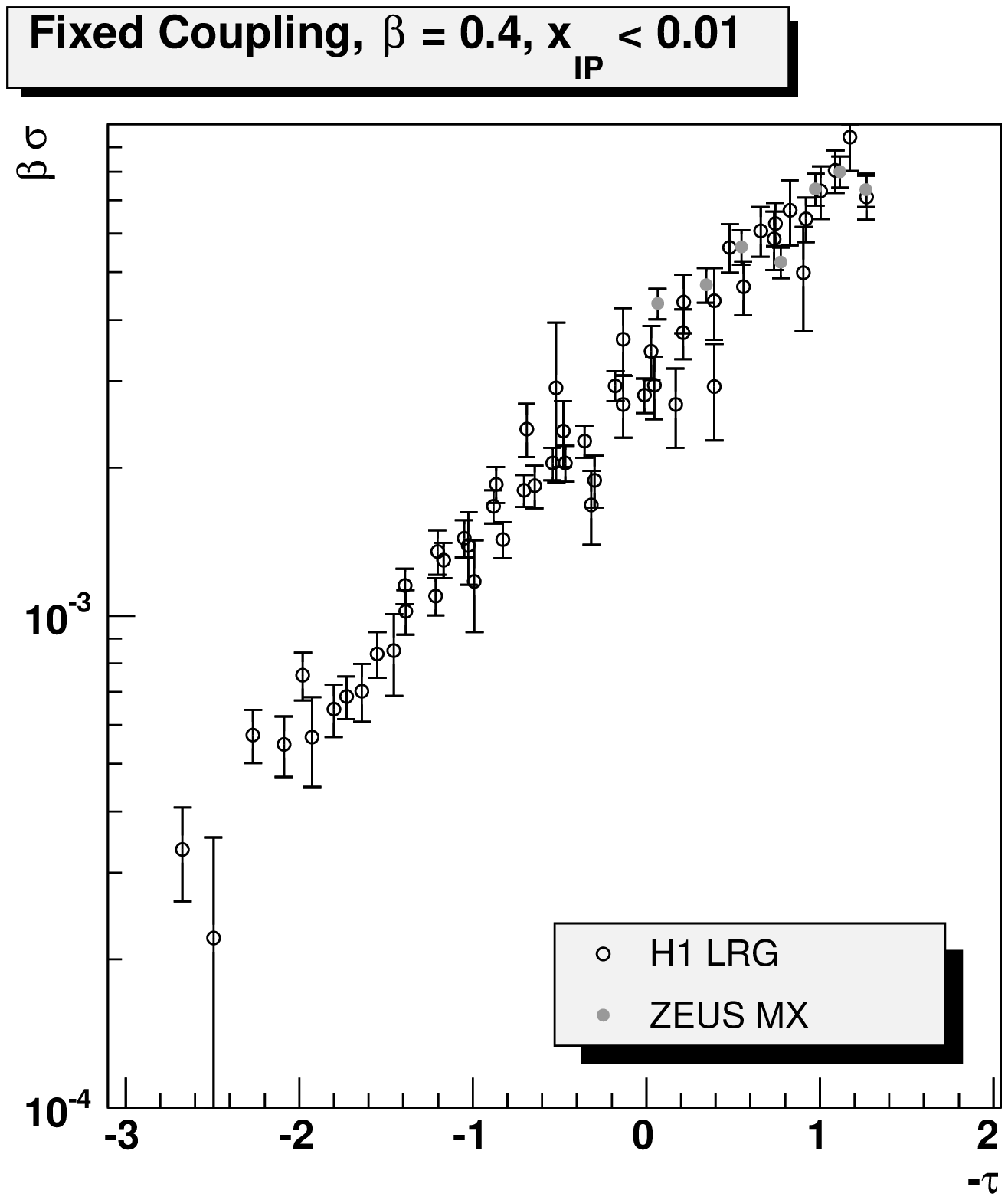,width=7.cm}\\
\epsfig{file=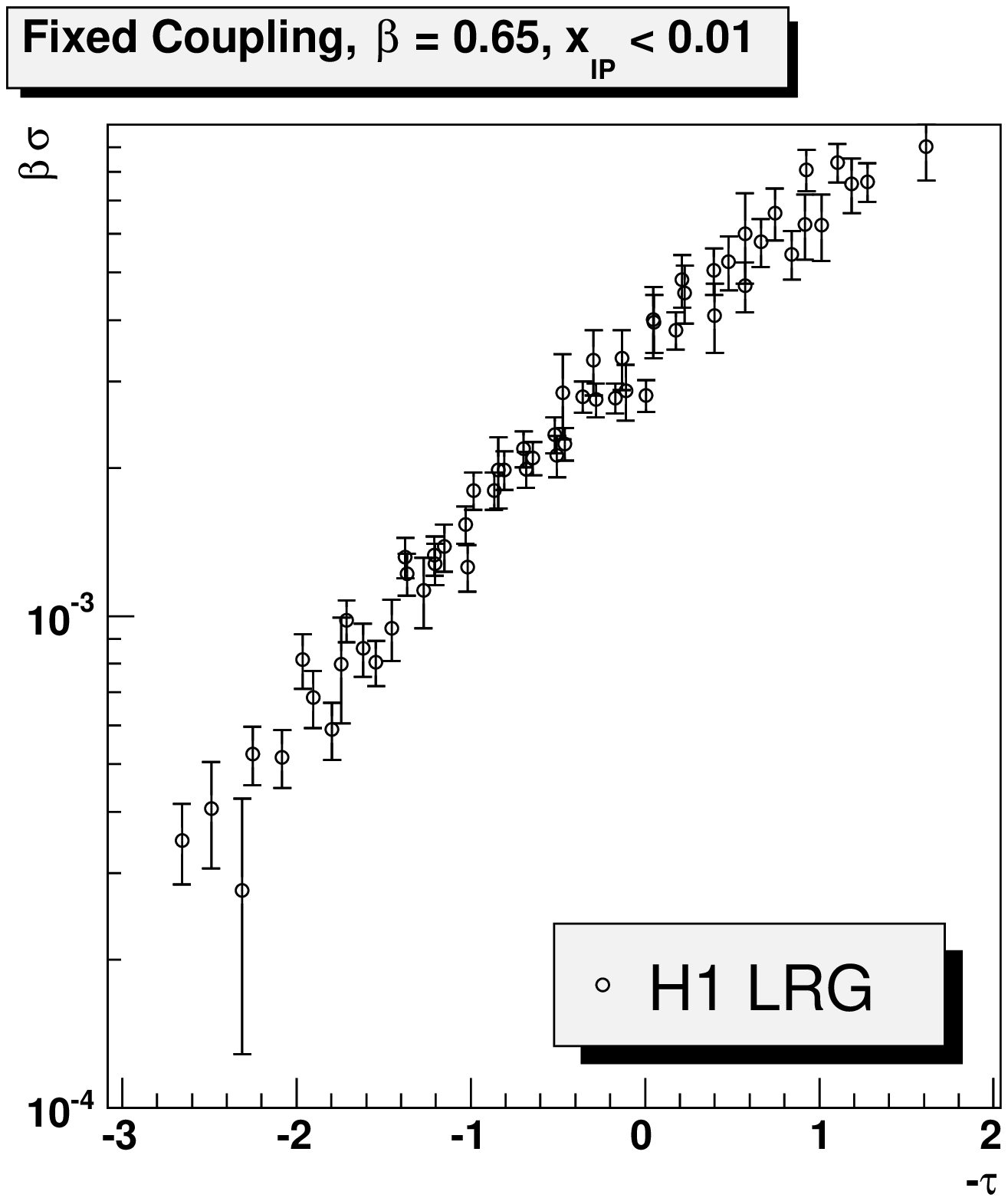,width=7.cm} &
\epsfig{file=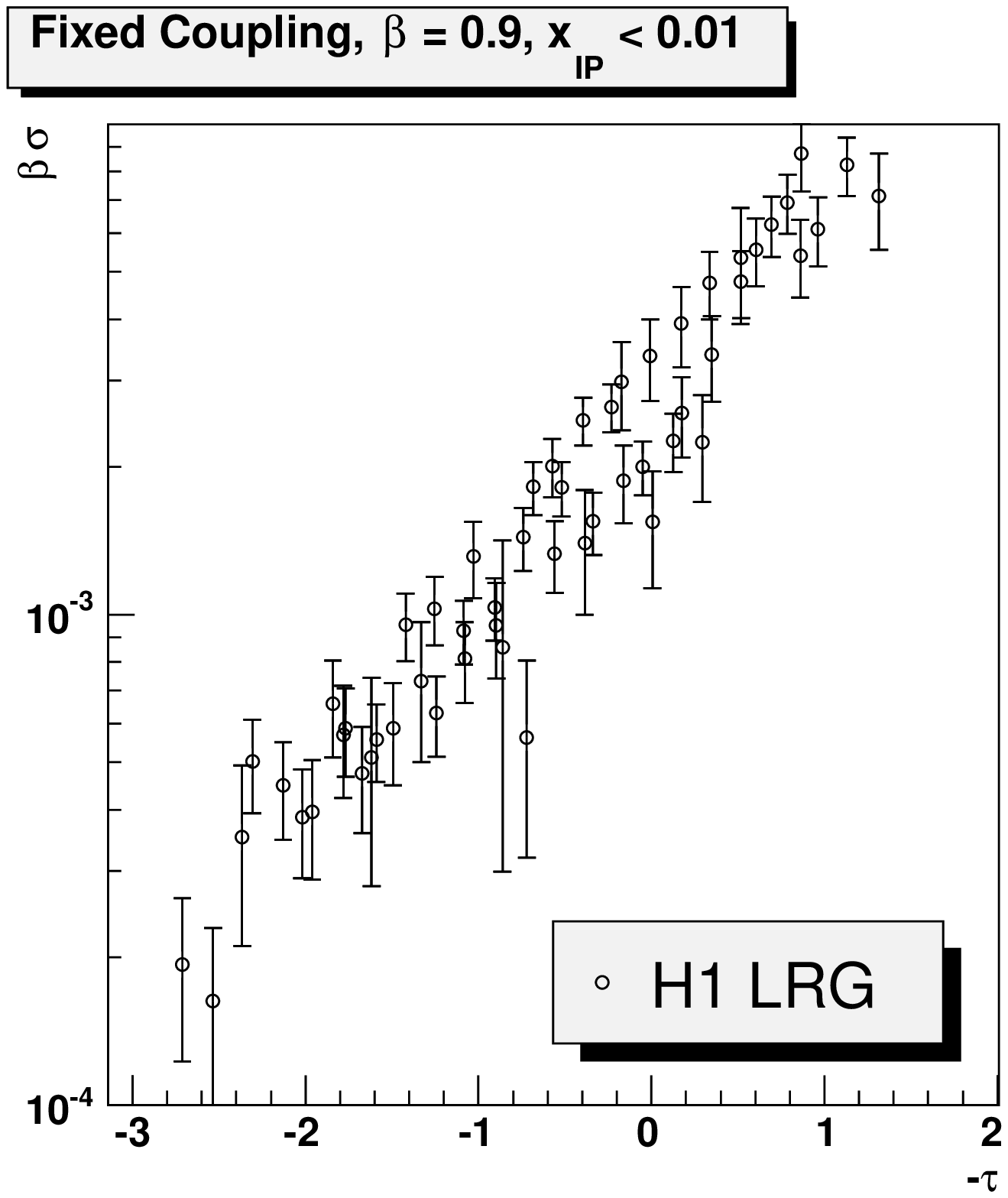,width=7.cm}\\
\end{tabular}
\caption{{\bf $F_2^D$ data:} Scaling curves obtained for fixed values of $\beta$ for $F_2^D$
and fixed coupling. The parameters
are fixed to the values obtained with a fit to $F_2$ data with $Q^2>3$ GeV$^2$.}
\label{beta_fixed}
\end{center}
\end{figure}

In figure~\ref{beta_fixed}, we show the H1 and ZEUS diffractive data
for $\beta d\sigma_{diff}^{\gamma^*p\rightarrow Xp}/d\beta$ as a function
of the fixed coupling scaling variable $\tau$ for six different fixed
values of $\beta$: 0.04, 0.1, 0.2, 0.4, 0.65, and 0.90. Only the data
points in the $Q^2$ region $[5;90]$~GeV$^2$ and with $x_{I\!\!P}<0.01$ are plotted.
The restriction to the mentioned fixed values implies that the data set
is dominated by H1 data. We just give in Fig.~\ref{beta_fixed} the results for ``Fixed Coupling"
since the other scalings give similar results which are not distinguishable given
the large uncertainties on $F_2^D$. The parameters fitted
to the H1 and ZEUS $F_2$ data are used to make the prediction
on $F_2^D$. Up to $\beta=0.65$ (the first five bins),
the data points lie on a unique curve and thus confirm the geometrical
scaling prediction. The data points at the highest $\beta$ values (low masses)
are more sensitive to low mass resonances and are not expected 
to show a perfect scaling. The values of the QF are also given in 
Table~\ref{diffbeta}. In many bins, ``Running Coupling IIbis" gives the best
QF.

\begin{figure}[t]
\begin{center}
\begin{tabular}{cc}
\epsfig{file=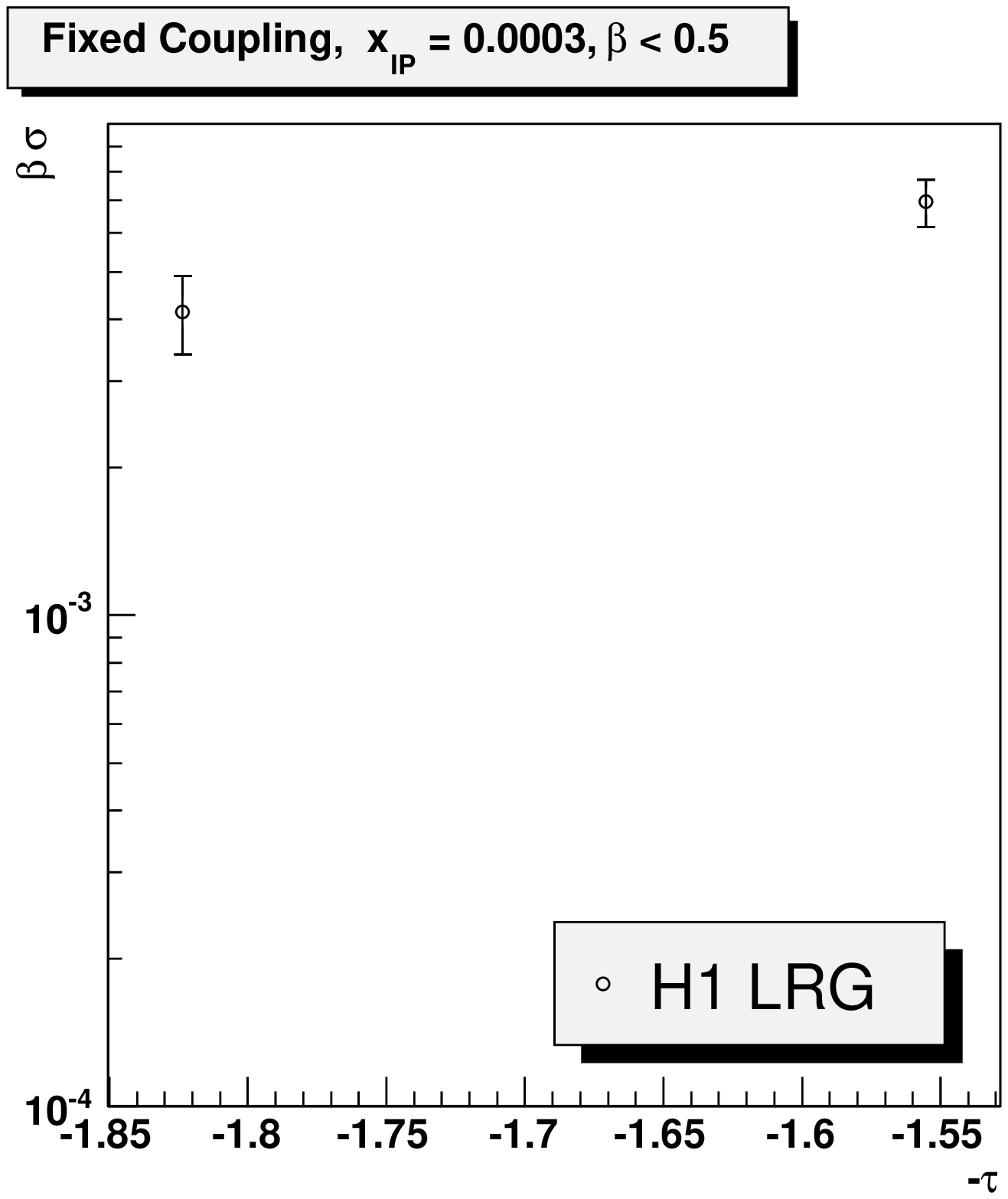,width=7.cm} &
\epsfig{file=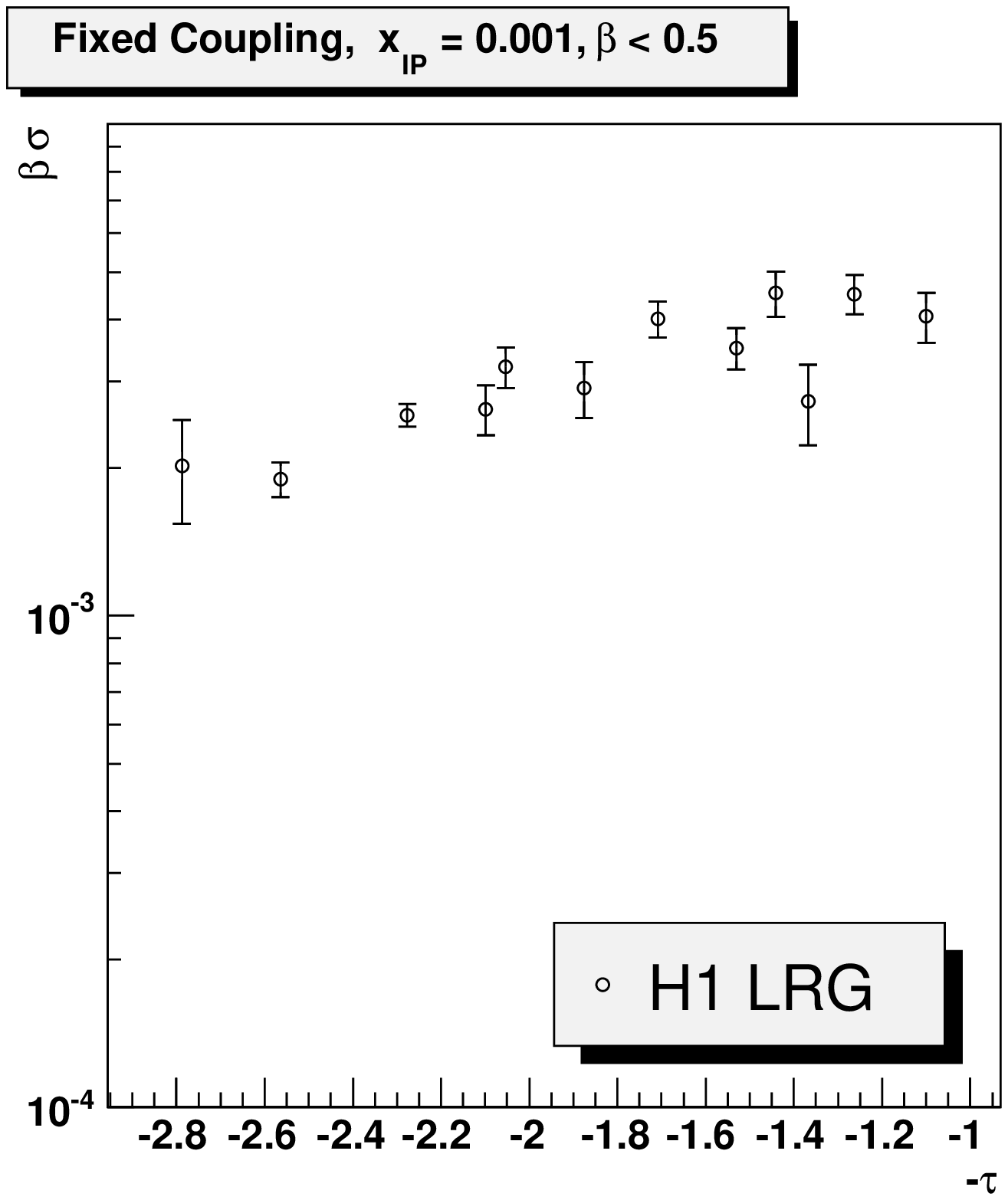,width=7.cm}\\
\epsfig{file=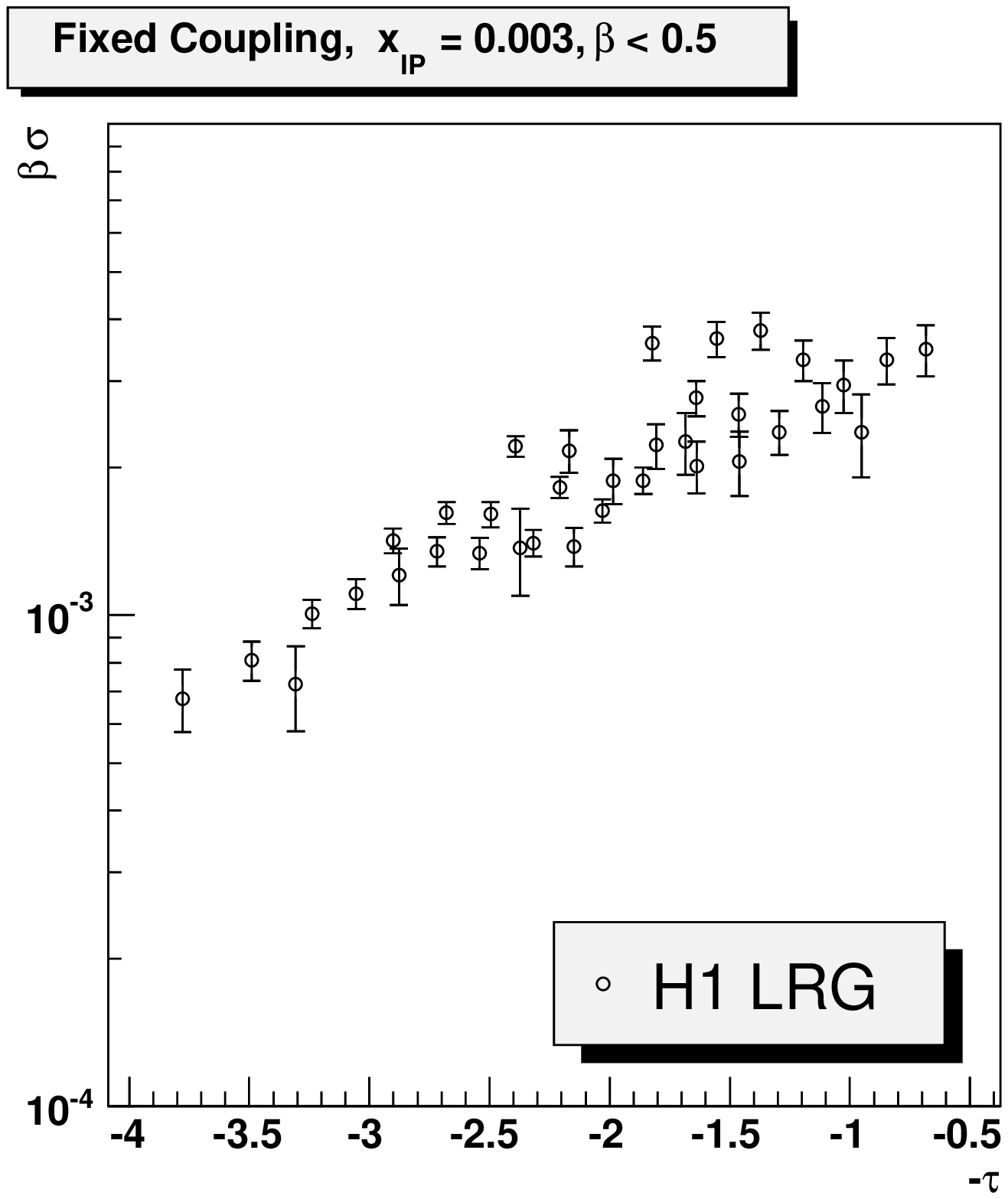,width=7.cm} &
\epsfig{file=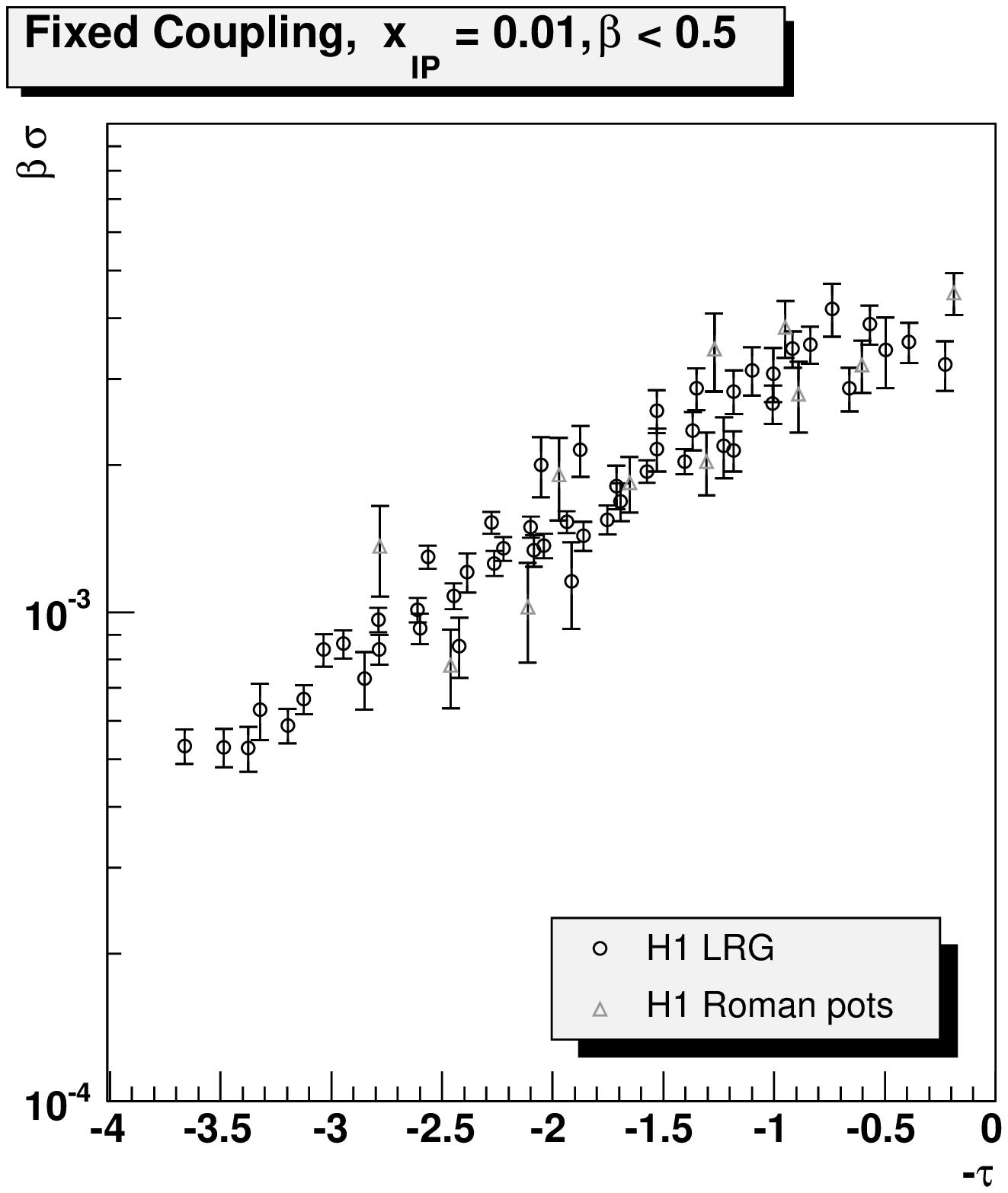,width=7.cm}\\
\epsfig{file=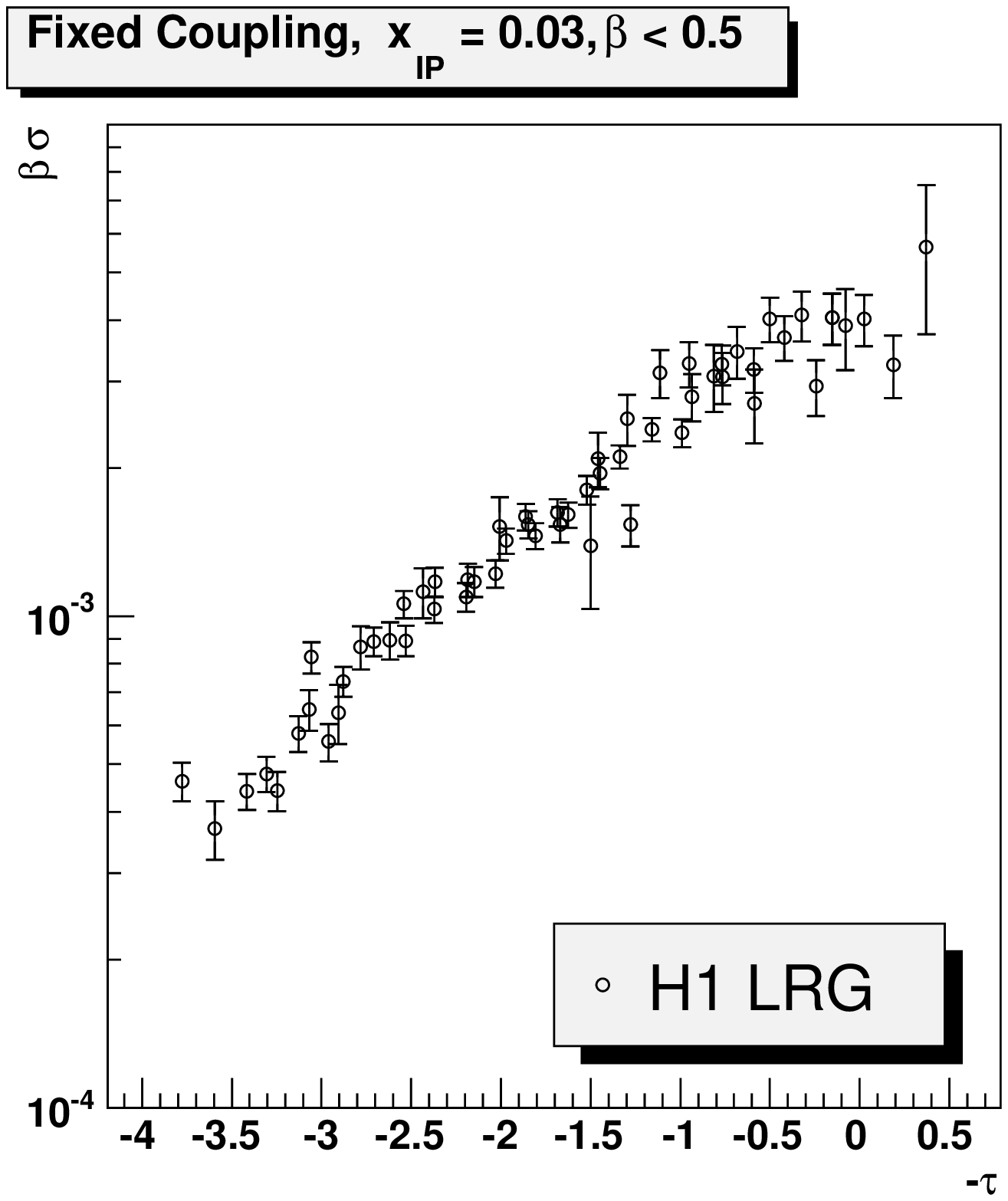,width=7.cm} &
\\
\end{tabular}
\caption{{\bf $F_2^D$ data:} Scaling curves obtained for fixed values of $\xp$ for $F_2^D$
and ``Fixed Coupling". The parameters
are fixed to the values obtained with a fit to $F_2$ data with $Q^2>3$ GeV$^2$.}
\label{xpom_fixed}
\end{center}
\end{figure}

Following the Ingelman-Schlein model of the Pomeron, we also tested the scaling
in $\beta$, $Q^2$ in the diffraction data.  The definition of the $\tau$ 
variable is the same as in Table I replacing the
variable $x$ by $\beta$, $Q^2$ remaining the same.
Figure \ref{xpom_fixed} presents the H1 and ZEUS diffractive data
in the same $Q^2$ region with $\beta<0.5$ at five different fixed
values of $x_{I\!\!P}$: 0.0003, 0.001, 0.003, 0.01, and 0.03. 
The restriction to these fixed values results in favouring the H1 LRG data.
Similarly as in the previous case, the parameter values from the fit to
H1 and ZEUS $F_2$ data are used. The scaling is definitely not as good as the one for a fixed
$\beta$.

\begin{figure}[t]
\begin{center}
\begin{tabular}{cc}
\epsfig{file=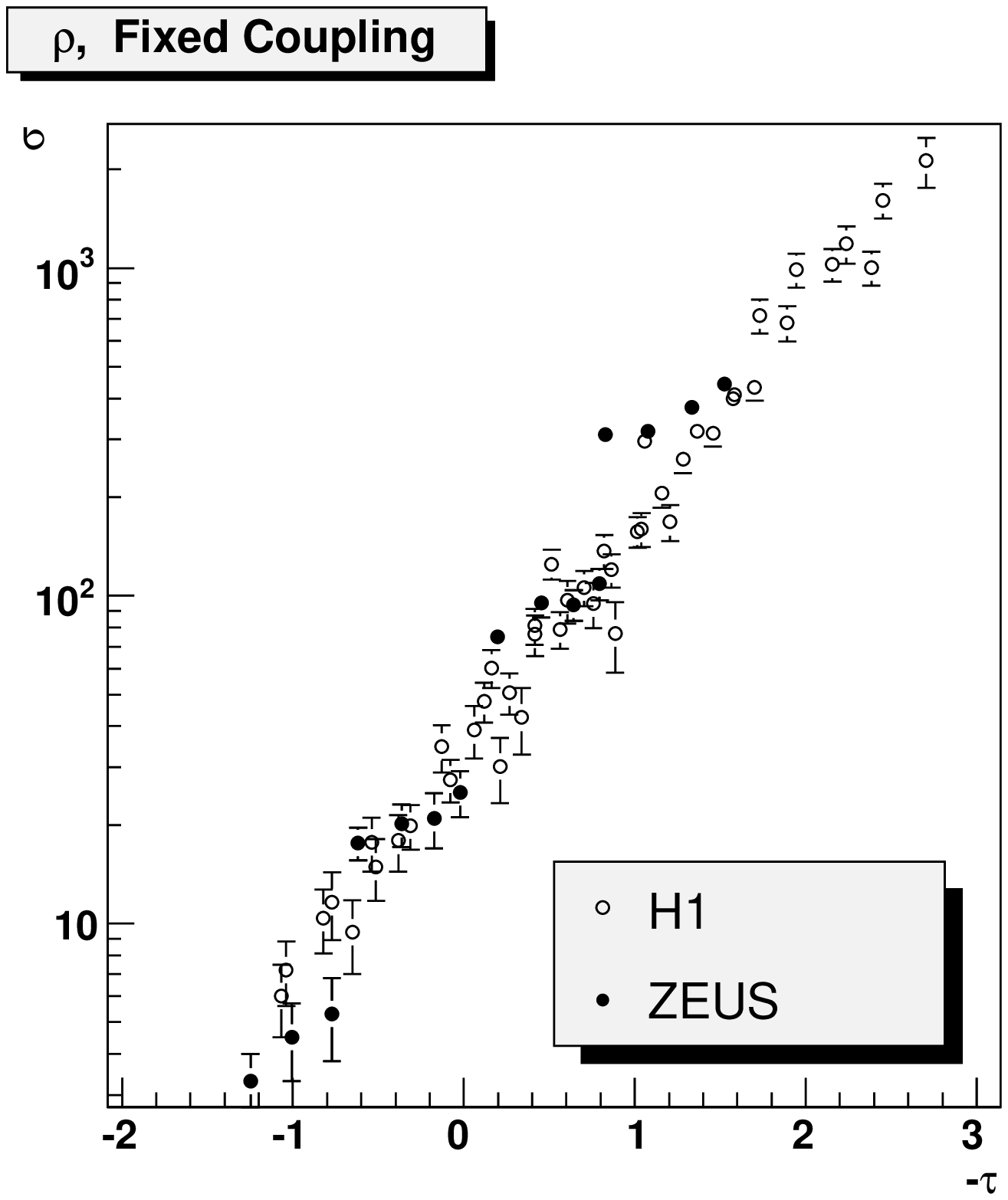,width=7.cm} &
\epsfig{file=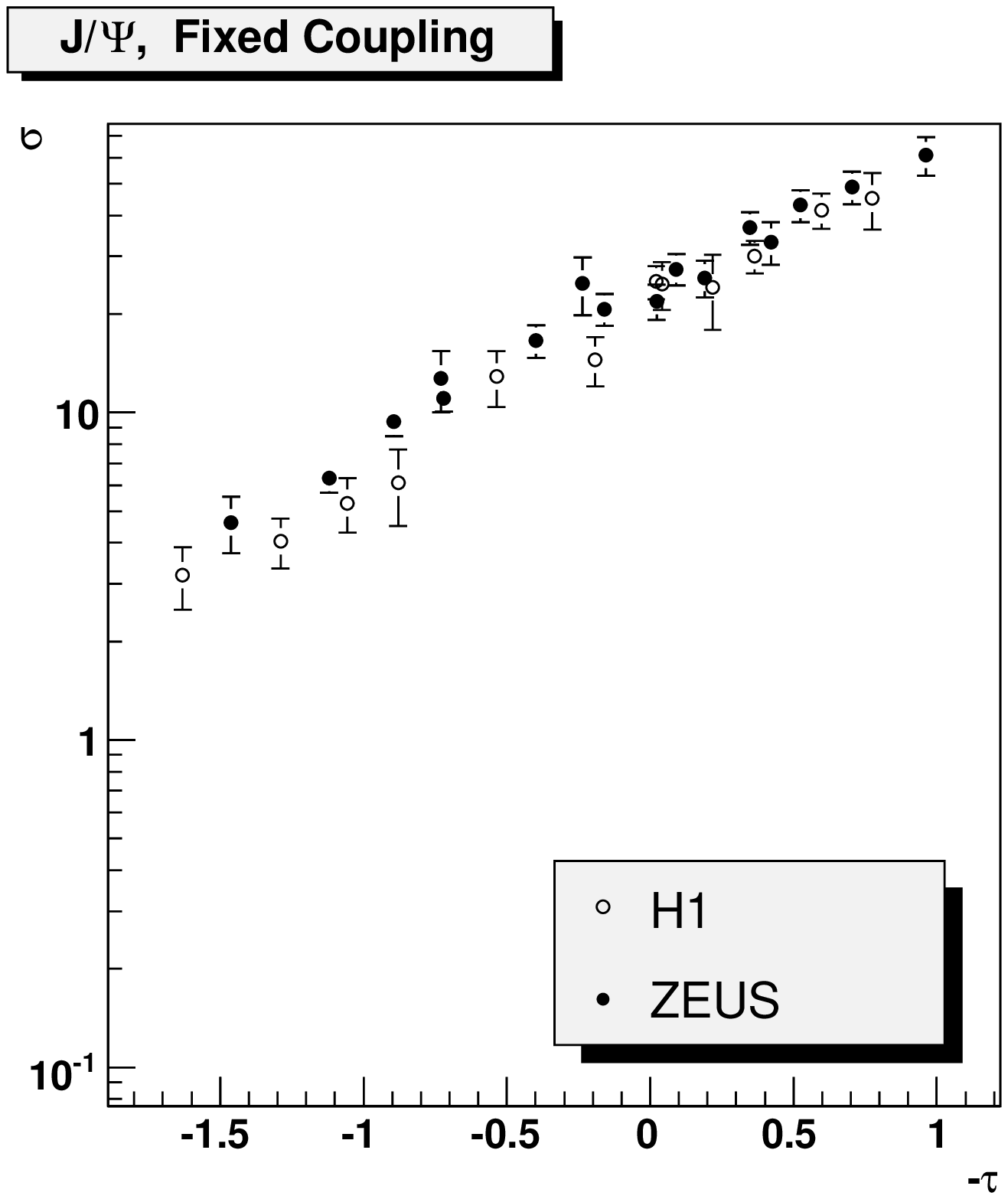,width=7.cm}\\
\epsfig{file=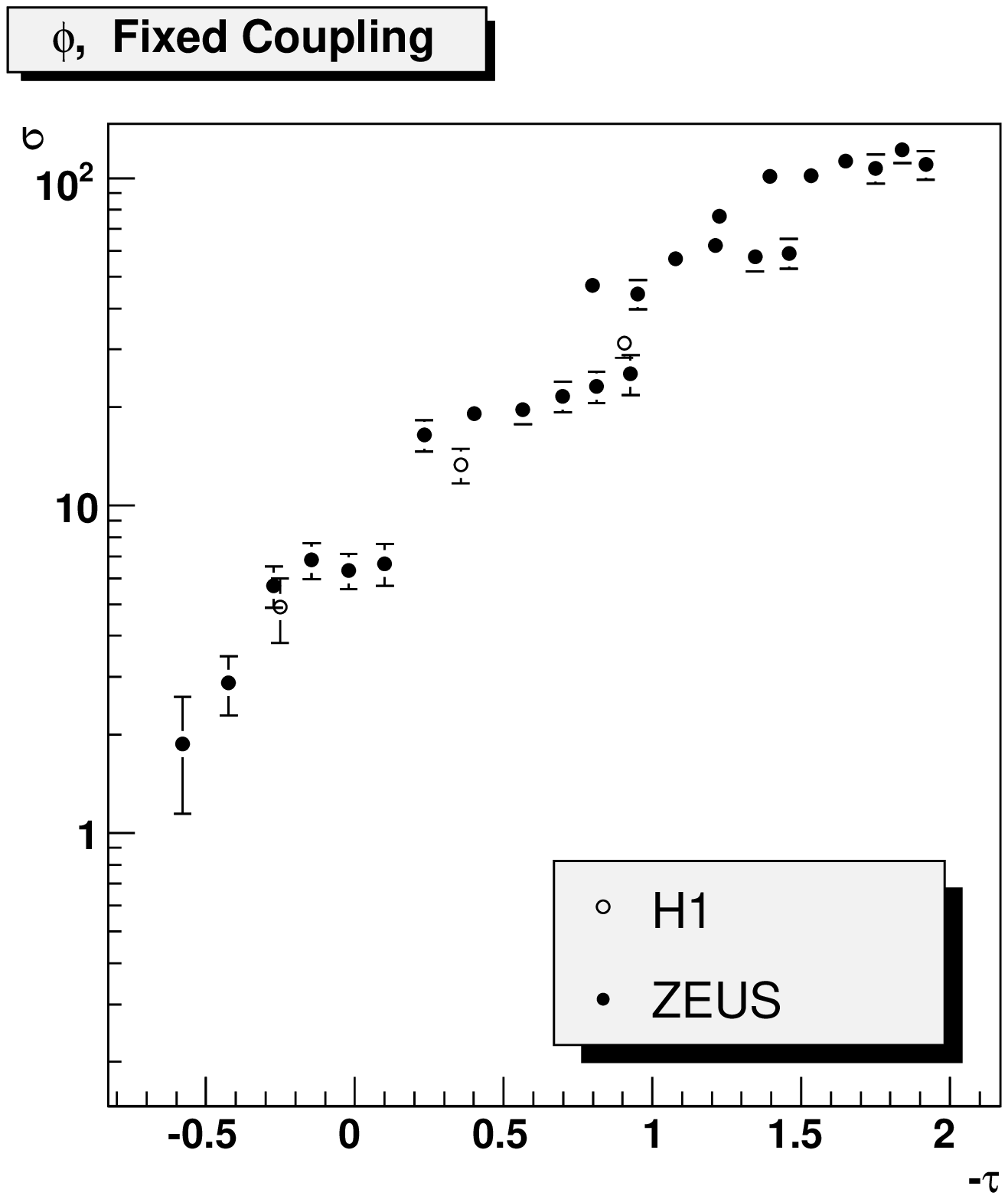,width=7.cm} &
\end{tabular}
\caption{{\bf Vector meson data:} Scaling curves obtained for fixed values of $\xp$ and for
``Fixed Coupling",
for vector meson data. 
The parameters
are fixed to the values obtained with a fit to $F_2$ data with $Q^2>3$ GeV$^2$.}
\label{vm_fixed}
\end{center}
\end{figure}

The fits to H1 and ZEUS $F_2$ data are also tested on vector meson data
from both experiments~\cite{vm}. Note the scale $M_V^2$ present in the scaling
function in (\ref{test2}). The definition of the $\tau$ variable is the same 
as in Table I replacing the
variable $Q^2$ by $Q^2+M_V^2$ $M_V^2$ being the mass of the vector meson,
$x$ remaining the same. The QFs are given in table \ref{vm_table}
and scaling curves for ``Fixed Coupling" are given in Fig.~\ref{vm_fixed}.
We only show the results for ``Fixed Coupling" since all scalings give similar 
results.
We see that the scaling is indeed verified for $\rho$, $J/\Psi$, and $\phi$ and the data
points show a tendency to be more dispersed for $\phi$. Interestingly enough, ``Diffusive
Scaling" leads to the best QF.

\section{Conclusions}
In this paper we analysed the scaling properties of deep inelastic observables 
such as the total cross section, DVCS, vector meson production, and the 
diffractive cross section, using all available low x data, \emph{i.e.} 
from the H1 \cite{H1}, ZEUS \cite{ZEUS}, NMC \cite{NMC} and E665 \cite{E665}
experiments. In order to be as model independant as possible, we used the 
quality factor method \cite{usgeom}, allowing to test the validity of scaling 
variables without any assumption concerning the shape of the scaling function. 
Four scaling variables (\ref{tau}-\ref{diffuse}), motivated by various 
approximations of QCD high energy evolution, were considered, including 
the new ``Running Coupling II'' one derived in \cite{gb}. Different versions 
of those scalings were tested, setting the rapidity shift $Y_0$ to $0$ and the 
reference scale $\Lambda$ to either
$0.2$ GeV or to $1$ GeV, or taking $\Lambda$ and 
$Y_0$ as free parameters to be fitted.

We showed that the ``Fixed Coupling'', ``Running Coupling I'' and 
``Running Coupling II'' scaling variables lead to a good scaling 
behavior of the total cross section from the $F_2$ data, with similar quality 
factors. The ``Diffusive scaling'' is disfavoured compared to them. More 
precisely, the ``Running Coupling IIbis'', with the free parameters $Y_0$ 
and $\Lambda$, gives the best scaling behavior in the range 
$3$ GeV$^2<Q^2<150$ GeV$^2$ and $x<0.01$, with reasonnable values of $Y_0$ 
and $\Lambda$ extracted from the fit of the quality factor. The ``Fixed 
Coupling'' scaling is less deteriorated than the ``Running Coupling II'' 
scaling when one adds the points with $1$ GeV$^2<Q^2<3$ GeV$^2$ to the fit, 
which corresponds to the boundary of the validity range of the 
\emph{perturbative}
scaling predictions. Dividing the data in four $Q^2$ bins, we also 
studied the $Q^2$ dependence of the optimal value for the $\lambda$ 
parameter. ``Running Coupling I'' gives the most stable values and 
``Diffusive scaling'' the less stable, and ``Running Coupling II'' gives 
stable values except at low $Q^2$. 

The quality factors of the scaling variables were fitted independantly on 
the DVCS data, and give similar results as on the total cross section. 
In particular, for each scaling variable, the 
values of the parameter $\lambda$ obtained from the fit in the two cases 
are very close. 

We used the values of the parameters obtained from a fit to $F_2$ data to test the 
various scaling variables on the diffractive cross section and vector meson 
production data. For the former observable we tested both the fixed 
$\beta$ scaling behavior in $\xp$ and the fixed $\xp$ scaling behavior in 
$\beta$. At fixed $\beta$, we found a scaling behavior up to $\beta=0.65$. For 
most $\beta$ bins the best results are given by ``Running Coupling '' scaling 
variables, and the ``Diffusive scaling'' is often disfavoured. 

At fixed $\xp$, the scaling behavior of the diffractive cross section as a 
function of $\beta$ and $Q^2$ is far less obvious. This is not a surprise, as 
not enough data is available in the genuine small $\beta$ region. A tendency 
of scaling is however observed for the $\xp=0.03$ bin. 
 
Concerning $\rho$, $J/\Psi$, and $\phi$ production, we found a 
reasonable scaling behavior for all tested scaling variables, with the hard 
scale $Q^2+M_V^2$, borrowed from vector mesons wave function studies. 
Surprisingly, the best scaling is for all three vector mesons the ``Diffusive 
scaling''.\\

As a phenomenological outlook, it seems useful to work out models dipole 
amplitude wich could incorporate the successful scaling laws following the 
example of \emph{e.g.} the Iancu Itakura Munier model \cite{IIM}. The quality 
factor method is a good tool to detect the scaling properties of data, and 
can guide the formulation of models with good $\chi^2$. Our study is a good 
incentive for the formulation of appropriate models with running coupling 
scaling. There exists phenomenological models based on fixed coupling 
diffusive scaling~\cite{Kozlov:2007wm}. The test of scaling and the 
formulation of model will be important to analyse future LHC data. 
In particular, the wider kinematical range open by the LHC could help 
disentangling the scaling laws in competition, or reveal a new scaling 
(such as the diffusive one) at higher energies.

On the theoretical ground, our phenomenological analysis can help to improve 
the theoretical analysis of scaling. Indeed, a mere comparison between the 
values of $\lambda$ and the corresponding prediction based on the leading 
order BFKL kernel shows that one should go beyond this approximation to get 
an agreement. Thus, our determination of the basic scaling parameters can help 
the theoretical QCD analysis.

\begin{table}[b]
\begin{center}
\begin{tabular}{|c||c|c||c|c|c|c|c|} \hline
 $Q^2$ & data & $n_{points}$ & FC & RC I & RC II & RC II bis & DS \\ 
\hline\hline
$Q^2 \ge 3$ & all data & 217 & $\lambda$=0.330 & $\lambda$=1.841 & 
   $\lambda$=3.436 & $\lambda$=3.905 & $\lambda$=0.362 \\
   & & & & & & $Y_0$=-1.200 & \\
   & & & & & & $\Lambda=0.300$ & \\
   & & & $QF$=1.63& $QF$=1.62 & $QF$=1.69
    & $QF$=1.82 
   & $QF$=1.44 \\
\hline
$Q^2 \ge 3$ & H1 & 87 & $\lambda$=0.331 & $\lambda$=1.736 & 
   $\lambda$=3.228 & $\lambda$=4.891 & $\lambda$=0.311 \\
   & & & & & & $Y_0$=-2.516 & \\
   & & & & & & $\Lambda=0.351$ & \\
   & & & $QF$=5.72 & $QF$=5.88 & $QF$=5.79
   & $QF$=6.31
   & $QF$=5.28 \\
\hline
$Q^2 \ge 3$ & ZEUS & 127 & $\lambda$=0.368 & $\lambda$=1.809 & 
   $\lambda$=3.395 & $\lambda$=4.327 & $\lambda$=0.366 \\
   & & & & & & $Y_0$=-1.917 & \\
   & & & & & & $\Lambda=0.203$ & \\
   & & & $QF$=3.03 & $QF$=3.08 & $QF$=3.00 
   & $QF$=3.28 
   & $QF$=2.32 \\
\hline
$Q^2 \ge 3$ & H1$+$ZEUS & 214 & $\lambda$=0.379 & $\lambda$=1.839 & 
   $\lambda$=3.436 & $\lambda$=4.147 & $\lambda$=0.321 \\
   & & & & & & $Y_0$=-1.182 & \\
   & & & & & & $\Lambda=0.333$ & \\
   & & & $QF$=1.99 & $QF$=1.76 & $QF$=1.83 
   & $QF$=2.02 
   & $QF$=1.54\\
\hline \hline
$Q^2 \ge 1$ & all data & 308 & $\lambda$=0.321 & $\lambda$=1.700 & 
   $\lambda$=2.932 & $\lambda$=3.154 & $\lambda$=0.369 \\
   & & & & & & $Y_0$=-0.199 & \\
   & & & & & & $\Lambda=0.440$ & \\
   & & & $QF$=1.30 & $QF$=1.20 & $QF$=1.07 
   & $QF$=1.27
   & $QF$=1.02 \\
\hline
$Q^2 \ge 1$ & H1 & 135 & $\lambda$=0.314 & $\lambda$=1.710 & 
   $\lambda$=3.073 & $\lambda$=3.159 & $\lambda$=0.353 \\
   & & & & & & $Y_0$=-0.367 & \\
   & & & & & & $\Lambda=0.201$ & \\
   & & & $QF$=3.43 & $QF$=3.56 & $QF$=3.51
   & $QF$=3.52
   & $QF$=2.79\\
\hline
$Q^2 \ge 1$ & ZEUS & 147 & $\lambda$=0.358 & $\lambda$=1.809 & 
   $\lambda$=3.331 & $\lambda$=3.747 & $\lambda$=0.313 \\
   & & & & & & $Y_0$=1.290 & \\
   & & & & & & $\Lambda=0.060$ & \\
   & & & $QF$=3.28 & $QF$=3.20 & $QF$=3.05
   & $QF$=3.29 
   & $QF$=2.22  \\
\hline
$Q^2 \ge 1$ & H1$+$ZEUS & 282 & $\lambda$=0.368 & $\lambda$=1.797 & 
   $\lambda$=3.226 & $\lambda$=3.918 & $\lambda$=0.367 \\
   & & & & & & $Y_0$=-1.201 & \\
   & & & & & & $\Lambda=0.225$ & \\
   & & & $QF$=1.69 & $QF$=1.71 & $QF$=1.61
   & $QF$=1.74 
   & $QF$=1.32 \\
\hline \hline
\end{tabular}
\end{center}
\caption{{\bf $F_2$ data:} QF (multiplied by $10^3$) and pamameters of the fixed coupling,
running coupling I and II, and diffusive scalings
for the different data sets. We distinguish the data sets for $Q^2>3$ and
$Q^2>1$~GeV$^2$, and we compare the fit results using the full data set,
or the H1 or ZEUS data only.}
\label{F2_table}
\end{table}

\begin{table}
\begin{center}
\begin{tabular}{|c||c||c|c|c|c|} \hline
 $Q^2$  & $n_{points}$ & FC & RC I & RC II & DS \\ 
\hline\hline
$1 \le Q^2 \le 3$  & 91 &  $\lambda$=0.279 & $\lambda$=1.603 & 
   $\lambda$=1.627 & $\lambda$=0.461 \\
   & & $QF$=0.594 & $QF$=0.575 & $QF$=0.600& 
   $QF$=0.571 \\
\hline
$3 < Q^2 \le 10$  & 98 &  $\lambda$=0.301 & $\lambda$=1.800 & 
   $\lambda$=3.219 & $\lambda$=0.357 \\
   & & $QF$=0.584 & $QF$=0.544 & $QF$=0.547 
   & $QF$=0.526 \\
\hline
$10 < Q^2 \le 35$  & 86 &  $\lambda$=0.367 & $\lambda$=1.794 & 
   $\lambda$=3.521 & $\lambda$=0.340 \\
   & & $QF$=3.53 & $QF$=3.06 & $QF$=3.22
   & $QF$=2.67 \\
\hline
$35 < Q^2 \le 150$  & 53 &  $\lambda$=0.397 & $\lambda$=1.877 & 
   $\lambda$=4.135 & $\lambda$=0.108 \\
   & & $QF$=8.33 & $QF$=8.00  & $QF$=8.26 
   & $QF$=6.37 \\
\hline

\hline \hline
\end{tabular}
\end{center}
\caption{{\bf $F_2$ data:} $\lambda$ dependence as a function of $Q^2$ - The QF
are multiplied by $10^3$ for simplification.}
\label{lambdaQ2_table}
\end{table}

\clearpage

\begin{table}
\begin{center}
\begin{tabular}{|c||c|c|c|c|c|c|} \hline
  $n_{points}$ & FC & RC I & RC II & RC II bis & DS \\ 
\hline\hline
 34 & $\lambda$=0.361 & $\lambda$=1.829 & 
   $\lambda$=3.481 & $\lambda$=5.717 & $\lambda$=0.335 \\
    & & & & $Y_0$=-1.89 & \\
    & & & & $\Lambda=0.01$ & \\
    & $QF$=3.75 & $QF$=3.62 & $QF$=3.24
    & $QF$=3.52
   & $QF$=3.38\\
\hline \hline
\end{tabular}
\end{center}
\caption{{\bf DVCS data:} Values of QF (multiplied by $10^3$) and fit parameters for the different
scalings. The paramters obtained for all scaling (except ``Running Coupling
IIbis") are close to those found for $F_2$ (see Table II).
We note that the small amount of data lead to a bad precision of the
fit for ``Running Coupling IIbis" when three parameters are used.}
\label{DVCS_table}
\end{table}

\begin{table}
\begin{center}
\begin{tabular}{|c||c||c|c|c|c|c|} \hline
 $\beta$  & $n_{points}$ & FC & RC I & RC II & RC II bis & DS \\ 
\hline\hline
0.04   & 14 &  2.54 & 2.80 & 2.91 &
 3.02 & 1.64 \\
0.1   & 30 &  0.610 & 0.579 & 0.600 &
 0.605 & 0.660 \\
0.2   & 40 &  0.951 & 1.13 & 1.20 &
 1.46 & 1.14 \\
0.4   & 64 &  1.05 & 0.952 & 0.984 &
 0.998 & 1.00 \\
0.65   & 60 &  1.34 & 1.47 & 1.93 &
 1.35 & 1.20 \\
0.9   & 59 &  0.380 & 0.510 & 0.572 &
 0.492 & 0.372 \\
\hline \hline
\end{tabular}
\end{center}
\caption{{\bf $F_2^D$ data:} Values of QF (multiplied by $10^3$) for the different scalings.
The parameters $\lambda$, $Y_0$ and $\Lambda$ are fixed
to the values obtained in the fits to $F_2$ data. Cuts on data
$\xp<0.01$, $5 \le Q^2 \le 90$ were applied.}
\label{diffbeta}
\end{table}

\begin{table}
\begin{center}
\begin{tabular}{|c||c||c|c|c|c|c|} \hline
 VM  & $n_{points}$ & FC & RC I & RC II & RC II bis & DS \\ 
\hline\hline
$J/\psi$ & 28 & 2.16 & 1.83 & 1.88 
& 1.93  & 2.50 \\
$\rho$ & 62 & 1.02 & 0.814 & 0.803 
& 0.964  & 1.28 \\
$\phi$  & 28 & 1.95 & 2.44 & 2.41 
& 3.17  & 3.18 \\
\hline \hline
\end{tabular}
\end{center}
\caption{{\bf Vector mesons data:} Values of QF (multiplied by $10^3$) for the different scalings.
The parameters $\lambda$, $Y_0$ and $\Lambda$ are fixed
to the values obtained in the fits to $F_2$ data.}
\label{vm_table}
\end{table}


\end{document}